\documentclass[12pt,letterpaper,amsmath,amssymb,final]{article}

\usepackage[top=1.0in, bottom=1.0in, left = 1.0in, right=1.0in]{geometry}

\usepackage{graphicx}
\usepackage{tabularx}
\usepackage{rotating}
\usepackage{subfig}
\usepackage{sidecap}
\usepackage{dcolumn}
\usepackage{bbm}
\usepackage{textcomp}
\usepackage{color}
\usepackage{gensymb}
\usepackage{wasysym}
\usepackage{amssymb}
\usepackage{bm}
\usepackage{array}
\usepackage{lineno}
\usepackage{hyperref}

 \def\begit{\begin{itemize}}
\def\endit{\end{itemize}}
\def\begc{\begin{center}}
\def\endc{\end{center}}
\def\begl{\begin{large}}
\def\endl{\end{large}}
\def\Begl{\begin{Large}}
\def\Endl{\end{Large}}
\def\begh{\begin{huge}\sf}
\def\endh{\end{huge}}
\def\non{\nonumber \\}

\def\begeq{\begin{equation}}
\def\endeq{\end{equation}}
\def\begeqar{\begin{eqnarray}}
\def\endeqar{\end{eqnarray}}  
\def\numu{\nu_\mu}
\def\nubar{{\overline\nu}}
\def\numubar{\overline\nu_\mu}
\def\nue{\nu_e}
\def\nuebar{\overline\nu_e}
\def\nutau{\nu_\tau}

\def\lsim{\mathrel{\rlap{\lower4pt\hbox{\hskip1pt$\sim$}}
    \raise1pt\hbox{$<$}}}         
\def\gsim{\mathrel{\rlap{\lower4pt\hbox{\hskip1pt$\sim$}}
    \raise1pt\hbox{$>$}}}         

\newcommand{\Nova}{NO\ensuremath{\nu}A}
\newcommand{\NOvA}{NO\ensuremath{\nu}A}

\begin{document}

\title{White Paper: Measuring the Neutrino Mass Hierarchy
\vspace{1.8in}}
\author{R. N. Cahn, D. A. Dwyer, S. J. Freedman$^{\dagger}$, W. C. Haxton, \vspace{0.2in}\\
R. W. Kadel, Yu. G. Kolomensky\footnote{Co-chair}, 
K. B. Luk, \vspace{0.2in}\\
P. McDonald, G. D. Orebi Gann\footnotemark[\value{footnote}], A. W. P. Poon\vspace{0.5in}\\
Lawrence Berkeley National Laboratory}


\date{\today\vspace {1.8in}}

\maketitle
\noindent $^{\dagger}$\begin{em}{Our colleague Stuart Freedman was a member of the Neutrino Hierarchy
Working Group.  His sudden passing, on November 10, 2012, was a
terrible loss to the Nuclear Science and Physics Divisions at LBNL and
to the entire international scientific community.  It was a great personal
loss to those of us who were Stuart's friends.  Stuart set
extraordinarily high standards for scientific inquiry and integrity.
We hope that these are reflected in our report.}\end{em}
\clearpage

\begin{abstract}
\em This whitepaper is a condensation of a report by a committee
appointed jointly by the Nuclear Science and Physics Divisions at
Lawrence Berkeley National Laboratory (LBNL). The goal of this study
was to identify the most promising technique(s) for resolving the
neutrino mass hierarchy.  For the most part,  we have relied on
calculations and simulations presented by the proponents of the
various experiments.  We have included evaluations of the
opportunities and challenges for these experiments based on what is
available already in the literature. 
\end{abstract}

\setcounter{tocdepth}{2}
\tableofcontents{}

\newpage
\section{Executive Summary}
The neutrinos remain the most enigmatic of the fundamental fermions and we still don't know the answers to several basic questions:  what is their absolute mass scale?  Do neutrinos violate CP?  Are neutrinos Dirac or Majorana?  
Knowledge of the neutrino mass hierarchy can help to inform each of these questions and is thus a fundamental step towards completion of the Standard Model of particle physics.
Moreover, it may lead to hints of physics beyond the Standard Model, since neutrinos may obtain their mass in a different way than other fundamental fermions.  The neutrino mass hierarchy has implications, as well,  for cosmology and for neutrinoless double beta decay.  
Though these are undeniably fundamental questions, it was outside the scope of this study to evaluate the importance of determining the neutrino mass hierarchy relative to other opportunities on a similar timescale.

Of the experiments considered, only the long baseline technique has demonstrated the ability to measure the mass hierarchy independent of oscillation parameters.  LBNE, in combination with T2K/\NOvA, promises to resolve the mass hierarchy with a significance of more than $3\sigma$ by 2030.   Hyper-Kamiokande can achieve a similar significance on a similar timescale by combining a shorter baseline measurement with atmospheric neutrino data.  The European LAGUNA-LBNO project promises an exceptional sensitivity of greater than ($5\sigma$) on a short timescale ($\approx 1$ year of data) due to its very long baseline, but the project status remains in question.

A variety of other experiments have been proposed with some sensitivity to the mass hierarchy. 
These are of interest primarily because some could be completed much more rapidly than long-baseline projects and careful attention to the design of the experiments could give them a reasonable chance of measuring the mass hierarchy.  

The most viable approaches appear to be reactor neutrinos (JUNO, formerly known as Daya Bay II) and  neutrinos in ice (PINGU at IceCube).  While requiring significant technological advances in detector design and performance, JUNO promises a potential sensitivity of more than  $3\sigma$ ($4\sigma$) assuming current (future $1.5\%$-level) uncertainties on $\Delta m^2_{32}$.  This challenging experiment appears to be on the fast track to approval in China.  PINGU offers excellent statistical sensitivity to the hierarchy, with the primary challenge lying in controlling and evaluating systematic effects.  Sensitivity estimates vary 
and are subject to the choice of oscillation parameters and hierarchy.  In a favorable scenario, a $4\sigma$ measurement could be achieved with 3 years of data; a more conservative analysis finds a $1-5\sigma$ range in sensitivity.
At the time of composing this report, these studies are still being refined.   

Future dark energy experiments such as MS-DESI (formerly BigBOSS),
Euclid, and LSST have the capability to measure the sum of the neutrino masses with precision relevant to the mass hierarchy.  Should the hierarchy be normal and the neutrino masses minimal, MS-DESI could provide an early indication and other dark energy experiments could discern this at a several-sigma level from the power spectrum on a timescale comparable to that for LBNE.

While none of these other experiments, nor current long-baseline oscillation measurements (T2K, \NOvA), is certain to be able to measure the mass hierarchy, one or more of them could do so if oscillation parameters are
 favorable.  With more probability, one might find an indication of the hierarchy at, say, a two-sigma level.   


\newpage
\section{Introduction}\label{s:intro}
\begin{figure}[!ht]
\begin{center}
\includegraphics[width=0.55\textwidth]{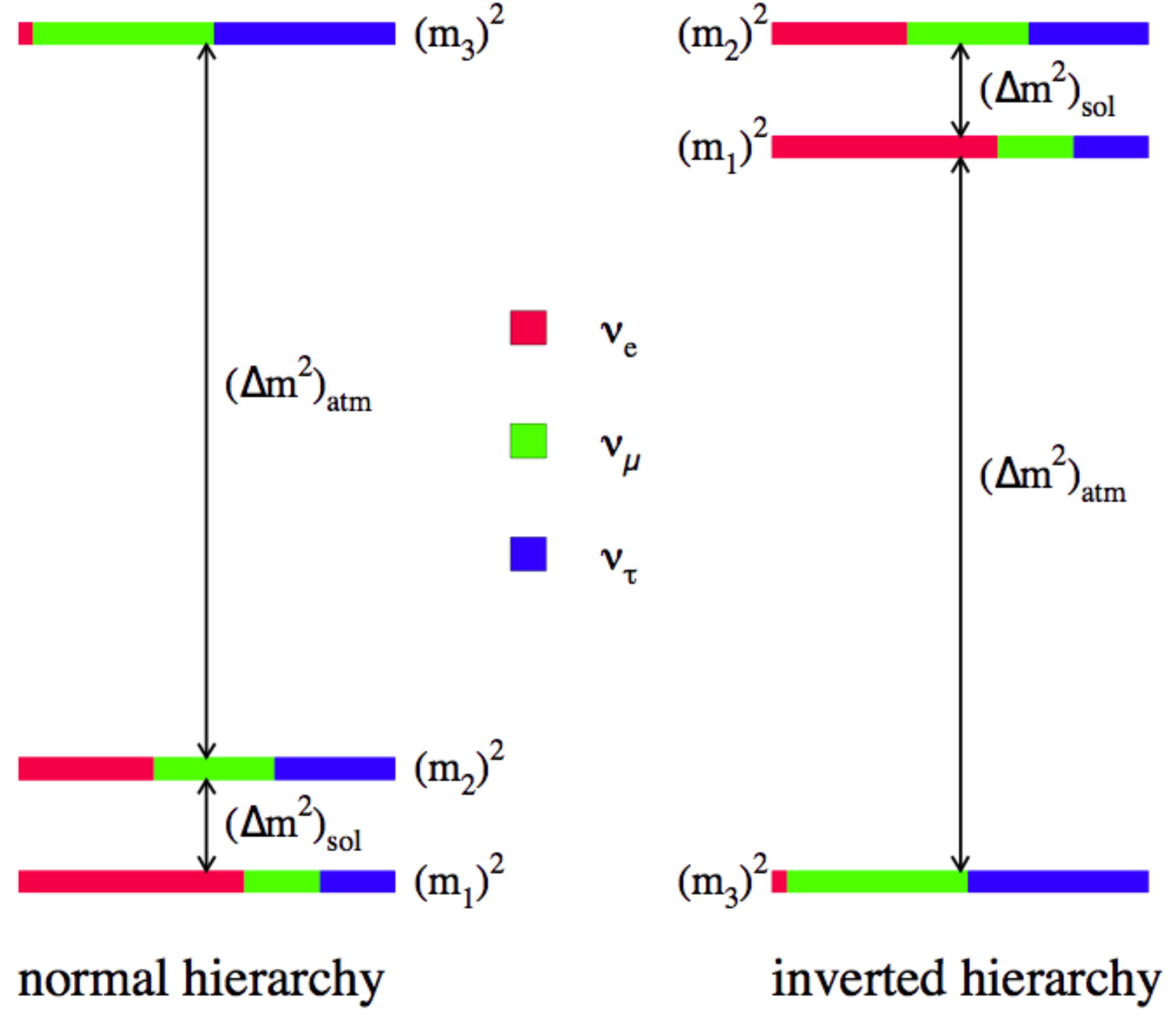}
\caption{ Pictorial representation of the possible neutrino mass hierarchies.  Note: $\Delta m^2_{atm}$ is equivalent to $\Delta m^2_{32}$ and  $\Delta m^2_{sol}$ is equivalent to $\Delta m^2_{21}$.~\cite{gouvea}.
\label{f:hierfig}}
\end{center}
\vspace{-0.2cm}
\end{figure}

The now well-accepted picture of neutrino mixing involves three underlying mass states, with three mixing angles defining the linear superpositions that make up each of the three weak, or flavor states.  The magnitude of the mass-squared splitting between states $\nu_1$ and $\nu_2$ is known from the KamLAND reactor experiment, and the much-larger splitting between  the third, $\nu_3$ state and the $\nu_1-\nu_2$ pair is known from atmospheric and long-baseline experiments.  However, pure neutrino oscillations are sensitive only to the magnitude of the mass splitting, not the sign.  
Defining the $\nu_1$ state as having the largest admixture of the electron flavor eigenstate, the sign of the mass splitting between states $\nu_2$ and $\nu_1$ is determined to be positive ($\Delta m^2_{21} > 0$) using the pattern of neutrino oscillations through the varying-density solar medium.
However, the corresponding sign of $\Delta m^2_{32}\approx \Delta m^2_{31}$ remains unknown.  That is, there are two potential orderings, or ``hierarchies'', for the neutrino mass states: the so-called ``normal hierarchy'', in which $\nu_3$ is the heaviest, and the ``inverted hierarchy'', in which $\nu_3$ is the lightest (as shown in Fig.~\ref{f:hierfig}).

\subsection{Status of Neutrino Mixing}\label{s:mix}

The relationship between neutrino flavor \{$\nu_e,\nu_\mu,\nu_\tau$ \} and mass \{ $\nu_1,\nu_2,\nu_3$ \} eigenstates is described 
by the PMNS mass matrix~\cite{Pontecorvo67,MNS}:
\begin{equation}
\left( \begin{array}{c} | \nu_e \rangle \\ | \nu_\mu \rangle \\ | \nu_\tau \rangle \end{array} \right) =
\left( \begin{array}{ccc} c_{12} c_{13} & s_{12} c_{13} & s_{13} e^{-i \delta_{CP}} \\
-s_{12}c_{23}-c_{12}s_{23} s_{13} e^{i \delta_{CP}} & c_{12}c_{23}-s_{12}s_{23}s_{13} e^{i \delta_{CP}} & s_{23} c_{13} \\
s_{12}s_{23}-c_{12}c_{23}s_{13} e^{i \delta_{CP}} & -c_{12}s_{23}-s_{12}c_{23}s_{13}e^{i \delta_{CP}} & c_{23} c_{13} \end{array}
\right) \left( \begin{array}{c} e^{i \alpha_1 /2} | \nu_1\rangle \\ e^{i \alpha_2 /2} | \nu_2 \rangle \\ | \nu_3 \rangle \end{array} \right)
\end{equation}
where $c_{ij} \equiv \cos{\theta_{ij}}$ and $s_{ij} \equiv \sin{\theta_{ij}}$.  
This matrix depends on:
three mixing angles $\theta_{12}$, $\theta_{13}$, and $\theta_{23}$, of which the first and last are the
dominant angles for solar and atmospheric oscillations, respectively; a Dirac phase $\delta_{CP}$ that
can induce CP-violating differences in the oscillation probabilities for
conjugate channels such as $\nu_\mu \rightarrow \nu_e$ versus
$\bar{\nu}_\mu \rightarrow \bar{\nu}_e$; and two Majorana phases $\alpha_1$ and $\alpha_2$
that will affect the interference among mass eigenstates in the effective neutrino mass probed
in the lepton-number-violating process of neutrinoless double $\beta$ decay.


The current best knowledge of the oscillation parameters is given in Table~\ref{tab:PDGparameters}.

\begin{table}[h]\begin{center}
\caption{Neutrino parameters from~\cite{pdg}.\label{tab:PDGparameters}}
\begin{tabular}{lr}\\ \hline \hline
Parameter & Best-fit value\\ \hline
$\sin^22\theta_{12}$&$0.857 \pm0.024 $\\
$\sin^22\theta_{23}$&$>0.95$\\
$\sin^22\theta_{13}$&$ 	0.098 \pm0.013 $\\ 
$\Delta m^2_{21}$&$7.50\pm 0.20\times 10^{-5}\,{\rm eV}^2$\\
$|\Delta m^2_{32}|$&$2.32^{+0.12}_{-0.08}\times 10^{-3}\,{\rm eV}^2$\\ \hline \hline
\end{tabular}
\end{center}
\end{table}

Using these values, we find that the first maximum of oscillation, which occurs
when
\begeq
\Delta=1.27\frac{\Delta M^2({\rm eV}^2) L({\rm km})}{E({\rm GeV})}=\frac\pi{2}
\endeq 
determines the oscillation distance according to
\begeqar
L({\rm km})&=&16\times 10^3 E({\rm GeV})\\ 
                  &=&16\  E({\rm MeV})
\endeqar in the ``solar'' case while
 \begeqar
L({\rm km})&=&540\ E({\rm GeV})
\endeqar
in the ``atmospheric'' case.  With reactor neutrinos, whose typical energy is 4 MeV, the preferred distance is about 65 km for ``solar'' oscillations, while for an accelerator experiment with a typical energy of 3 GeV, we would want a distance of about 1500 km for ``atmospheric'' oscillations, or those needed to determine the sign of the $\Delta m_{31}^2$ mass splitting.


\subsection{Motivation for Determining the Neutrino Mass Hierarchy}\label{s:mot}
Progress in neutrino physics has been impressive over recent decades,
with the discovery of atmospheric neutrino disappearance by the
Super-Kamiokande experiment, observation of solar neutrino flavor
change by the Sudbury Neutrino Observatory (SNO), and final
confirmation of the phenomenon of neutrino oscillation by the KamLAND
reactor neutrino experiment, which observed both disappearance and
reappearance.  More recently, the final mixing angle, $\theta_{13}$, 
has been determined by the Daya Bay collaboration to be large:
$\sin^22\theta_{13}=0.089\pm0.010\,\textrm{(stat)}\,\pm0.005\,\rm(syst)$~\cite{DB},
an observation subsequently confirmed by RENO~\cite{Reno} and others. 


Once we understand the ordering of the neutrino mass states, the uncertainty on a measurement of the  CP-violating phase, $\delta_{CP}$, is significantly reduced.   
Knowledge of the mass hierarchy would define the scope for future neutrinoless double beta decay ($0\nu\beta\beta$) experiments, seeking to resolve the mass nature of the neutrino, by limiting the domain for observation of a signal.  In combination with cosmological measurements, which are sensitive to the sum of neutrino masses, knowledge of the mass hierarchy could also be used to determine the absolute mass scale of neutrinos.
The mass hierarchy could also further understanding of core-collapse supernovae.  
For these many reasons, determination of the neutrino mass hierarchy is thus a fundamental step towards completion of the Standard Model of particle physics.

\section{Related Experiments that will not Resolve the Hierarchy}\label{s:not}
There are a variety of experiments subtly influenced by the neutrino
mass hierarchy.  However, the anticipated sensitivity of these
experiments is unlikely to provide useful information about the
hierarchy.  These include solar neutrino experiments, measurements of
galactic supernovae, neutrinoless double beta decay, and direct
measurement of neutrino masses.   These are summarized in Table~\ref{t:MH1}.

Solar neutrino measurements are sensitive to the sign of $\Delta m^2_{21}$. 
While the sensitivity of solar neutrino oscillations
to the sign of the remaining mass splitting, $\Delta m^2_{31}$, is negligible, precision probes of the solar
sector can provide a more detailed understanding of the interaction of neutrinos with matter,
and thus inform terrestrial experiments.

The observation of neutrinos from core-collapse supernovae would be sensitive to the mass
hierarchy via MSW effects in two ways: ``standard'' adiabatic level-crossing effects, and
collective effects in the core. The former is well understood, as evidenced by observations
in the solar neutrino sector, and predicts large rate and systematic shape
distortions, affecting the relative rates observed in different detectors. The collective effects
are less well understood, with large uncertainties and many open questions. Further work
is needed to fully understand these effects, and to determine with confidence that they do
not smear out the standard MSW effects to the point at which a hierarchy determination
becomes impossible. As such, due to the large model uncertainties and the non-predictive
nature of the source, there is an inherent difficulty in using these observations to measure the
hierarchy. Because the rate of supernovae in the galaxy is about one per century, it is also
difficult to anticipate the next experimental opportunity in this field. Instead, knowledge of
the neutrino mass hierarchy would provide valuable input to supernova modeling.

 Current direct neutrino mass measurements do not have sufficient  sensitivity to reach either the
normal or the inverted hierarchy regimes. It is not anticipated that these experiments could
reach mass scales below 0.1 eV in the foreseeable future. The primary purpose for current
experiments is to probe the mass scale in the quasi-degenerate region in a model independent
way. They will also complement neutrinoless double beta decay experiments and cosmological studies down to the $0.2$~eV level.
 
Neutrinoless double beta decay experiments are sensitive to the
neutrino mass hierarchy, but taken alone cannot provide a definitive
measurement unless a clear observation is made in an unambiguous
region of parameter space. The unambiguous region, unfortunately,
occurs at an extremely low value of the decay rate, orders of
magnitude below that accessible to current experiments.  As a
consequence, this is not considered a viable method by which to
determine the neutrino mass hierarchy; instead, the hierarchy would
provide a valuable input to this field, defining the scope for future
experiments. The next-generation experiments, planned for late 2010s
to early 2020s, will aim to have sensitivity sufficient to completely
explore the inverted hierarchy region, independently of nuclear
effects. 

\begin{table}[!htdp]
\begin{center}
\caption{Comparison of mass hierarchy (MH) experiments. \label{t:MH1}}
\begin{tabular}{ p{3.5cm }p{3cm }p{3.cm}p{4cm}p{0cm}}
\hline \hline
{\bf Technique} $\qquad\qquad$ Experiment  & \centering{MH sensitivity} & \centering{Timescale for results}    &  \centering{Major concerns} &  \\
\hline 
\bf Solar \\
All   & \centering{Zero} & \centering{Ongoing}  &  \centering{No sensitivity to sign of $\Delta m^2_{32}$} &\\
\hline
\bf Supernova \\
 Liquid argon TPC, $\quad\quad\quad$ large-scale LS, $\quad$ water Cherenkov & \centering{Model dependent} & \centering{Unpredictable}  & \centering{ Unpredictable timescale, astrophysical uncertainties} & \\
\hline
\bf Direct mass \\
 All & \centering{Zero (unless degenerate)} & \centering{$\sim$2020}  &  \centering{Only sensitive in degenerate region }& \\ 
\hline
\multicolumn{3}{l}{\bf Neutrinoless double beta decay }\\
 All & \centering{Limited by Nature}  & \centering{$\sim$2025}   &  \centering{No scope for definitive mass hierarchy measurement}& \\
\hline \hline
\end{tabular}
\end{center}
\end{table}%






\section{Long Baseline Experiments}\label{s:lb}

The long-baseline experiments have sensitivity to the neutrino mass
hierarchy, due to the interaction of neutrinos with matter as they
pass through the Earth.  The baseline itself is a critical factor in
the hierarchy sensitivity, and thus we consider the experiments in
multiple categories: near term and relatively short baseline (T2K and
\NOvA); the US-based long-baseline experiment (LBNE); and alternatives
outside the US (primarily HyperK and LBNO).  While T2K has very little
sensitivity to the hierarchy, due to the short baseline, \NOvA\ has the
potential to make a measurement at the 2--3$\sigma$ level, dependent
on the value of the CP phase parameter, $\delta_{CP}$. LBNE has
demonstrated 2$\sigma$ (3$\sigma$) sensitivity over 100\% (80\%) of
the values of $\delta_{CP}$ with 10 years of data.  LBNO has the
potential to achieve a high sensitivity on a shorter timescale, due to
the longer baseline (therefore increased sensitivity to the matter
effects).  HyperK can achieve a similar sensitivity to LBNE on a
similar timescale by combining the somewhat shorter baseline
measurement with an independent atmospheric measurement in the same
detector.  A combination of several
experiments at different baselines (e.g. T2K+\NOvA , or T2K/\NOvA +LBNE,
etc) can disentangle the competing effects of CP violation and
matter-induced neutrino-antineutrino differences, and thus improve the
constraints in the hierarchy significantly beyond any single
measurement. 

\subsection{T2K and \NOvA }\label{s:tn}

\subsubsection{Introduction}
Nearly all the neutrino parameters are already well measured, as shown in Table~\ref{tab:PDGparameters}.                 
In the absence of the matter effect, the amplitude for $\numu\to\nue$ oscillation is given by
\begeq
\langle \nu_e|\nu_\mu(t)\rangle=\Delta_{21}\sin 2\theta_{12}\cos\theta_{23}+e^{-i(\Delta_{31}+\delta)}\sin\Delta_{31}\sin 2\theta_{13}\sin\theta_{23}
\endeq
where we have assumed that $\Delta_{31}\approx \pi/2$ so that $\Delta_{21}\approx \Delta_{31}/30$ is small.  

\begin{eqnarray}
 P(\nu_\mu\to\nu_e)
&=&\sin^2 \theta_{23}\,\sin^22\theta_{13}\,\sin^2 \Delta_{31}\non
 &&\qquad\qquad +\Delta_{21}\sin 2\theta_{13}\sin 2\theta_{12}\,\sin 2 \theta_{23}\sin \Delta_{31} \cos(\Delta_{31} 
 +\delta_{CP})\non
 &&\qquad\qquad\qquad\qquad   +\Delta_{21}^2\cos^2\theta_{23}\sin^2 2\theta_{12} 
 \end{eqnarray}
 where $\Delta_{ij}=\Delta m_{ji}^2 L/(4E)$ and where  $\sin 2\theta_{13}$, $\Delta_{12}$ and 
$|\Delta m_{21}^2/\Delta m_{31}^2|$ are treated as small. In practice, experiments are designed so that oscillations are maximal, i.e. $\Delta_{31}\approx \pi/2$. For ${\overline \nu}_\mu \to {\overline\nu}_e$ the sign of $\delta_{CP}$ is reversed.  Notice that changing $\Delta_{31}$ to $-\Delta_{31}$  and $\delta_{CP}$ to $\pi-\delta_{CP}$ leaves $P$ unchanged so measuring $P(\nu_\mu\to\nu_e)$ and the corresponding probability for $\nubar$ cannot alone determine the hierarchy.  
 
If we adopt as polar coordinates
\begeq
r=\sin 2\theta_{13}; \quad \theta=\delta_{CP}
\endeq
then a result $P_\nu$ for $P(\numu\to\nue)$ gives a circle with radius squared proportional to $P_\nu$.

Similarly, the result $P_\nubar$ for $\numubar\to \nuebar$ gives a circle with a different center and radius squared proportional to $P_\nubar$.

Perfect measurements of $P(\nu_\mu\to\nu_e)$ and $P(\nubar_\mu\to\nubar_e)$ would give two intersecting circles.  The plots for normal and inverted hierarchies would be mirror images of each other.

When the matter effect is included, the result is 
\begin{eqnarray}
 &&P(\nu_\mu\to\nu_e)
=\sin^2 \theta_{23}\,\sin^22\theta_{13}\,\frac{\sin^2 (1-x)\Delta_{31}}{(1-x)^2}\non
 &&\quad +\frac{\Delta m_{21}^2}{\Delta m_{31}^2}\sin 2\theta_{13}\sin 2\theta_{12}\,\sin 2 \theta_{23}\frac{\sin[(1-x) \Delta_{31}]}{1-x}  \frac{\sin x\Delta_{31}}x\cos(\Delta_{31}+\delta_{CP})\non
 &&\qquad\qquad  +\left(\frac{\Delta m_{21}^2}{\Delta m_{31}^2}\right)^2
 \cos^2\theta_{23}\sin^2 2\theta_{12} \frac{\sin^2( x\Delta_{31})}{x^2} 
 \end{eqnarray}
where $x=2\sqrt 2 G_F N_e E/\Delta m_{31}^2$ and where non-leading terms in $\Delta m_{21}^2/\Delta m_{31}^2$ and $\theta_{13}$ have been neglected.  



For antineutrino scattering, again the sign of $\delta_{CP}$ changes, but also the sign of $x$ is reversed because the effective potential for neutral current scattering of $\nuebar$ is the negative of that for $\nue$.

For any given experiment with data for both $P(\nu_\mu\to\nu_e)$ and $P(\nubar_\mu\to\nubar_e)$, we can draw two circles to represent these data.  Of course, in practice the circles would have  thicknesses indicative of the uncertainties  in the measurement.  In addition, a circle could be drawn, with center at the origin, for the measurement of $\sin^2 2\theta_{13}$ from reactor experiments.  Two plots would be made, one on the assumption of normal hierarchy, the other assuming inverted hierarchy.  In at least one of the two there should be a solution where the three circles intersect.  

As we see in the examples below,  the centers of the circles are such that $\delta_{CP}=\pi/2$ and $\delta_{CP}=3\pi/2$ form the extreme situations, making either maximal or minimal separation between the circles.  
 With increasing length $L$, the discriminating power increases as the radii of the circles grow and shrink in the different hierarchies.

\subsubsection{T2K}
\begin{figure}\begin{center}
\includegraphics[width=2.75in,angle=90]{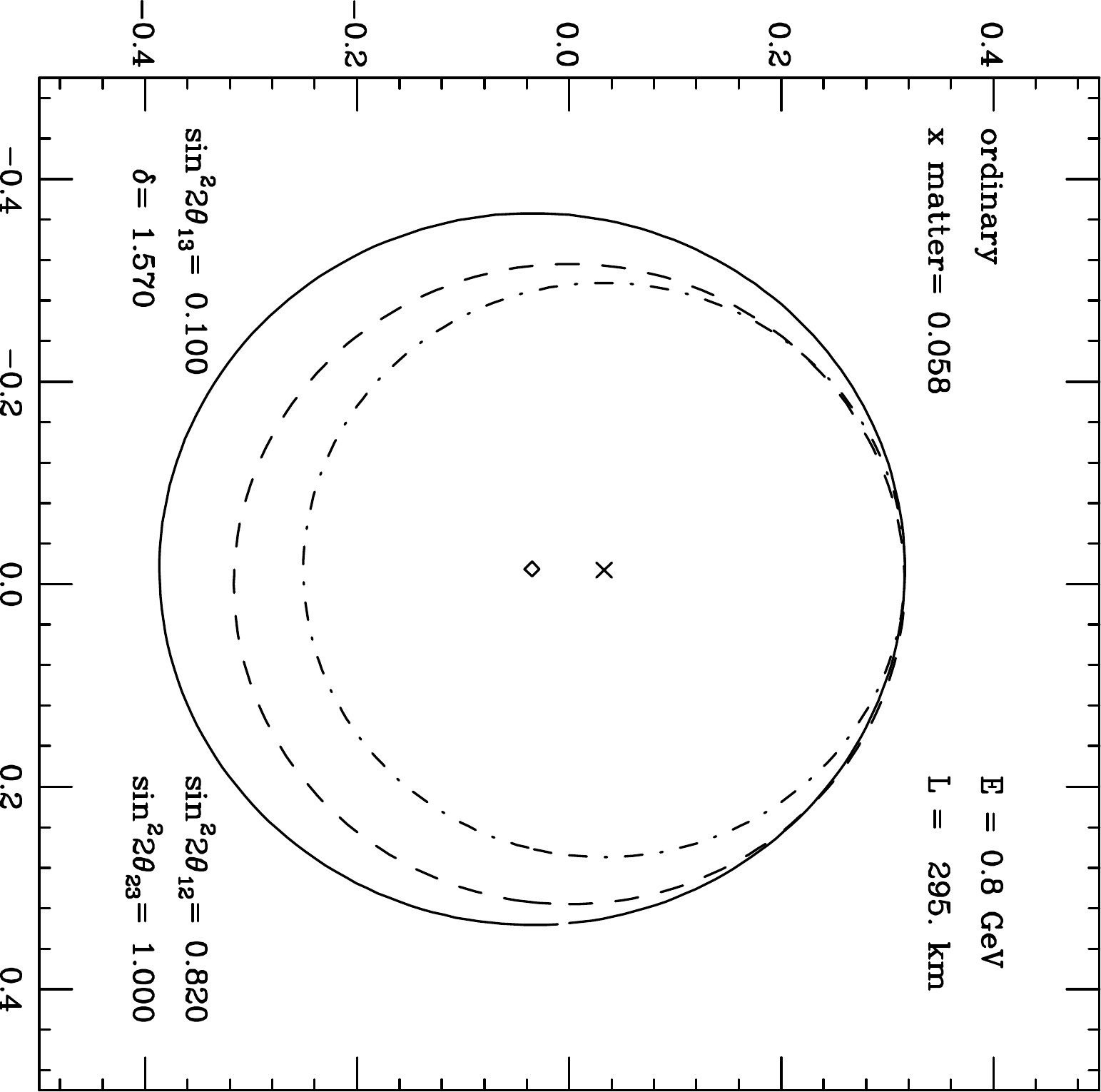}
\includegraphics[width=2.75in,angle=90]{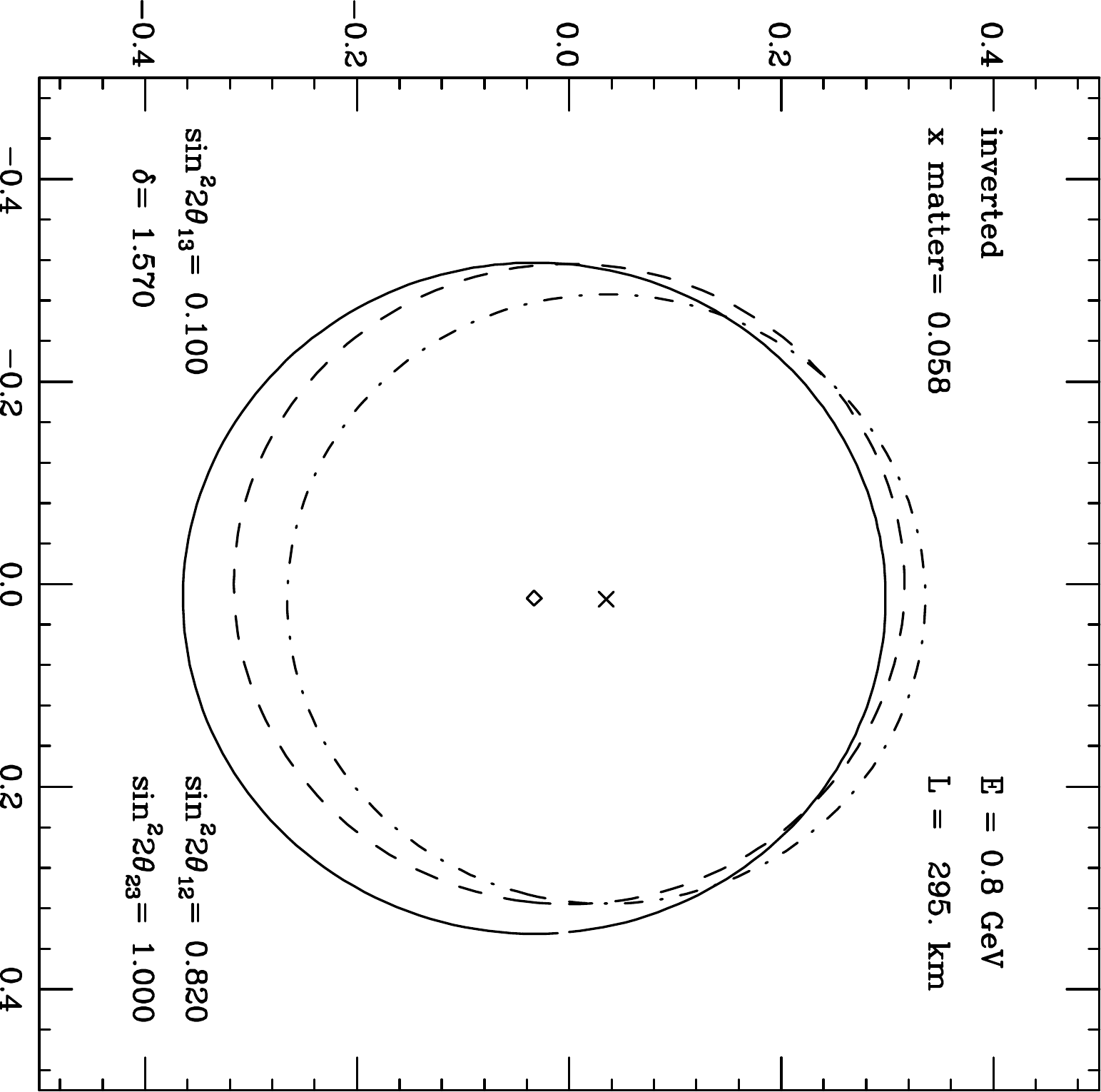}

\includegraphics[width=2.75in,angle=90]{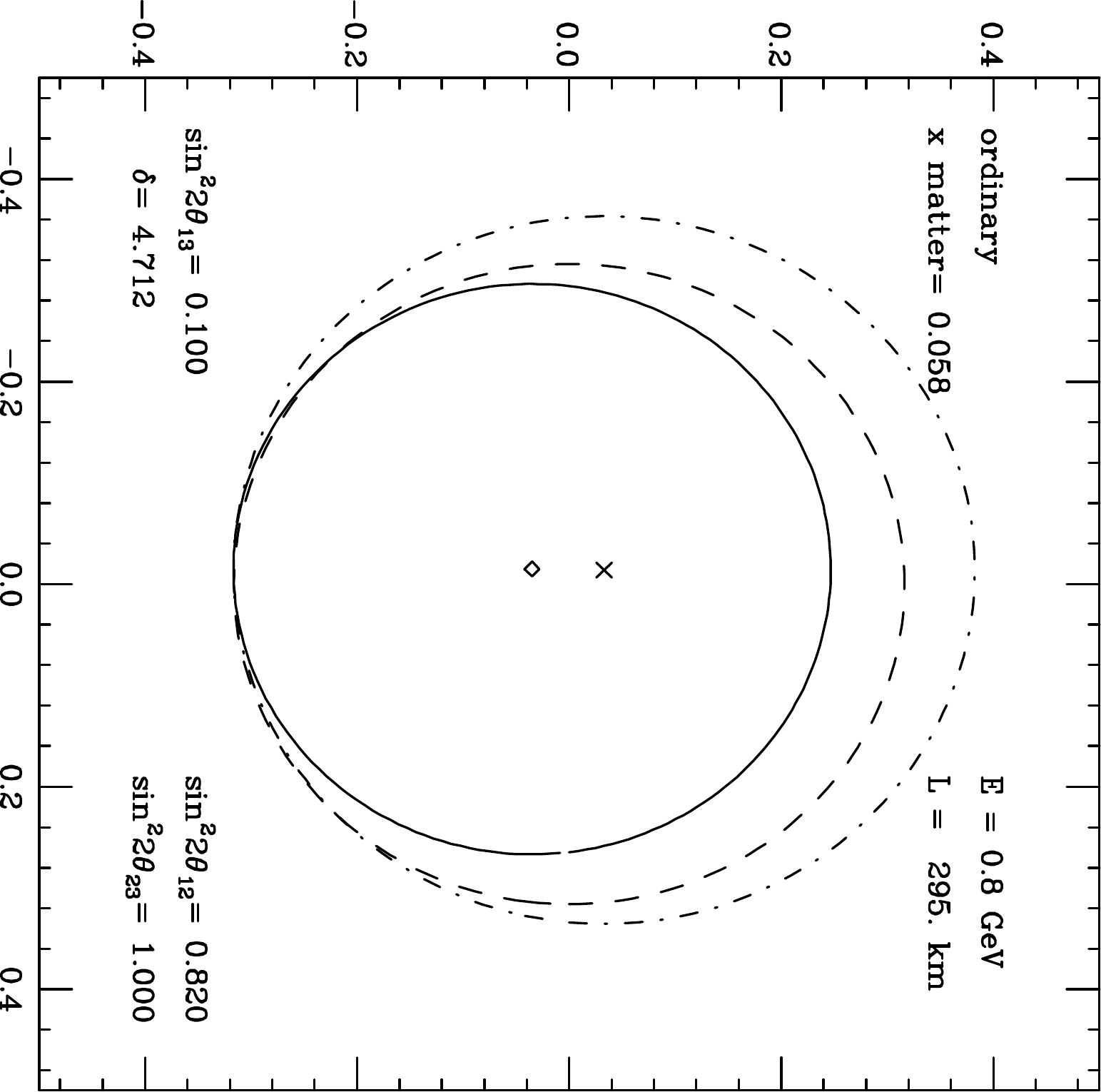}
\includegraphics[width=2.75in,angle=90]{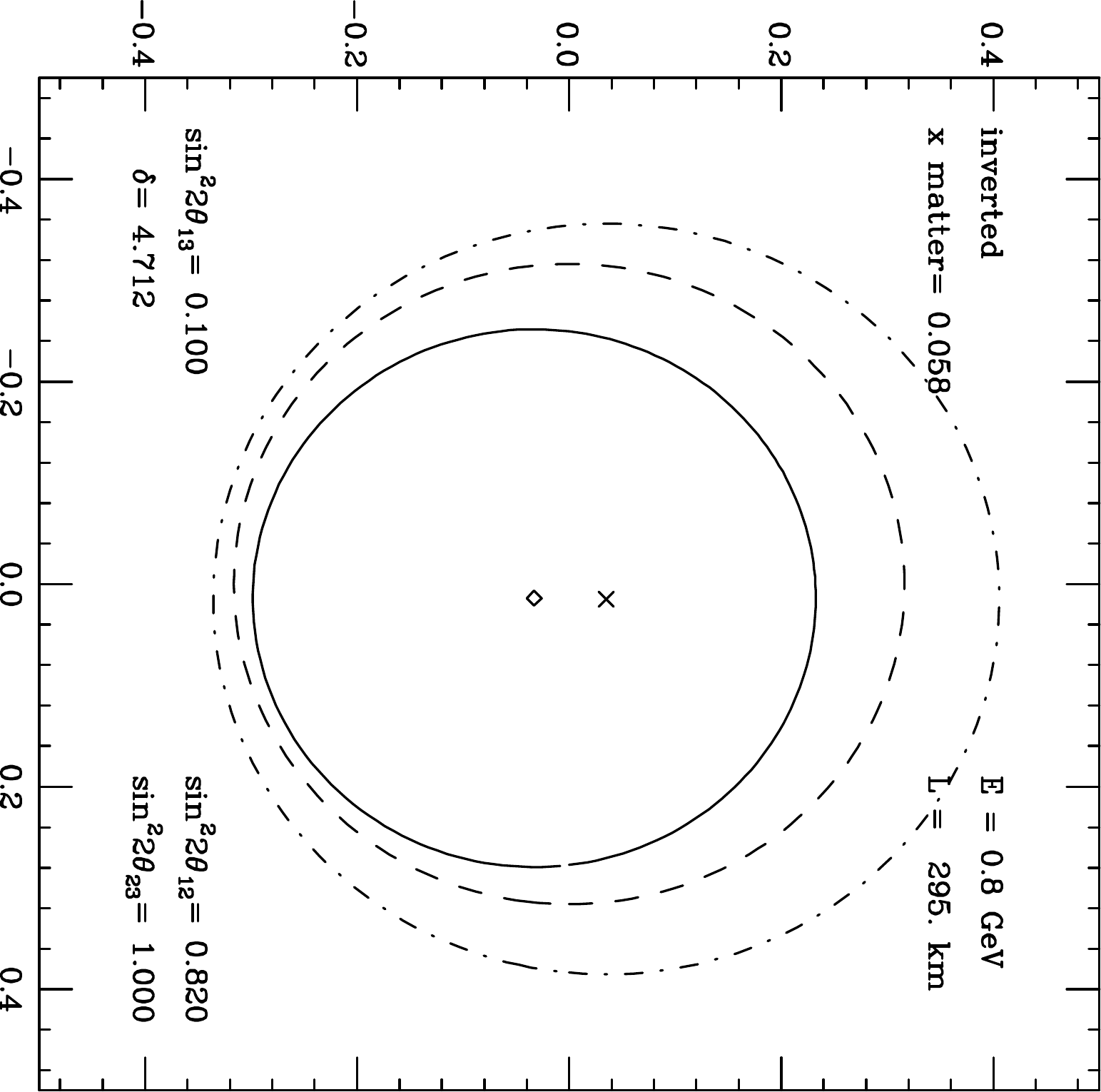}
\caption{ Plots for T2K.  Upper, with true hierarchy being normal, with $\delta_{CP}=\pi/2$.  Lower, again with true hierarchy being normal, but now with $\delta_{CP}=3\pi/2$. The dot dash circles show the constraints from $\nu_\mu$ scattering, while the solid circles are for $\nubar_\mu$.  The dashed circles show the constraints from a hypothetical perfect measurement of $\sin^22\theta_{13}$.\label{fig_four}}
\end{center}\end{figure}

T2K runs 295 km from J-PARC in Tokai to Super-Kamiokande.   In Fig.~\ref{fig_four} we see that the low energy of the T2K beam makes it impossible to determine the hierarchy if $\delta_{CP}=\pi/2$, and even for $\delta_{CP}=3\pi/2$, superb precision would be required.

\subsubsection{\NOvA}

\begin{figure}\begin{center}
\includegraphics[width=2.75in,angle=90]{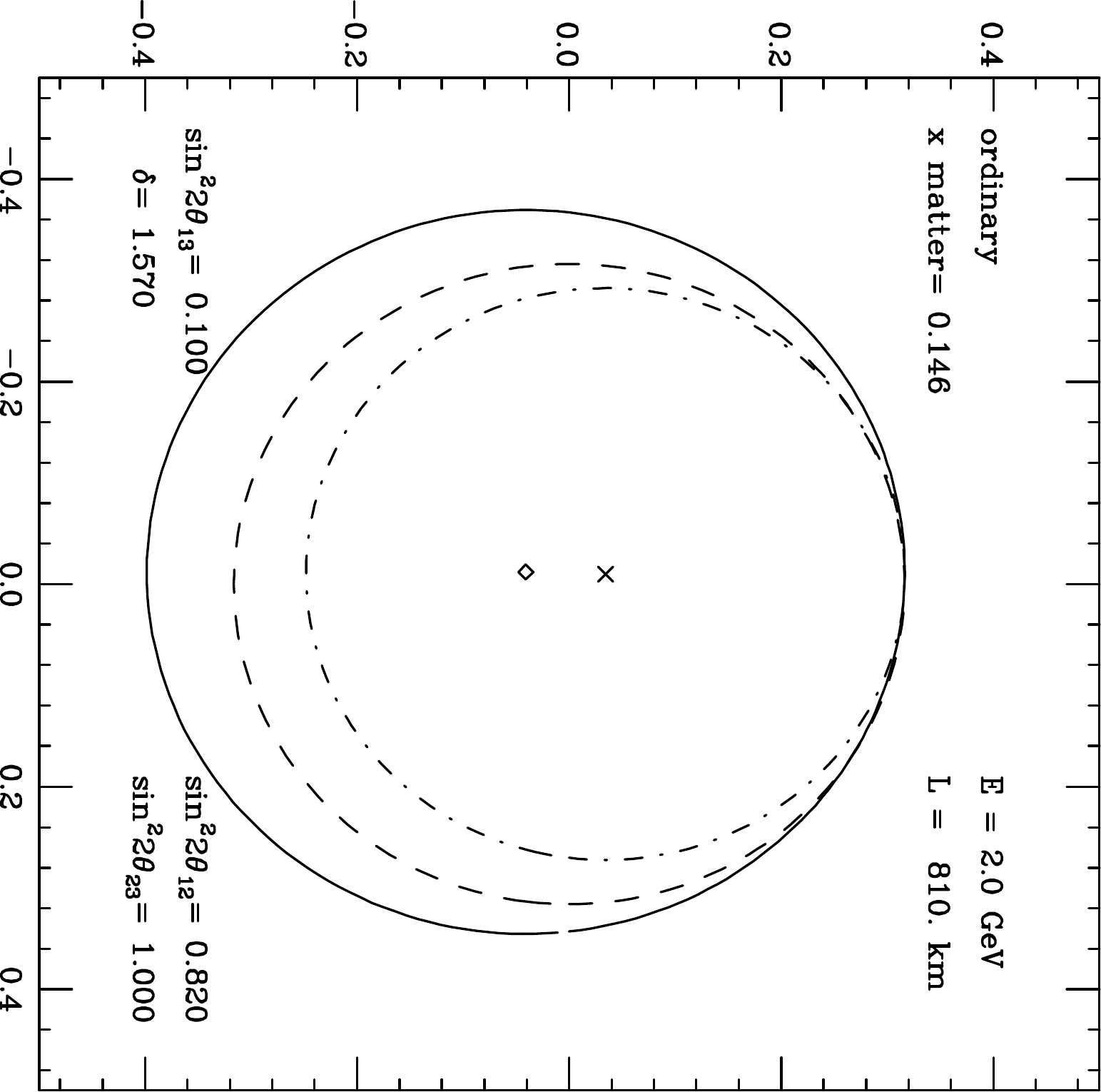}
\includegraphics[width=2.75in,angle=90]{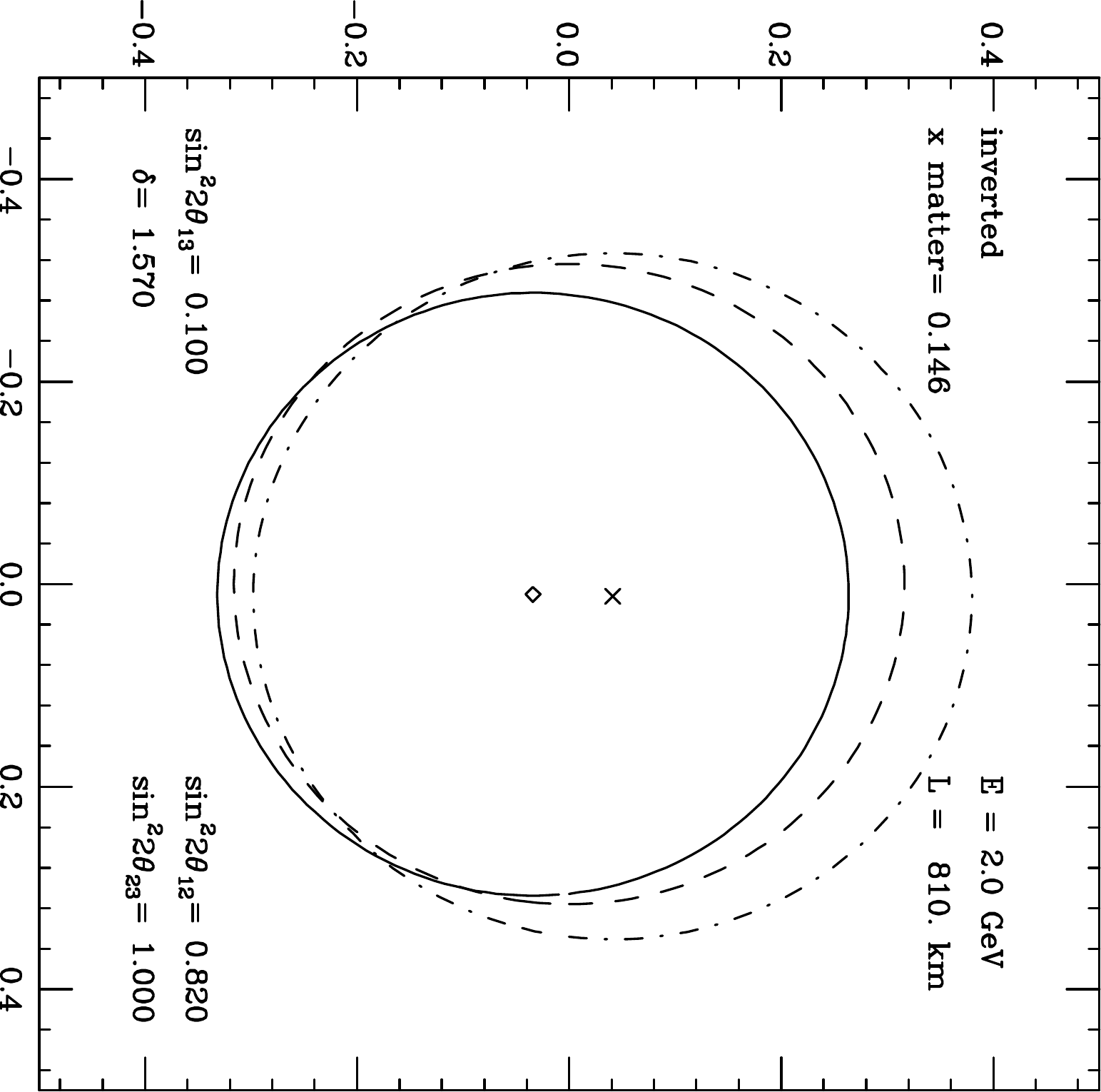}

\includegraphics[width=2.75in,angle=90]{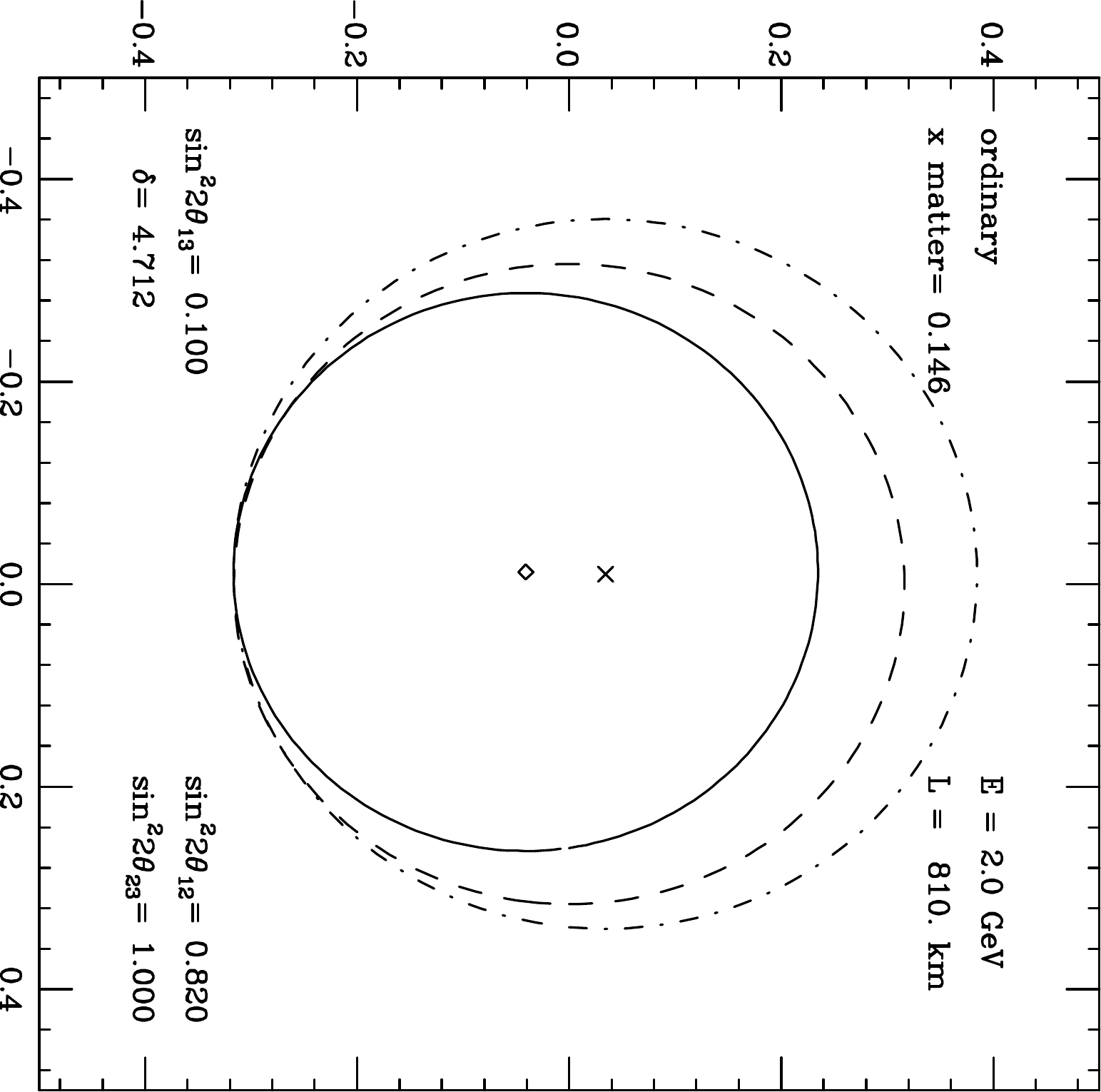}
\includegraphics[width=2.75in,angle=90]{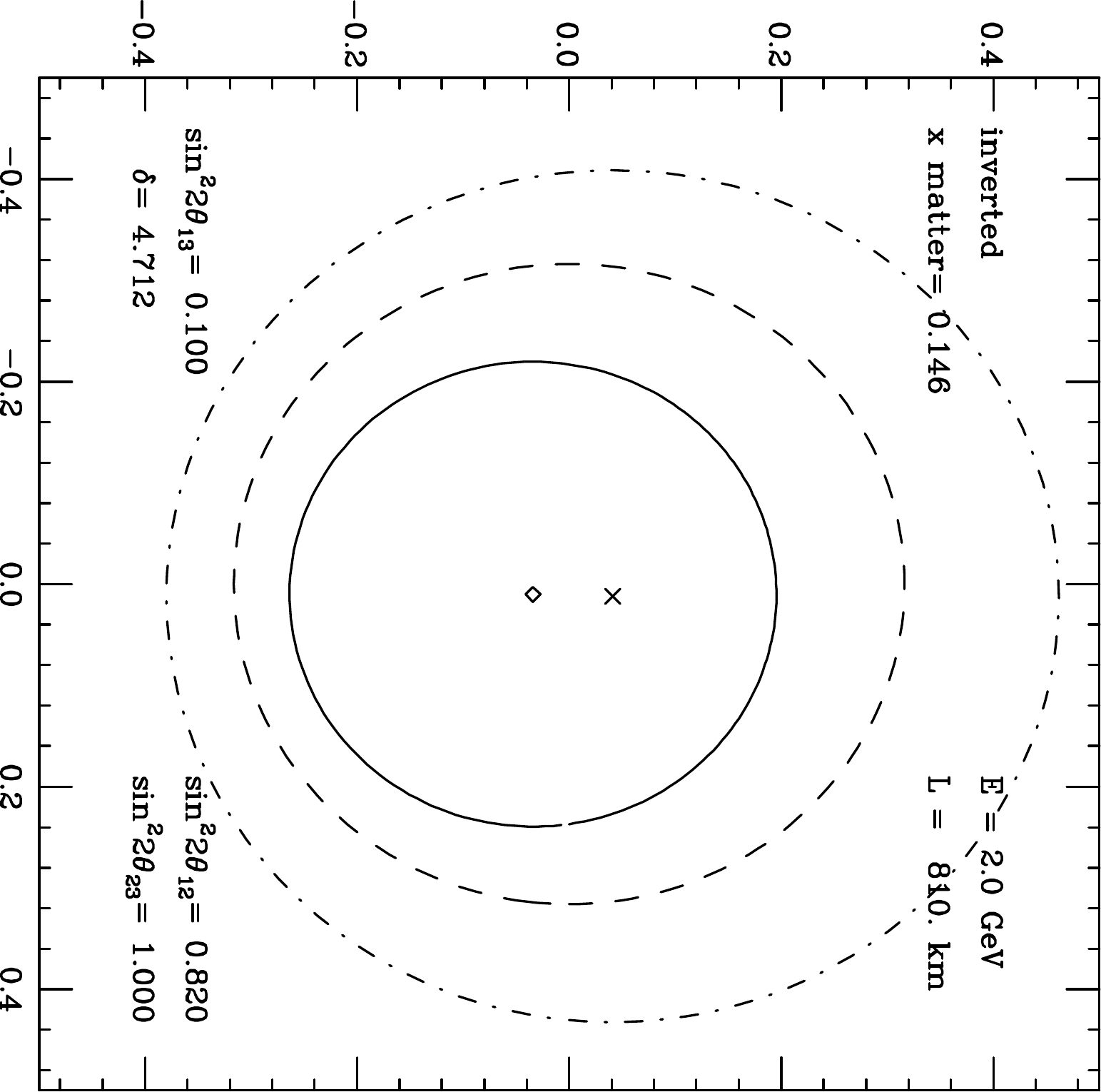}
\caption{Plots for \NOvA.  Upper, with true hierarchy being normal, with $\delta_{CP}=\pi/2$.  Lower again,  with true hierarchy being normal, but now with $\delta_{CP}=3\pi/2$. The dot dash circles show the constraints from $\nu_\mu$ scattering, while the solid circles are for $\nubar_\mu$. The dashed circles show the constraints from a hypothetical perfect measurement of $\sin^22\theta_{13}$\label{fig_two}}
\end{center}
\end{figure}

\begin{figure}\begin{center}
\includegraphics[width=0.44\textwidth]{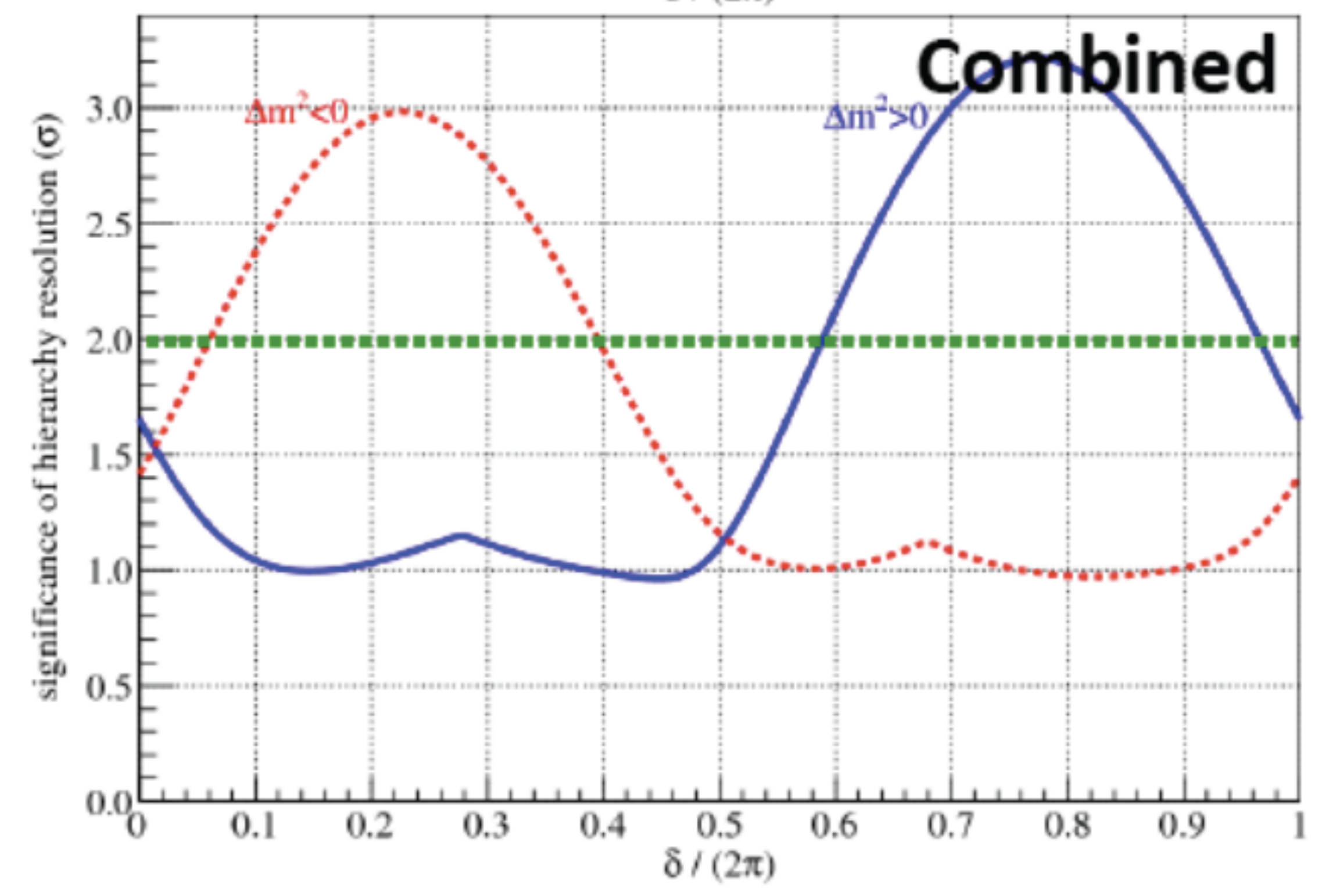}
\caption{Significance with which (Left) \NOvA\ and (Right) T2K+\NOvA\ can resolve the mass hierachy for  $\sin^22\theta_{13}=0.095$ and $\sin^22\theta_{23}=1$, as a function of $\delta_{CP}$. This assumes a nominal 3+3 year run plan. The blue/solid (red/dashed) curve shows the sensitivity given a normal (inverted) hierarchy. 
Figures from \NOvA\ document \cite{patterson}. \label{fig_seven}}
\end{center}
\end{figure}

\NOvA\ uses an off-axis beam, which is rather monochromatic.  The far detector has a mass of 14 kt and is located at Ash River, Minnesota, 810 km from the beam source at Fermilab.  The experiment is scheduled to take data starting in 2014 and to run for six years.  In addition to the far detector, there is a 330-t near detector.   The \NOvA\ website describes the detectors as being ``made up of 344,000 cells of extruded, highly reflective plastic PVC filled with liquid scintillator. Each cell in the far detector measures 3.9 cm wide, 6.0 cm deep and 15.5 meters long.'' 
 Fig.~\ref{fig_two} shows the statistical separation power of \NOvA; we see that the separation is easiest when $\delta_{CP}\approx 3\pi/2.$   
The reach as determined by the \NOvA\ collaboration is shown in Fig.~\ref{fig_seven}.
These figures are consistent with Fig.~\ref{fig_two} and show that 
for about a third of a range of $\delta_{CP}$, centered around
 $\delta_{CP}=3\pi/2$, \NOvA\ can determine the hierarchy with a
 confidence of $2-3\sigma$. A combination of \NOvA\ and T2K
 (Fig.~\ref{fig_seven}, right) has an ability to resolve the hierarchy at
 $1-3\sigma$ significance for all values of $\delta_{CP}$. This comes
 about because of the difference in baselines: T2K is primarily
 sensitive to the CP asymmetry between neutrinos and antineutrinos
 (effect of $\delta_{CP}$), which can be subtracted from the \NOvA\
 asymmetries to extract the matter-induced hierarchy signal. 

\subsubsection{Prospects}

\NOvA\ has a much better chance of measuring the hierarchy than T2K, but this
would only be possible for a very favorable value of $\delta_{CP}$.
It could find an indication of the hierarchy and this could reduce the
impact of later experiments if they ended up confirming, say, a
two-sigma \NOvA\ result.   
The sensitivity of T2K and \NOvA\ could be improved with additional run
time, or upgrades to the detectors or the beam
intensity~\cite{atm:Winter}.

\clearpage

\subsection{LBNE}\label{s:lbne}
\def\nubar{\ensuremath{\overline\nu}}
\def\nue{\ensuremath{\nu_e}}
\def\nuebar{\ensuremath{\overline{\nu}_e}}
\def\nutaubar{\ensuremath{\overline{\nu}_\{tau}}
\def\nuone{\ensuremath{\nu_1}}
\def\nutwo{\ensuremath{\nu_2}}
\def\nuthree{\ensuremath{\nu_3}}
\def\deltam{\ensuremath{\Delta m^2}}
\def\deltamthreeone{\ensuremath{\Delta m^2_{31}}}
\def\deltamtwothree{\ensuremath{\Delta m^2_{32}}}
\def\deltamonetwo{\ensuremath{\Delta m^2_{21}}}
\def\thetaonethree{\ensuremath{\theta_{13}}}
\def\sinthetaonethree{\ensuremath{sin \theta_{13}}}
\def\sintthetaonethree{\ensuremath{sin \theta_{13}^2}}
\def\sinsqtthetaonethree{\ensuremath{sin 2\theta_{13}^2}} 
\def\thetatwothree{\ensuremath{\theta_{23}}}
\def\sinthetatwothree{\ensuremath{sin \theta_{23}}}
\def\sintwothetatwothree{\ensuremath{sin \theta_{23}^2}}
\def\sinsqtwothetatwothree{\ensuremath{sin 2\theta_{23}^2}} 
\def\thetaonetwo{\ensuremath{\theta_{12}}}
\def\sinthetaonetwo{\ensuremath{sin \theta_{12}}}
\def\sintwothetatonetwo{\ensuremath{sin \theta_{12}^2}}
\def\sinsqtwothetaonetwo{\ensuremath{sin 2\theta_{12}^2}} 
\def\deltacp{\ensuremath{\delta_{CP}}}
\DeclareGraphicsRule{.tif}{png}{.png}{`convert #1 `dirname #1`/`basename #1 .tif`.png}

\newcommand{\mutoe}{\ensuremath{\nu_{\mu} \rightarrow \nu_e}}
\newcommand{\mubartoebar}{\ensuremath{\overline{\nu}_{\mu}\rightarrow\overline{\nu}_e}}
\newcommand{\ev}{\,\mathrm{eV}}
\newcommand{\mev}{\ensuremath{\,\mathrm{MeV}}}
\newcommand{\gev}{\ensuremath{\,\mathrm{GeV}}}
\newcommand{\km}{\ensuremath{\,\mathrm{km}}}
\newcommand{\kw}{\ensuremath{\,\mathrm{kW}}}
\newcommand{\MW}{\ensuremath{\,\mathrm{MW}}}
\newcommand{\kt}{\ensuremath{\,\mathrm{kT}}}



\subsubsection{Introduction}

The purpose of the Long Baseline Neutrino Experiment is to measure all the neutrino oscillation parameters including the currently unknown values of the the CP violating  phase \deltacp\ and the neutrino mass hierarchy.  Matter effects compete with the 
CP violation effects, and the baseline needs to be optimized to be able to disentangle them. 

\begin{figure}[hbp]\begin{center}
\includegraphics[width=2.5in,angle=90]{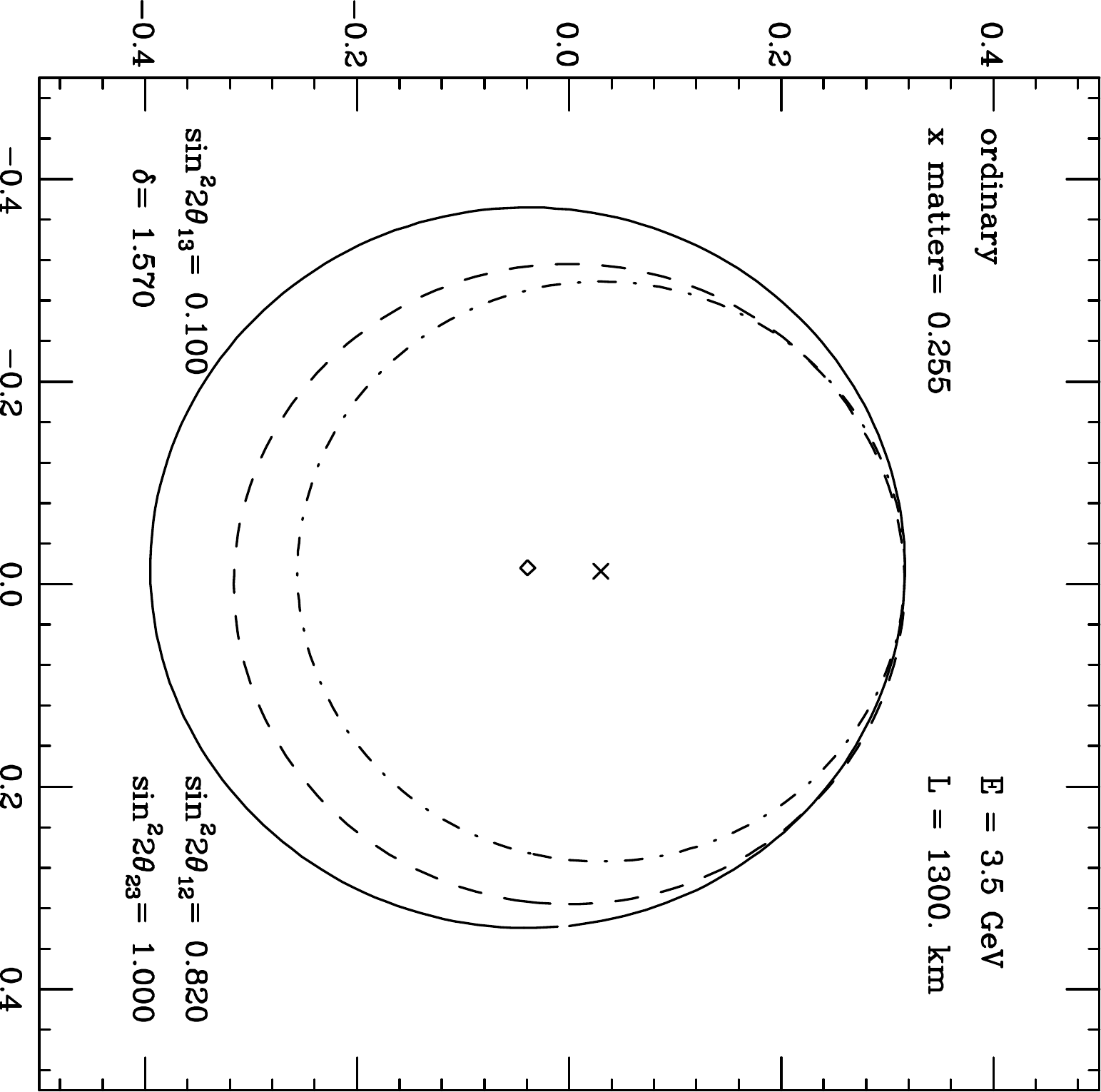}
\includegraphics[width=2.5in,angle=90]{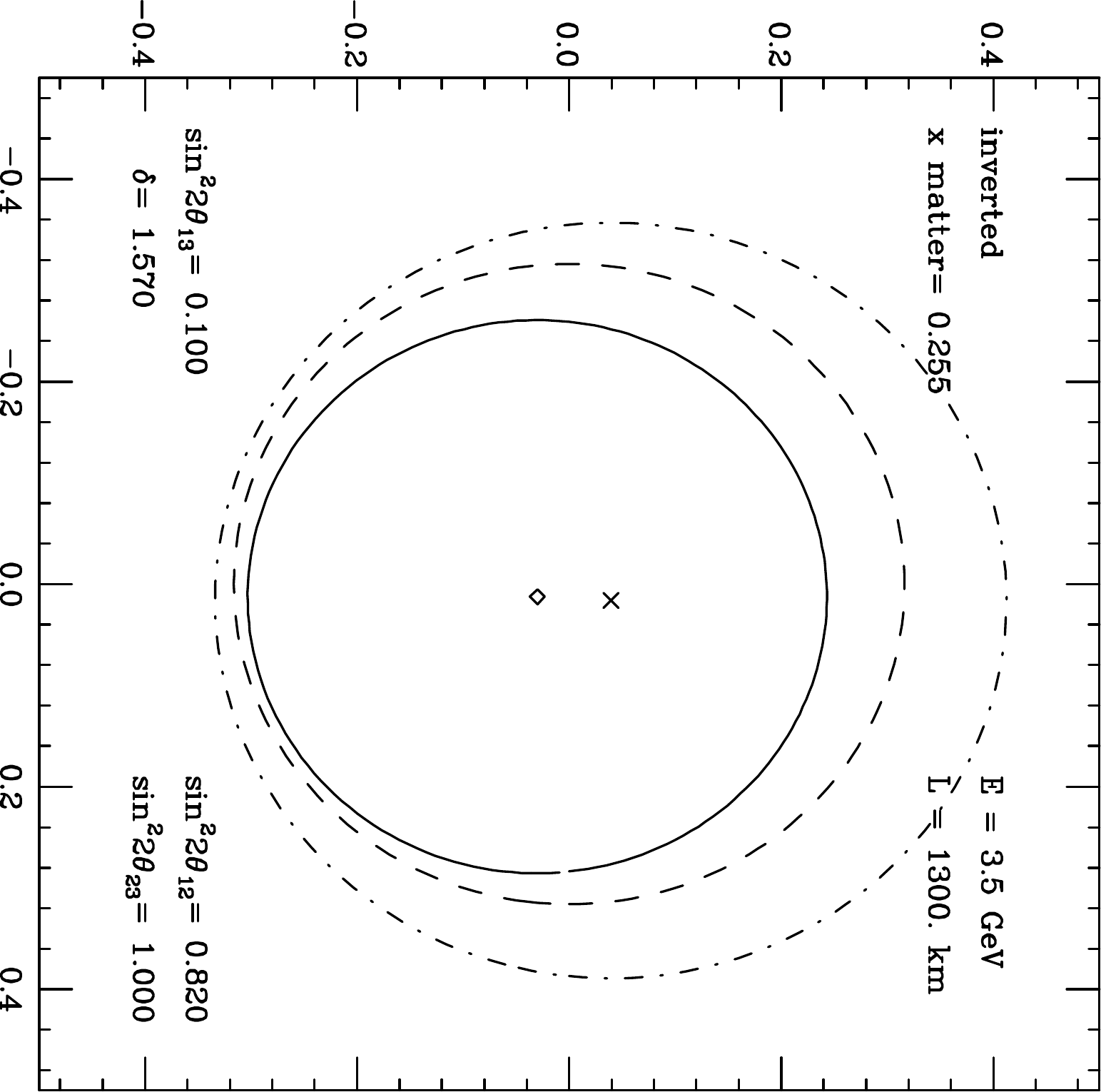}
\includegraphics[width=2.5in,angle=90]{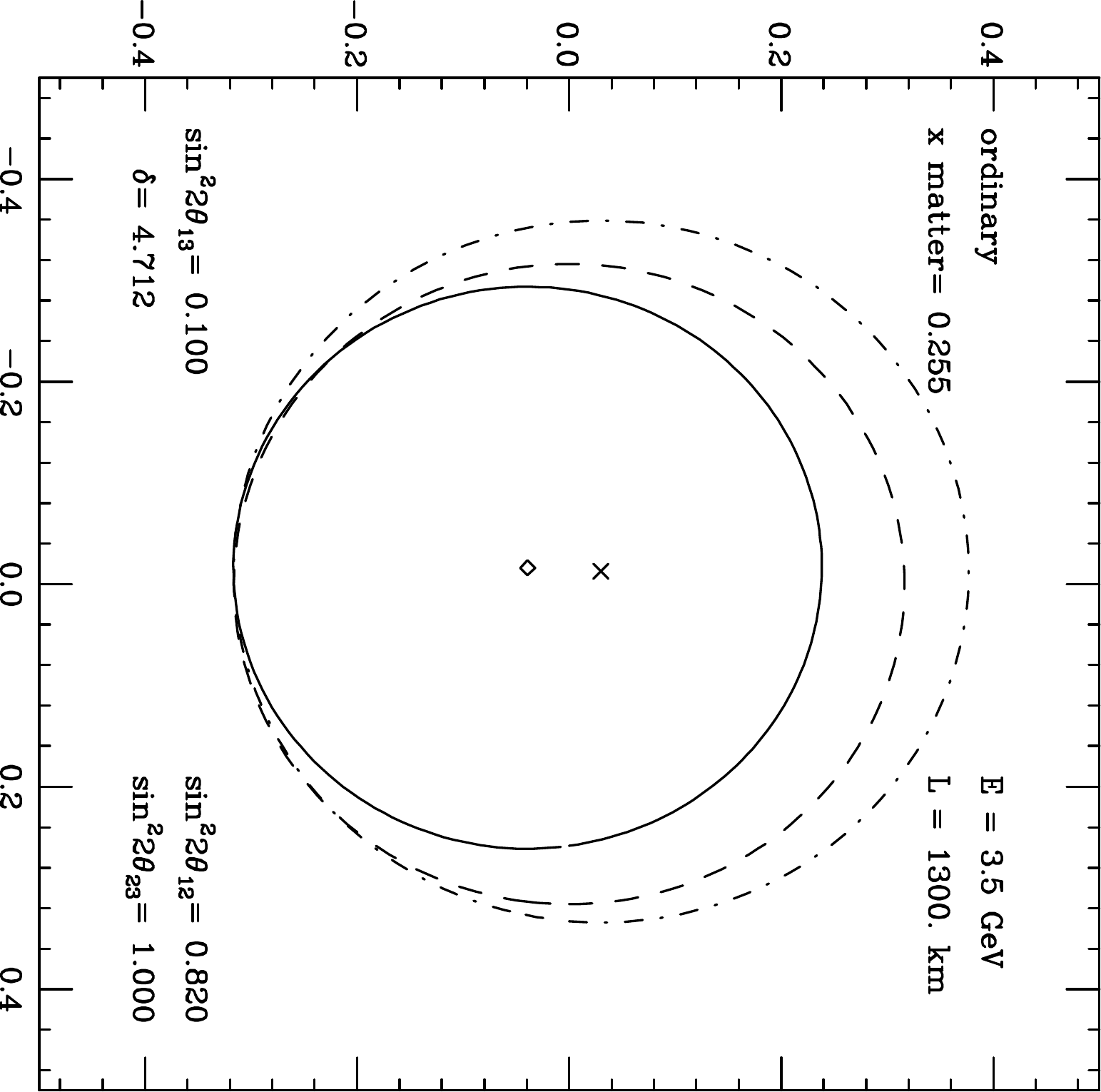}
\includegraphics[width=2.5in,angle=90]{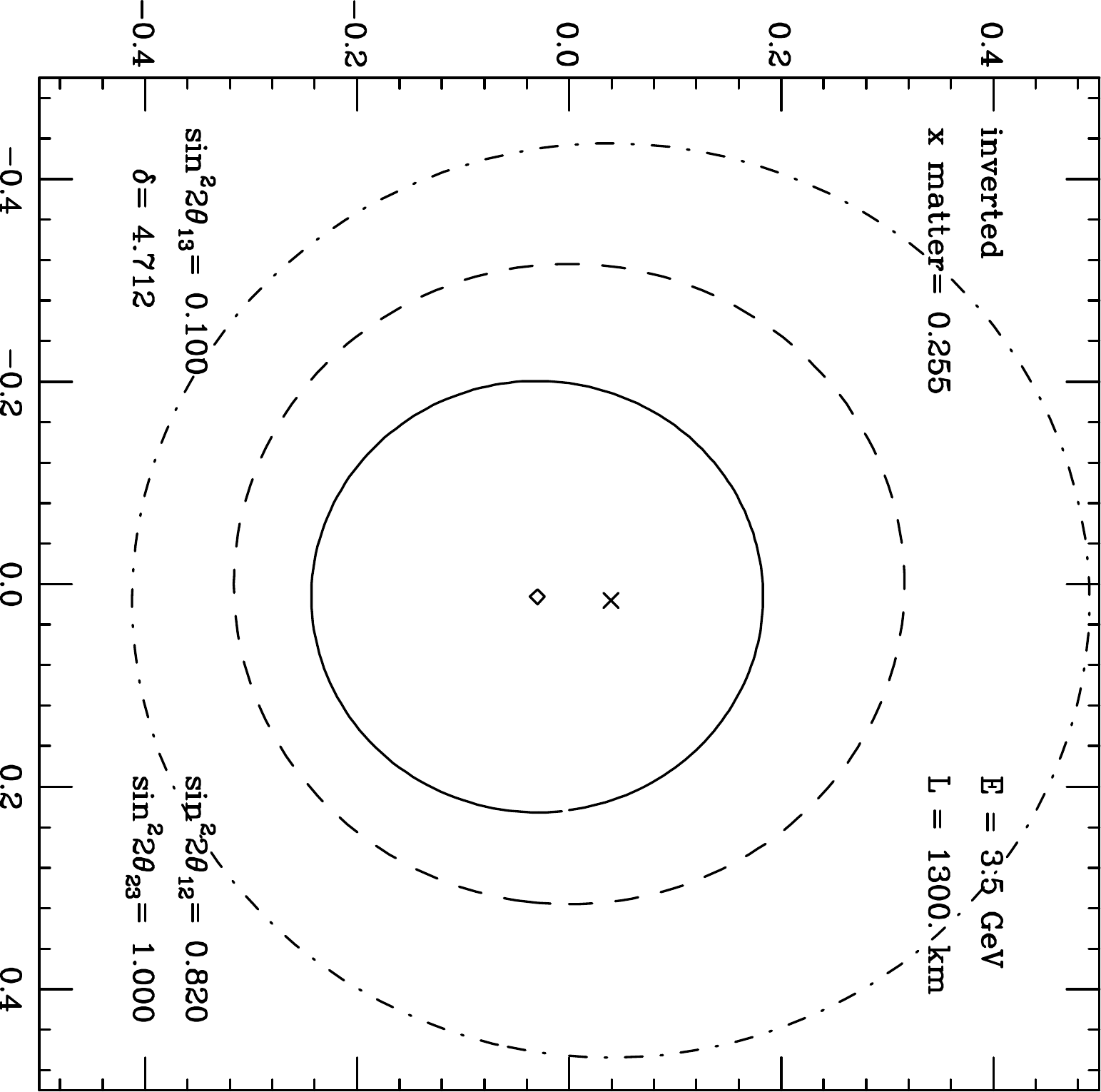}
\caption{Illustrative plots for a LBNE.  Upper, with true hierarchy
being normal, 
with $\deltacp=\pi/2$.  Lower, again with true hierarchy being normal,
but now with $\deltacp=3\pi/2$. The dot-dash circles show the
constraints from $\nu_\mu$ scattering, while the solid circles are for
$\nubar_\mu$.  The dashed circles show the constraints from a
hypothetical perfect measurement of $\sin^22\theta_{13}$.  The power
of LBNE is enhanced by combining with results from T2K (and to lesser
extent \NOvA) because the ``wrong'' solution occurs at different values
of $\deltacp$ for the three experiments. Note that the circles
correspond to a fixed beam energy of $3.5$~GeV, whereas the planned
neutrino beam is broadband. \label{fig_lbne}} 
\end{center}\end{figure}

To detect these neutrino oscillations, a  broad-band neutrino beam  of
mean energy 3.5\gev\ is proposed to originate at FNAL and to be
directed underground 1300\km\ away to the Sanford Underground Research
Facility in South Dakota.  The advantage of the long baseline and
correspondingly high neutrino energy at maximal mixing is apparent in
Fig.~\ref{fig_lbne}, analogous to those shown for T2K and \NOvA. 

\begin{figure}[tp]
\begin{center}
\includegraphics[width=0.7\textwidth]{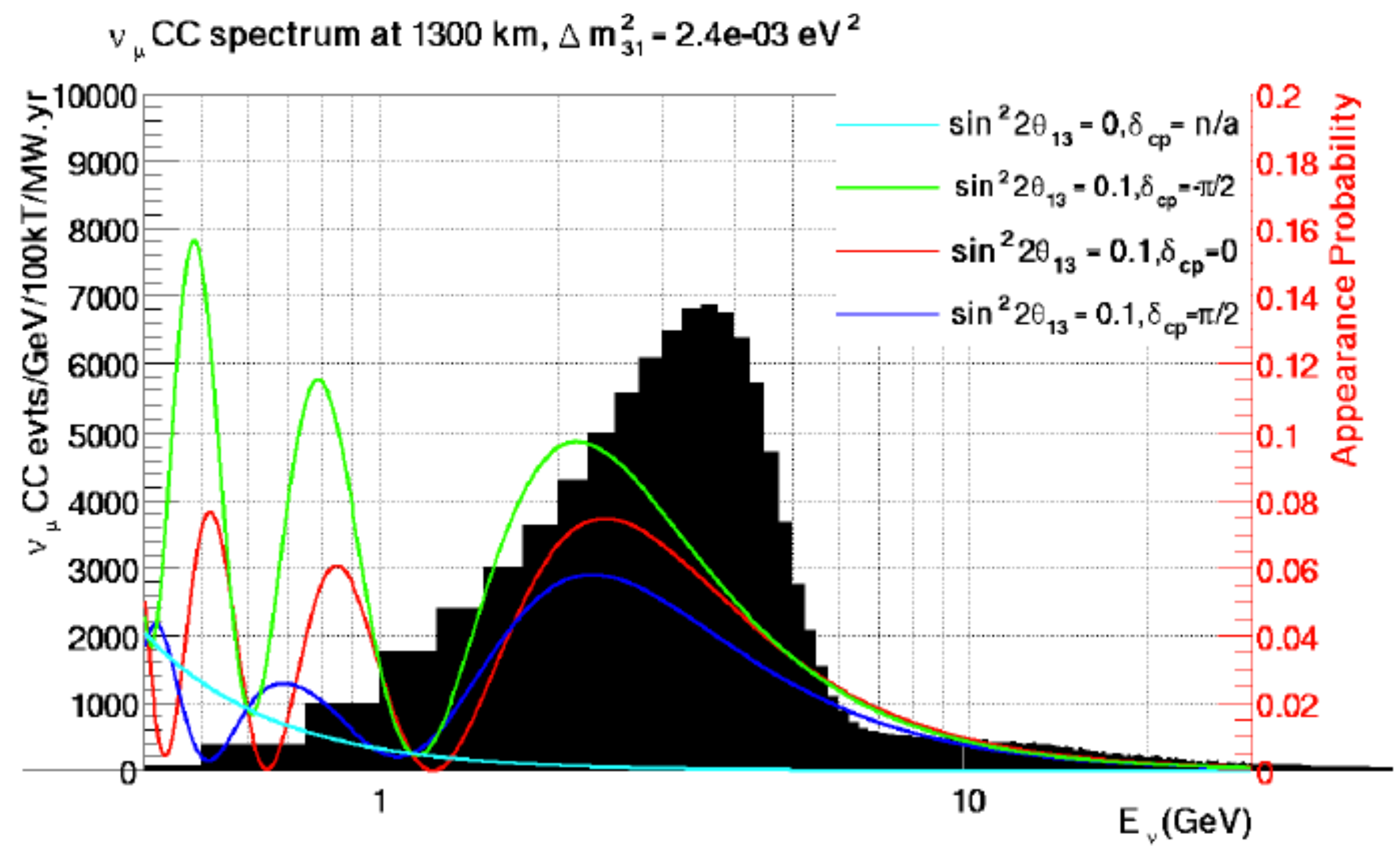}
\includegraphics[width=0.7\textwidth]{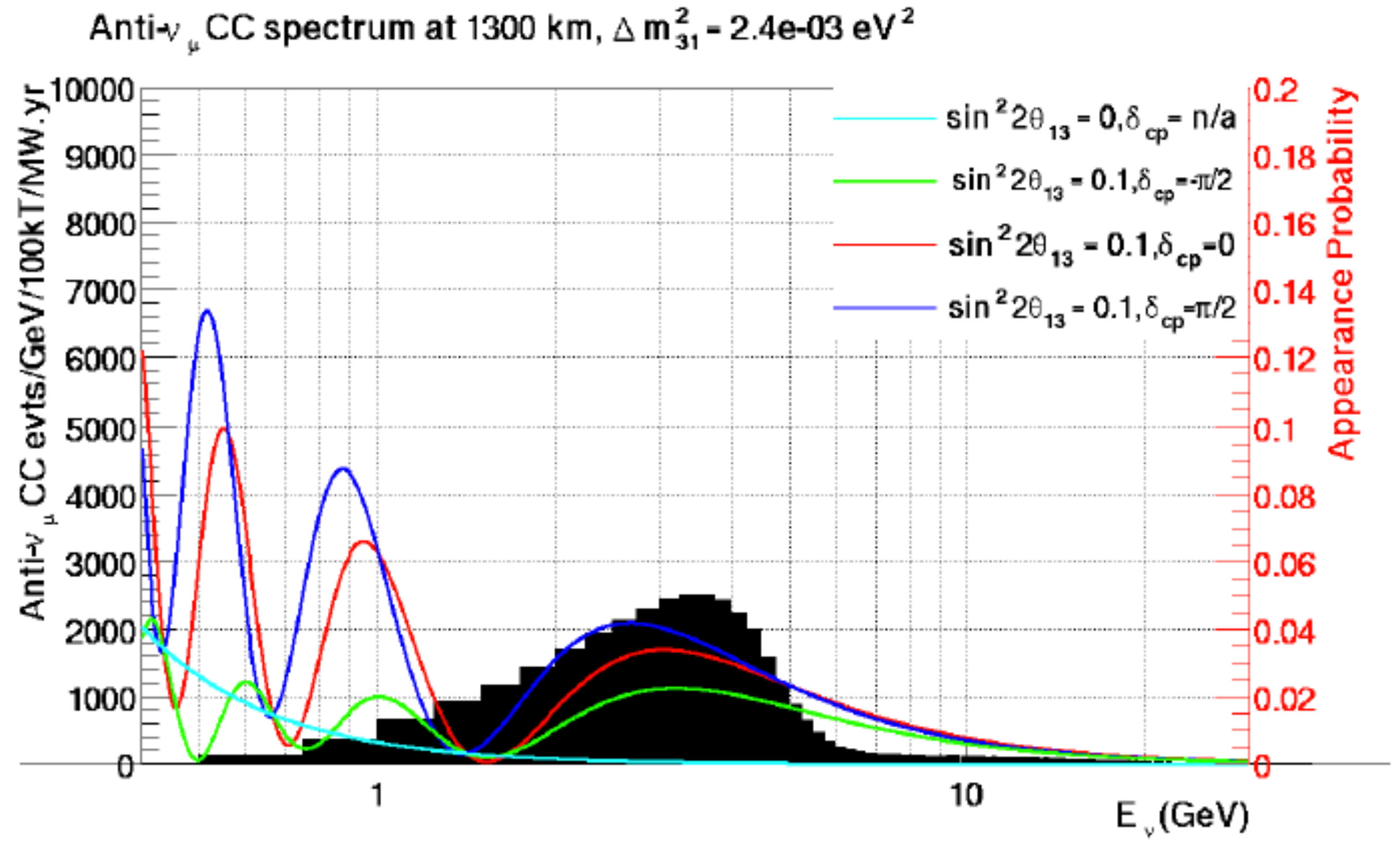}
\caption{The black shaded areas are the unoscillated charged current spectra for the Homestake site (left-hand y-axis scale) at 1300 \km\ for \mutoe\ (top) and \mubartoebar\ transitions (bottom).  The colored lines are oscillation probabilities for different oscillation parameters, as given in the legend (right-hand y-axis scale). The maxima of the beam energy has been optimized to correspond to the  maximum of the first ``node,''  { \it{i.e.}} maximum of the probability of \mutoe\ transitions \cite{PWGReconfigurationReport}.}
\label{fig:BeamProfileNHandIH}
\end{center}
\end{figure}

To maximize the event rate, the beam is typically tuned so that the
peak of the neutrino energy is at the first oscillation maximum for a
given distance, see Fig.~\ref{fig:BeamProfileNHandIH}. This is a
broad-band beam, optimized to cover both the first and second oscillation
maxima. Narrow band beams pick out a single energy and typically do
not sample the full distribution, but only a single  energy bin.  A
broad-band beam allows a measurement of the shape of the spectra
(including multiple maxima) favorable for the measurement
of \deltacp\, while a narrow band beam  only allows a counting
experiment at a single energy.

\subsubsection{Phasing the Long Baseline Neutrino Experiment}

Due to budgetary constraints, the experiment is to be phased. 
The stated goal of the first stage of LBNE, as summarized in the recent HEPAP Major Facilities report, is to determine the hierarchy to greater than $3 \sigma$ for all values of the CP phase $\delta_{CP}$ (when constraints from other experiments are included), and to set the stage for a world-class US neutrino program that will include Project X, and physics goals such as a comprehensive search for CP violation.

The Phase I experiment consists of the beamline and a 10~kT liquid argon (LAr)
detector 
located on the surface at the SURF facility. The beamline at FNAL
includes a target, horn and decay tunnels consistent with accepting a
2.3\MW\ beam, and a set of instruments (an array of high pressure gas
muon Cherenkov counters, four layers of stopped muon detectors, and
two or three layers of hadron ionization detectors) at the absorber
complex to measure the flux, angle and centerline of the beam. There
is no near neutrino detector in the first phase.  The final phase
includes a 34~kT liquid argon detector, underground at the far site
and a substantial near detector most likely with  both a liquid argon
detector and a magnetized, high pressure Ar  [straw] tracker, and EM
calorimeter to measure the absolute neutrino flux via
neutrino-electron scattering. 

\subsubsection{Detector Performance}

The detector performance is evaluated in the context of the ``GLoBES''
software package~\cite{GLoBES}, which includes the energy spectrum of
the beam, interaction cross sections, the corresponding detector
efficiencies as well as specific  backgrounds and uncertainties {\it
{e.g.}} $\nutau$  backgrounds have recently been added, see
below. Table~\ref{fig:LArDetectorEfficiency} summarizes the properties
of the detector as coded into GLoBES. 

\begin{table}[!bp]
\caption{Estimated range of the LAr-TPC detector performance parameters for the primary oscillation physics. The expected
range of signal efficiencies, background levels, and resolutions from various studies (middle column) and the value chosen for
the baseline LBNE neutrino-oscillation sensitivity calculations (right column) are shown. For atmospheric neutrinos this is
the misidentification rate for events below 2 GeV; the misidentification rate is taken to be zero events  above 2 GeV \cite{PWGReconfigurationReport}.}\begin{center}
\includegraphics[trim = 0cm 0cm 0cm 2cm, clip=true, width=1.0\textwidth]{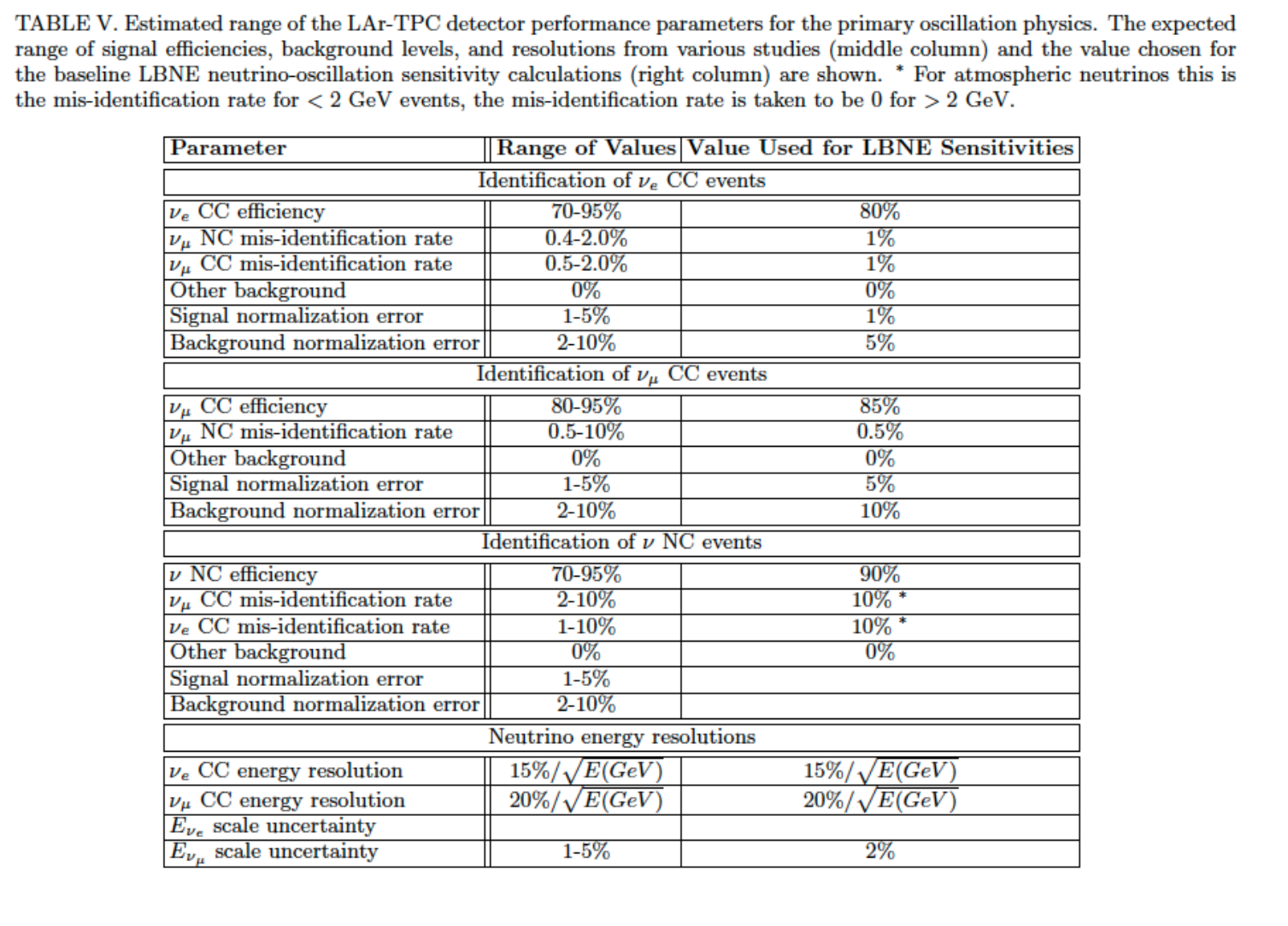}
\label{fig:LArDetectorEfficiency}
\end{center}
\vspace{-1.2cm}
\end{table}

\begin{figure}[htbp]
\begin{center}
\includegraphics[width=0.9\textwidth]{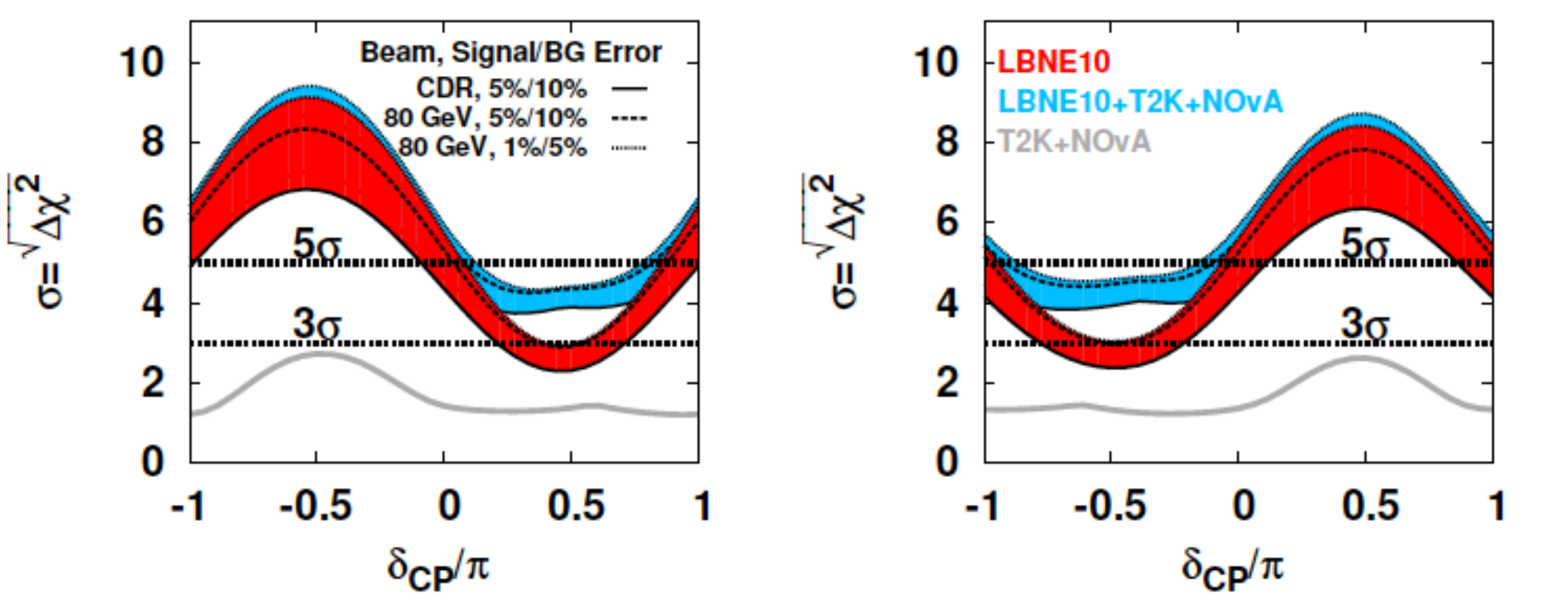}
\caption{Mass hierarchy sensitivity for Phase I of LBNE alone (red band) and in combination with T2K+NO$\nu$A (blue band) for (Left) normal and (Right) inverted mass hierarchies~\cite{lbne:sm}.}
\label{fig:SMMH}
\end{center}
\end{figure}

\begin{figure}[htbp]
\begin{center}
\includegraphics[width=0.5\textwidth]{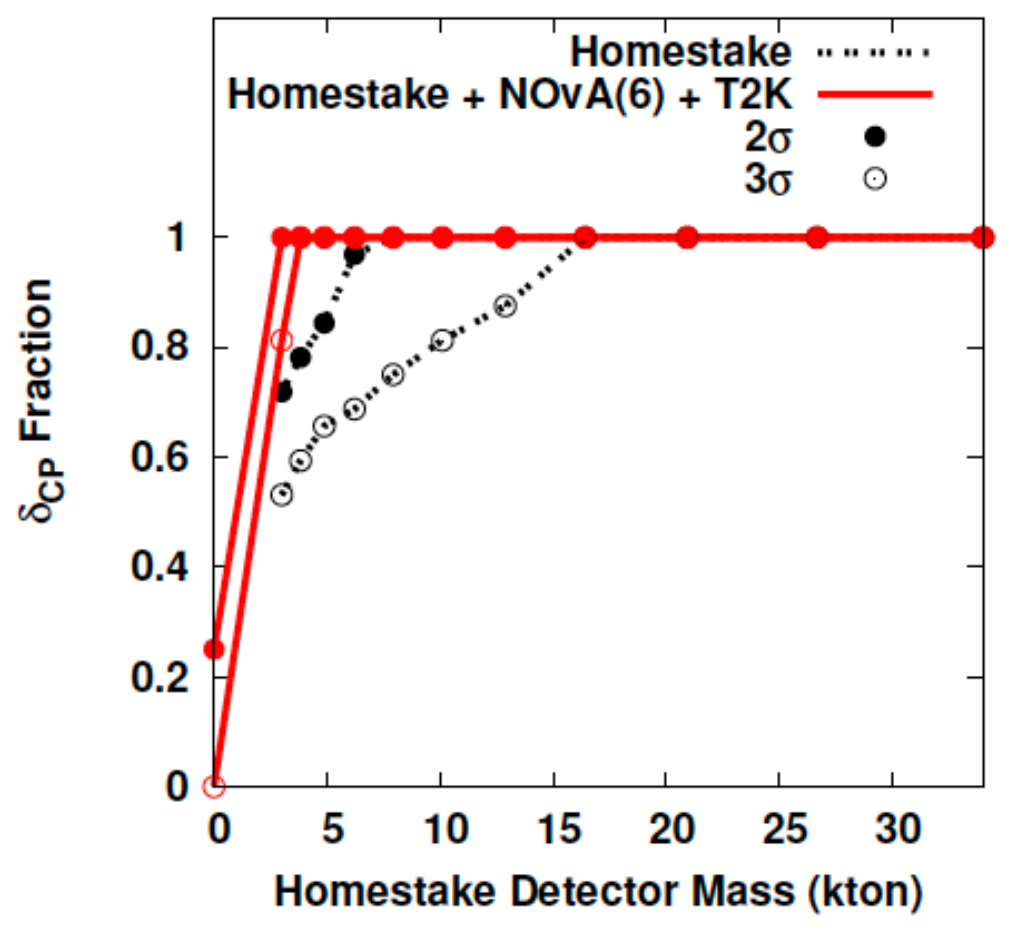}
\includegraphics[width=0.45\textwidth]{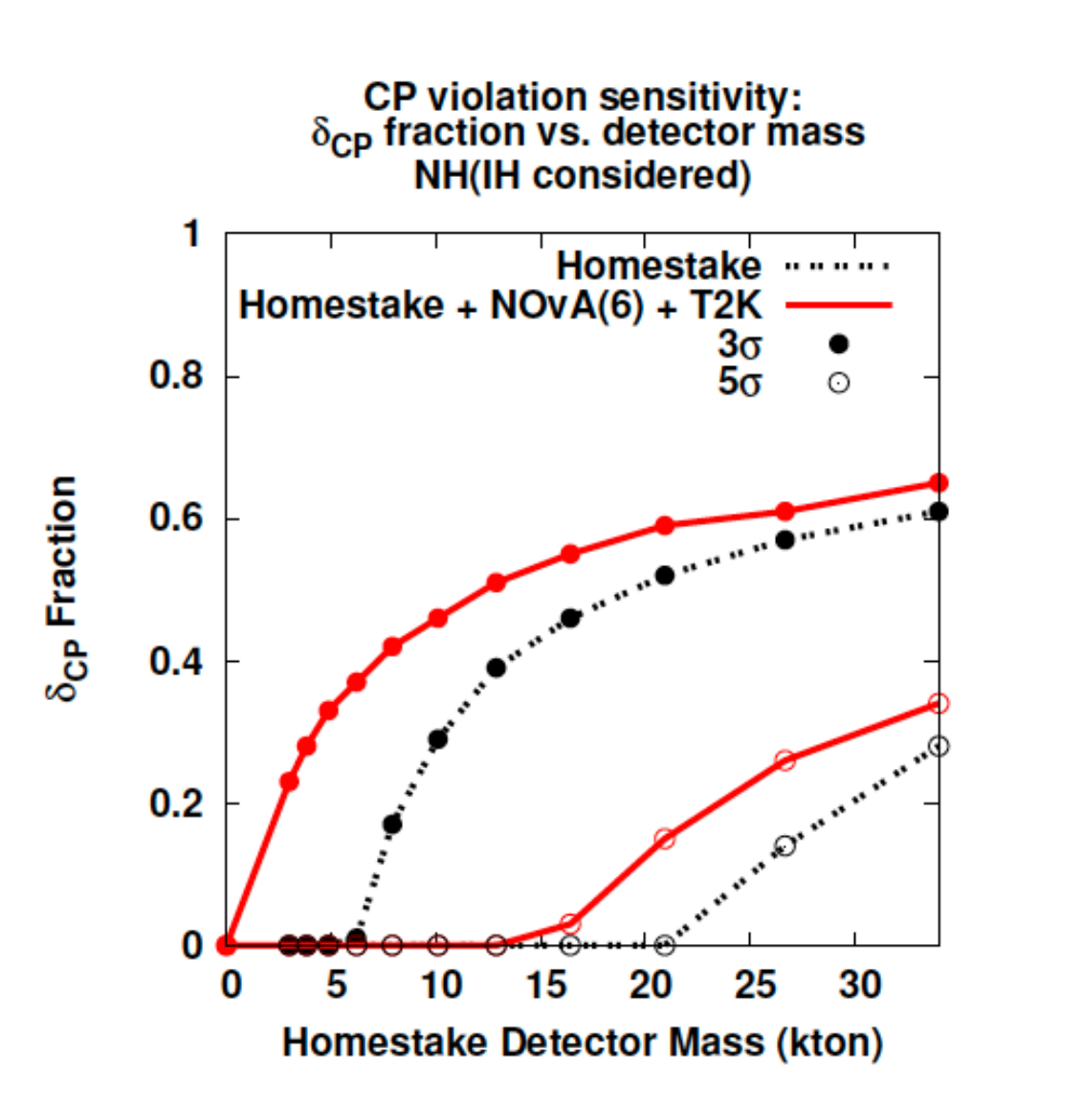}
\caption{(Left) 2- and 3-$\sigma$ sensitivity to the mass hierarchy versus the fraction of  \deltacp\ coverage  and detector mass when located at SURF, under the assumption of a normal hierarchy. Solid points are  2$\sigma$ limits, while open circles are 3$\sigma$ limits. Black is for LBNE alone, with $5\nu+5\bar{\nu} $ years at 700 kW, and red is combined with results from \Nova\ and T2K. \cite{PWGReconfigurationReport}. \\ (Right) 3- and 5-$\sigma$ sensitivity to $\delta_{CP}$ versus the fraction of the CP phase \deltacp\ as a function of the mass of the detector at the SURF site, under the assumption of a normal hierarchy. Solid points are  3$\sigma$ limits, while open circles are 5$\sigma$ limits. Black is for LBNE alone, with $5\nu+5\bar{\nu} $ years at 700 kW, and red is combined with results from \Nova\ and T2K. \cite{PWGReconfigurationReport}.}
\label{fig:MHsensitivityVsSize}
\end{center}
\end{figure}

Fig.~\ref{fig:SMMH} shows the range of sensitivity to the hierarchy for LBNE Phase I, both alone and in combination with T2K+NO$\nu$A.  
The left panel of Fig.~\ref{fig:MHsensitivityVsSize} shows the $2\sigma$ and $3\sigma$ sensitivity limits on the hierarchy as a function of  detector mass at the SURF site, and the right panel shows the 3$\sigma$ and 5$\sigma$ sensitivities to \deltacp.  

\begin{figure}[htbp]
\begin{center}
\includegraphics[width=0.7\textwidth]{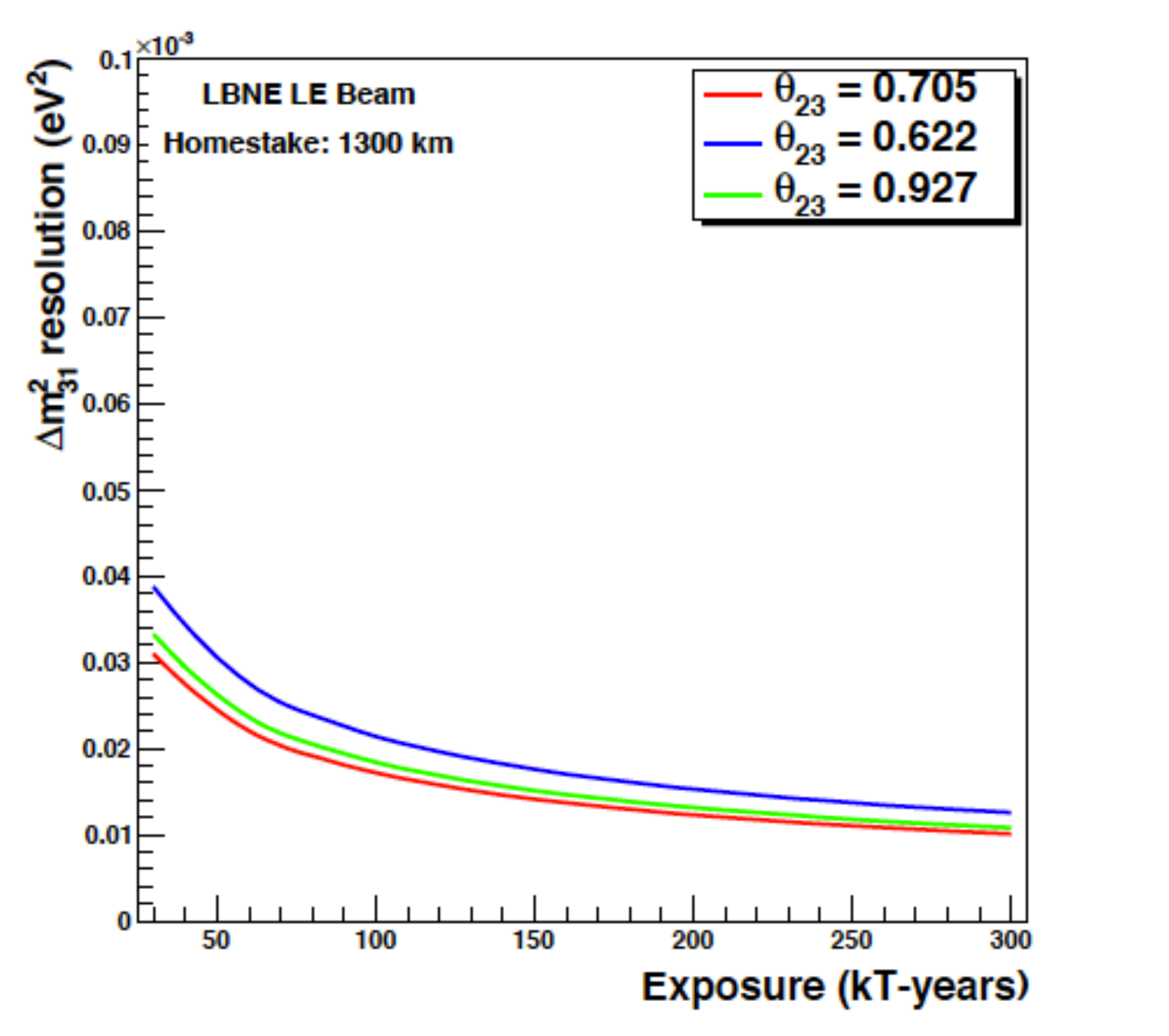}

\caption{ Estimated error in \deltamthreeone\ $\sim$ \deltamtwothree\ versus detector exposure at 1300 km for three values of $\sin^2 2\thetatwothree$ [indicated in the legend only as \thetatwothree], for a 700\kw\ beam. 
Note the vertical scale (\deltamtwothree\  resolution) has a scale factor of $10^{-3}$.
The plot assumes a near detector for these precision measurements. Neutrino and anti-neutrino running are combined in the ratio of 1:1, The mass hierarchy is assumed to be known \cite{PWGReconfigurationReport}.}
\label{fig:DeltaM31sqMeasumentError}
\end{center}
\end{figure}

The value of \deltamtwothree\ is one of the limiting factors in the
ability of second generation reactor experiments to measure the mass
hierarchy. 
As a further example of the
capabilities of the experiment, 
LBNE will have an unprecedented capability to
measure \deltamtwothree. 
The precision with which \deltamthreeone\
$\sim$ \deltamtwothree\ can be measured is shown in
Fig.~\ref{fig:DeltaM31sqMeasumentError}, approaching $10^{-5}\ev^2$ for
a 10 year exposure of a 34\kt\ detector.

\clearpage

\subsubsection{Open Concerns}

There are four major concerns for the Phase I LBNE detector program:
\begin{enumerate}
\item {The surface location of the LAr detector;}
\item{The lack of a near detector;}
\item{The small mass (10\kt) of  the far detector;}
\item{Uncertainties in nuclear effects.}
\end{enumerate}

With a 10$\mu\mathrm{sec}$ beam window, the detector is live for approximately 100\,sec/year and assuming an overburden of 3m of rock the background from cosmogenically induced events is negligible,  less than $1\%$ in the \nue\ appearance experiment. Measuring \deltacp\ and the mass hierarchy can be accomplished by a LAr detector on the surface. Monte Carlo studies have shown that these measurements are statistics limited, and a simplified set of flux monitors at the end of the decay pipe at FNAL are sufficient to monitor the beam intensity.
For longer exposures (above 100 \kt\ $\times$ years), a near detector is needed, and collaborators from India have expressed interest in building such a detector.

A larger detector is attractive, and an excellent opportunity for
foreign investment. Some potential foreign collaborators have also
expressed interest in technical components of the beamline that would
help offset portions of the total project cost, and free DOE funds to increase
the scope of the US project. A smaller detector can also be offset via
higher beam intensity, and current plans at FNAL call for Phase I of
Project X to be completed by the start of LBNE, boosting the beam
power from 700\kw\ to $\sim1.1\MW$. This would effectively remove all
ambiguity in the 3$\sigma$ mass hierarchy measurement for all values
of \deltacp\  for a 10 year run of a 10kT LBNE LAr detector. 

Putting the detector underground would permit a broader physics
program, including proton decay searches and the detection of
neutrinos from supernova events. As a LAr
detector can easily  
identify charged kaons (which are below threshold in a water Cherenkov
detector), its sensitivity to proton decay quickly exceeds that of
SuperK by an order 
of magnitude summed over all decay modes, despite SuperK's
extrapolated 30yr integrated running history.

The CD1 approval letter states that  LBNE may increase the scope of the project if additional resources can be found.
It is important to recognize that unlike other DOE/HEP projects where
the total project cost  (TPC) is usually capped, {\it including both
non-DOE domestic or foreign contributions, }in the LBNE case the
project has been given a ``hunting license'' to seek outside funding
to increase the {\it scope }of the project. Such negotiations are
ongoing.  

Construction of an unprecedentedly massive liquid-argon detector
presents significant technological challenges.  
It will require a  rigorous R\&D effort before the technology
is mature enough for this endeavor. 
If cost is not an issue, one approach in realizing this scheme is to modulize
the detector so that it can be staged. 

Nuclear effects will smear the reconstruction of the incident neutrino
energy \cite{Mosel}. A LAr detector is a ``tracking detector'' and
this may help in identifying low energy protons or gammas ejected from
the nucleus, but properly identifying neutrons may be problematic,
and extracting elementary reaction amplitudes for meson production is
difficult. The issue of the energy reconstruction in the presence of
these effects needs to be addressed.

\subsection{Non-US Options}\label{s:nonus}
Both the European community and Japan have proposed deep-underground
accelerator-based neutrino oscillation experiments, which overlap
significantly in motivation, technology, and physics scope with
the US neutrino 
program. There are two main proposals on the table: LAGUNA-LBNO,
a European initiative with the CERN proton beam and several potential
sites for the far detector; and T2HK, an oscillation experiment with
the JPARC beam and the HyperK target. It is important to compare the sensitivity and timescale of
these experiments with the US options. 

\subsubsection{Hyper-Kamiokande}

The Japan Subcommittee on Future HEP projects
(2012)~\cite{nonus:HEPfuture} has recommended 
that:
``Japan should aim to realize a large-scale neutrino detector ...This
new large-scale neutrino detector should have sufficient sensitivity
to allow the search for proton decays''. 
The emphasis of the program is on CP violation and proton decay, not
determination of the mass hierarchy.

The neutrinos for Japan's long-baseline program are produced at the
50 GeV proton synchrotron at J-PARC 
accelerator laboratory, in Tokai 
on the east coast of Japan. The J-PARC development plan calls for a
staged increase in proton beam power, from 300 kW presently to 1-2 MW
by the middle of the next decade~\cite{nonus:JapanMasterPlan}. 

Two primary options have been considered for the far detector:
Hyper-Kamiokande (HyperK)~\cite{nonus:HyperK}, a Megaton-scale water 
Cherenkov detector in the Kamioka region~\cite{nonus:HyperK}, 295 km
away from J-PARC; and a 100 kTon liquid argon detector at Okinoshima
island at a baseline of 660 km. As of this writing, HyperK appears
to be the front-runner. 

HyperK is a proposed very large water Cherenkov detector with a total
mass of 1 MTon. It is proposed to be located in a new detector cavity
in the Tochibora mine, about 8 km from the Super-Kamiokande location,
under an overburden of about 1750 mwe. 

The HyperK design is based on the very successful Super-Kamiokande
(SuperK) experiment, and represents a factor of 20 increase in the
fiducial mass. The extrapolation to the Megaton-scale detector will
certainly present technological challenges; however, the detector
technology and, in particular, systematic effects in  water
Cherenkov detectors are well understood. Water Cherenkov technology
offers excellent particle ID capabilities and high reconstruction
efficiency for the sub-GeV electrons and muons relevant for the
oscillation program. 

The total cost of the detector is estimated to be 50-75 Billion yen
($\approx \$500-700$M)~\cite{nonus:JapanMasterPlan}. The cost of the
neutrino beamline upgrades to provide 1-2 MW of beam power at J-PARC
is estimated at 38B yen. Thus, the long-baseline neutrino program
falls in the medium range of the future projects being considered at
Japan, between SuperKEKB/SuperBelle and the ILC. 

At the baseline of 295 km, the matter-induced neutrino mixing is
relatively small. This means a weak sensitivity to the
neutrino mass hierarchy (see Fig.~\ref{fig_four} in
Section~\ref{s:tn}). On the other hand, the sensitivity to the CP
angle $\delta_{CP}$ is enhanced compared to the matter-induced
effects. Thus, CP measurements in HyperK would require an independent
determination of the mass hierarchy. Such determination is possible
with an atmospheric neutrino campaign in HyperK, as discussed below. 

The short baseline also requires a lower neutrino energy to maintain
the detector at the first oscillation maximum. The T2HK (Tokai to HyperK)
project will utilize an off-axis, narrow-band beam similar to that of
T2K, with the 
$\nu_\mu$ energy peaking at around 600~MeV. At that energy,
backgrounds from $\nu_e$ and in particular $\nu_\tau$ are
suppressed. This configuration requires large beam power and long run
time. The HyperK proposal assumes 10 years of running at a beam power of
750~kW (3 years in neutrino and 7 years in anti-neutrino modes).

The very large mass and significant overburden of HyperK provides the
potential for  a measurement of the neutrino mass hierarchy
with atmospheric neutrinos. 

The sensitivity of atmospheric neutrino oscillations to the neutrino
mass hierarchy is discussed in detail in Section~\ref{s:atm}. 
The sensitivity comes from the $\nu_\mu\to \nu_e$
appearance induced by the Earth matter, and thus the effect is 
strongest for upward-going neutrinos. 
In a
charge-symmetric detector like HyperK, the measurement requires
excellent identification of neutrino flavor and
precise reconstruction of neutrino energy and azimuthal angle, at
the energies of a few GeV. A water Cherenkov detector like HyperK is
well matched to these requirements. 

As discussed in
Section~\ref{s:atm}, understanding of the energy and angular
resolutions, as well as the systematics of the overall energy scale
are extremely important for extracting the mass hierarchy signal from
the atmospheric neutrino data. Unlike PINGU and other large
atmospheric neutrino detectors, with SuperK experience these
systematic effects are fairly well understood in a water Cherenkov
detector. Most importantly, HyperK will have a beam-based calibration
signal (the beam energy spectrum) and the ability to detect both
upward- and downward-going neutrinos. Those factors make the potential
for observing the neutrino mass hierarchy with atmospheric neutrinos
in HyperK quite robust. 

\begin{figure}[bp]
\begin{center}
\includegraphics[width=0.9\textwidth]{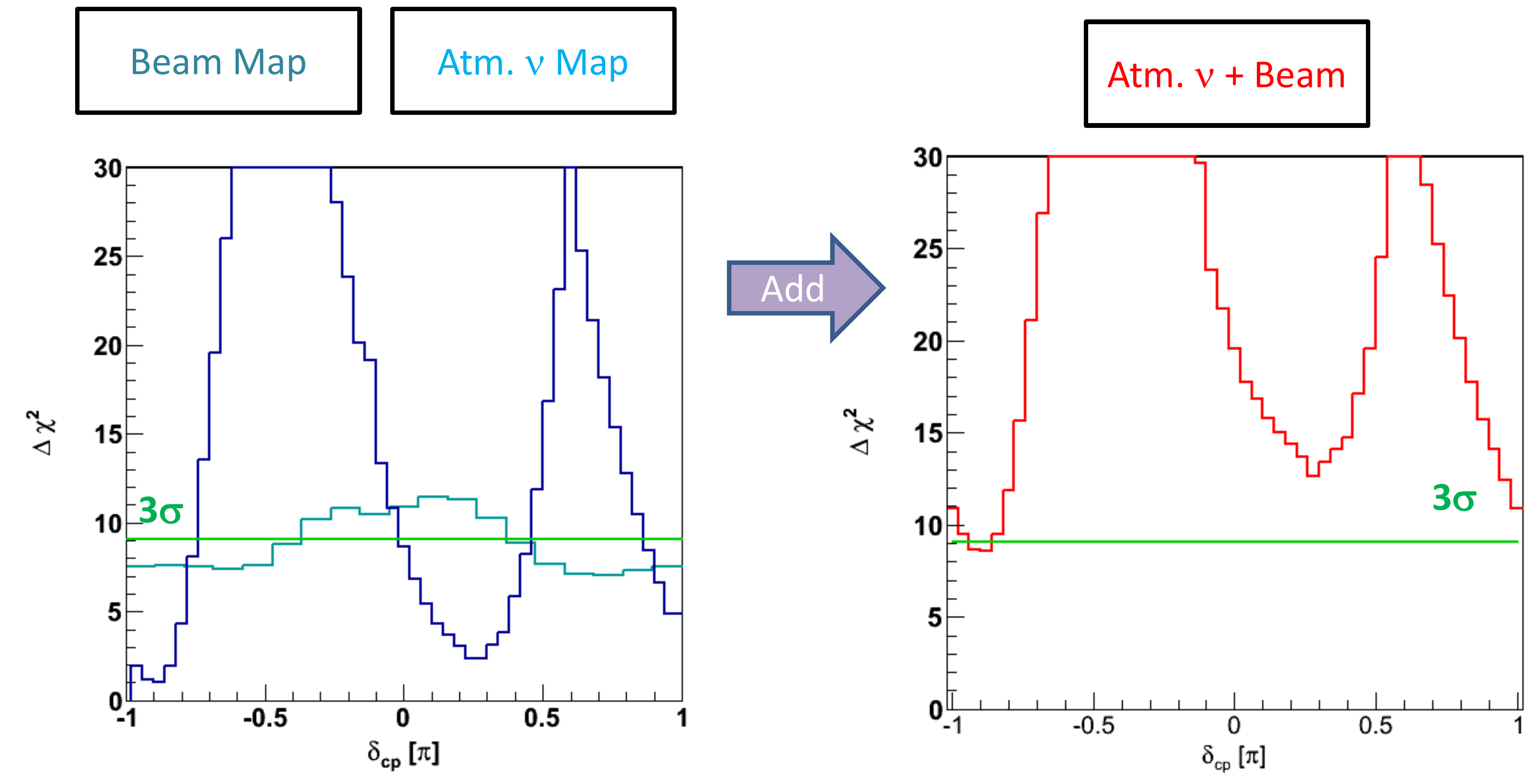}
\caption{Projected sensitivity in units of $\Delta\chi^2$ of HyperK to
  the neutrino mass 
  hierarchy from the long-baseline beam measurements (left plot, blue
  histogram) and from the atmospheric measurements (left plot,
  dark-green histogram). The combined sensitivity is shown in the
  right plot. The plot is from Ref.~\cite{nonus:HyperKsens}. } 
\label{fig:HyperK_sens}
\end{center}\end{figure}

After 10 years of operation starting at the end of this decade
(i.e. results ready by 2028-2030), HyperK
projects to be able to determine 
the mass hierarchy with a confidence level of $2-3\sigma$ with the atmospheric
measurements, and greater than $1\sigma$ with the 
beam-based measurements alone (Fig.~\ref{fig:HyperK_sens}). The sensitivity
of the beam-based measurement depends strongly on the value of
$\delta_{CP}$ (Section~\ref{s:tn}), while the atmospheric sensitivity
depends somewhat on the value (octant) of $\theta_{23}$ and, to a
small extent, $\delta_{CP}$. 
The two
measurements are complementary, and adding them together allows HyperK
to project greater than $3\sigma$ sensitivity to the mass hierarchy for the entire
range of $\delta_{CP}$. 

\subsubsection{LAGUNA-LBNO}

\begin{figure}[!h]\begin{center}
\includegraphics[width=2.45in,angle=90]{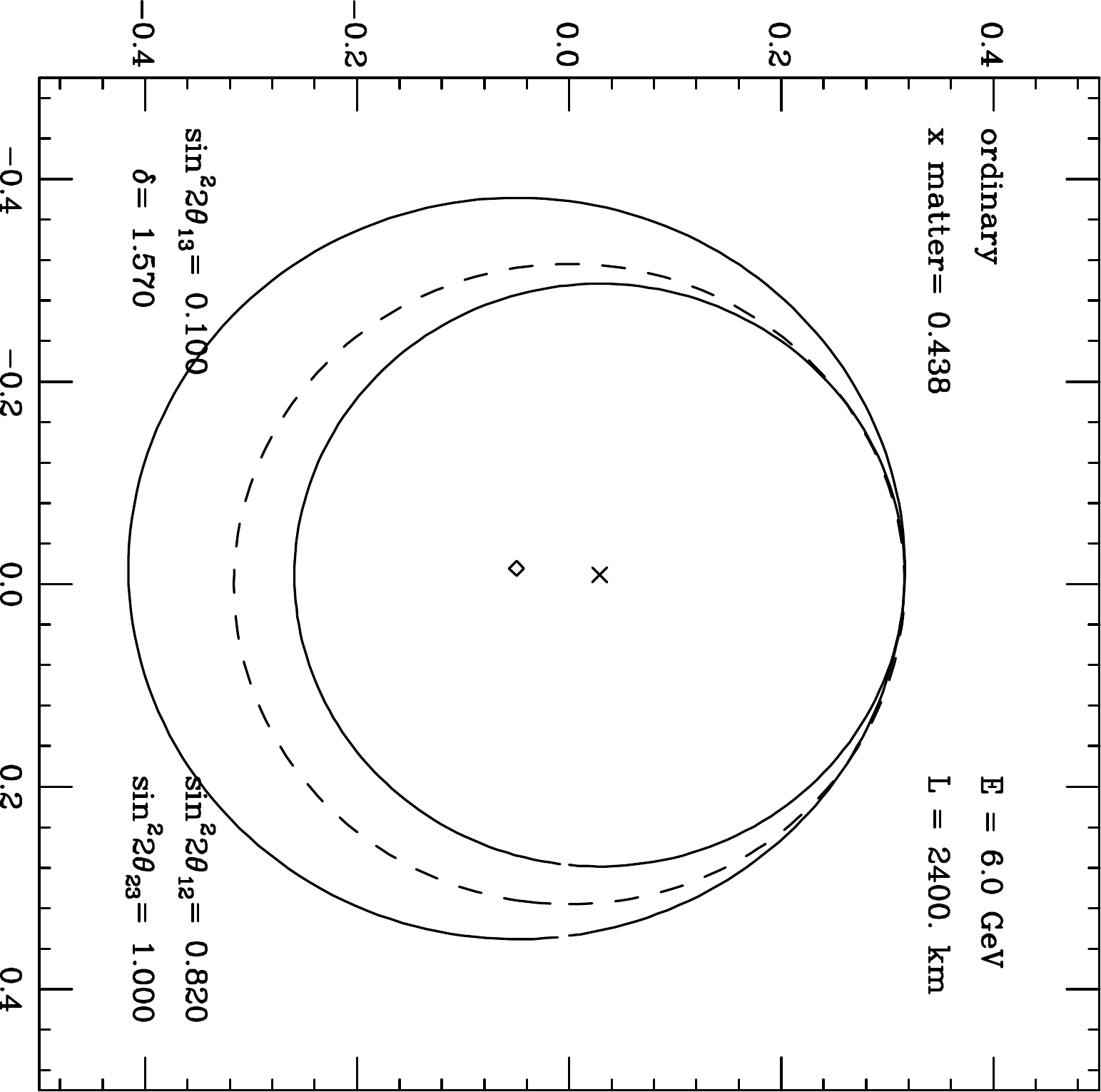}
\includegraphics[width=2.45in,angle=90]{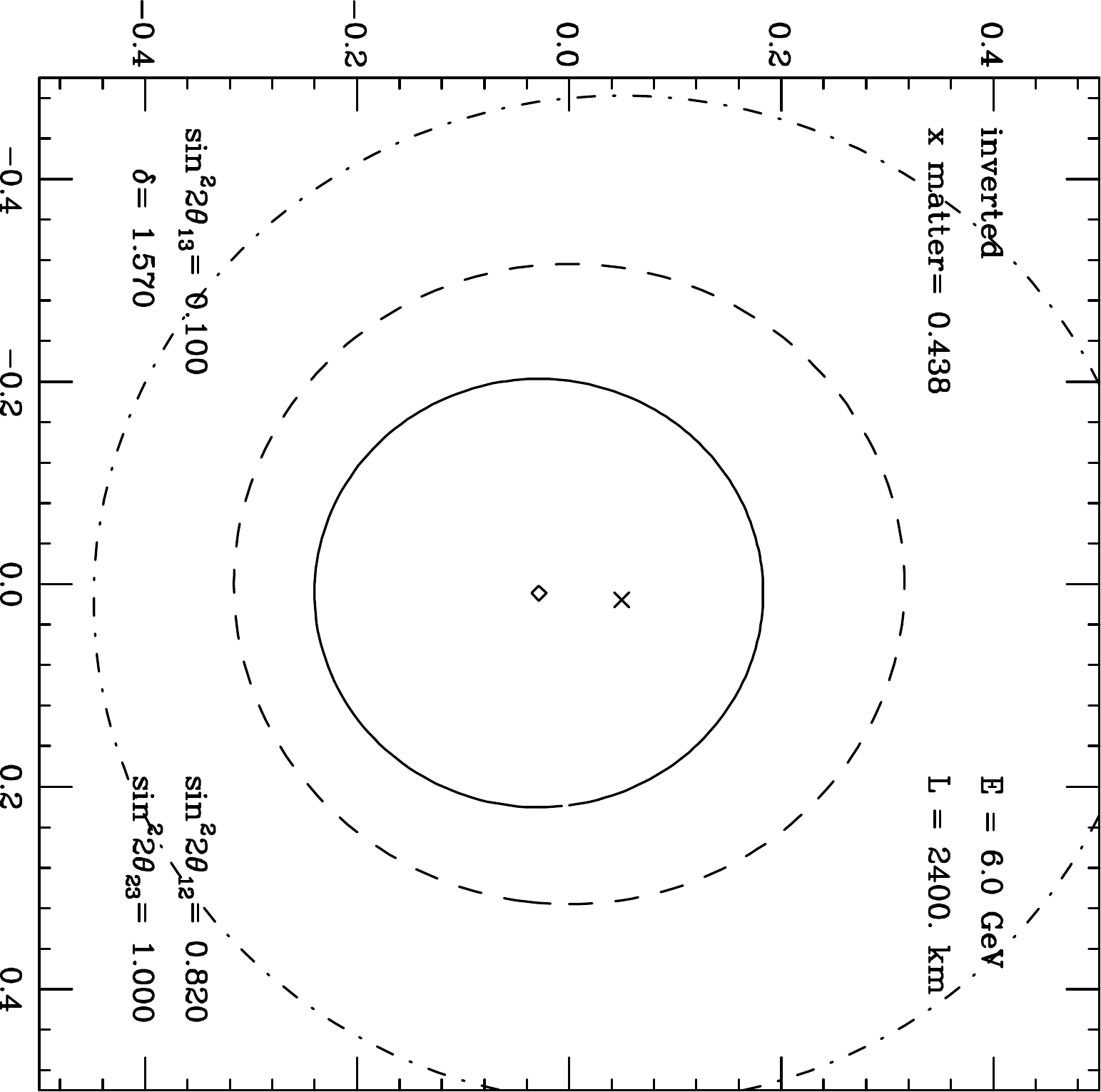}
\includegraphics[width=2.45in,angle=90]{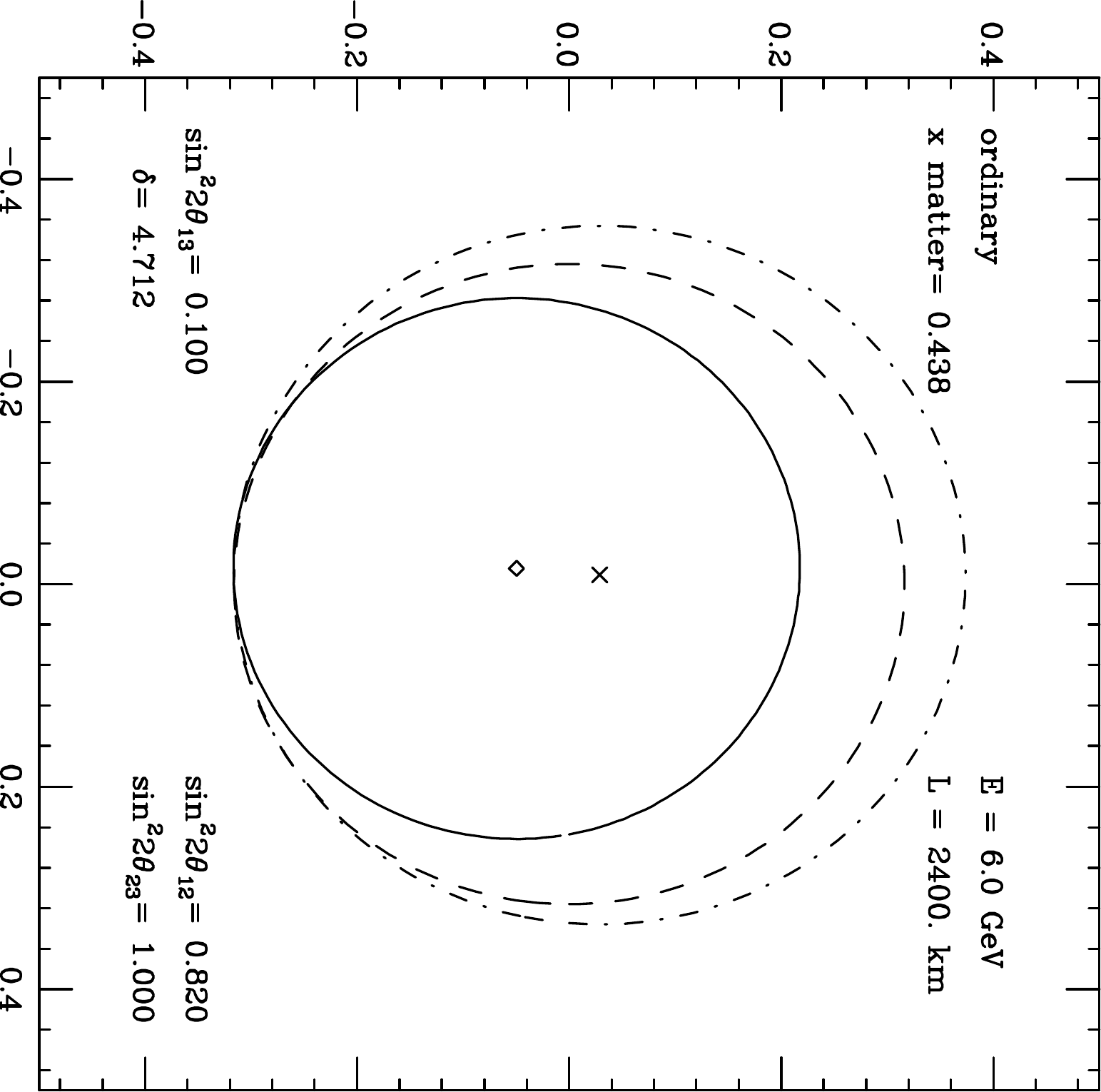}
\includegraphics[width=2.45in,angle=90]{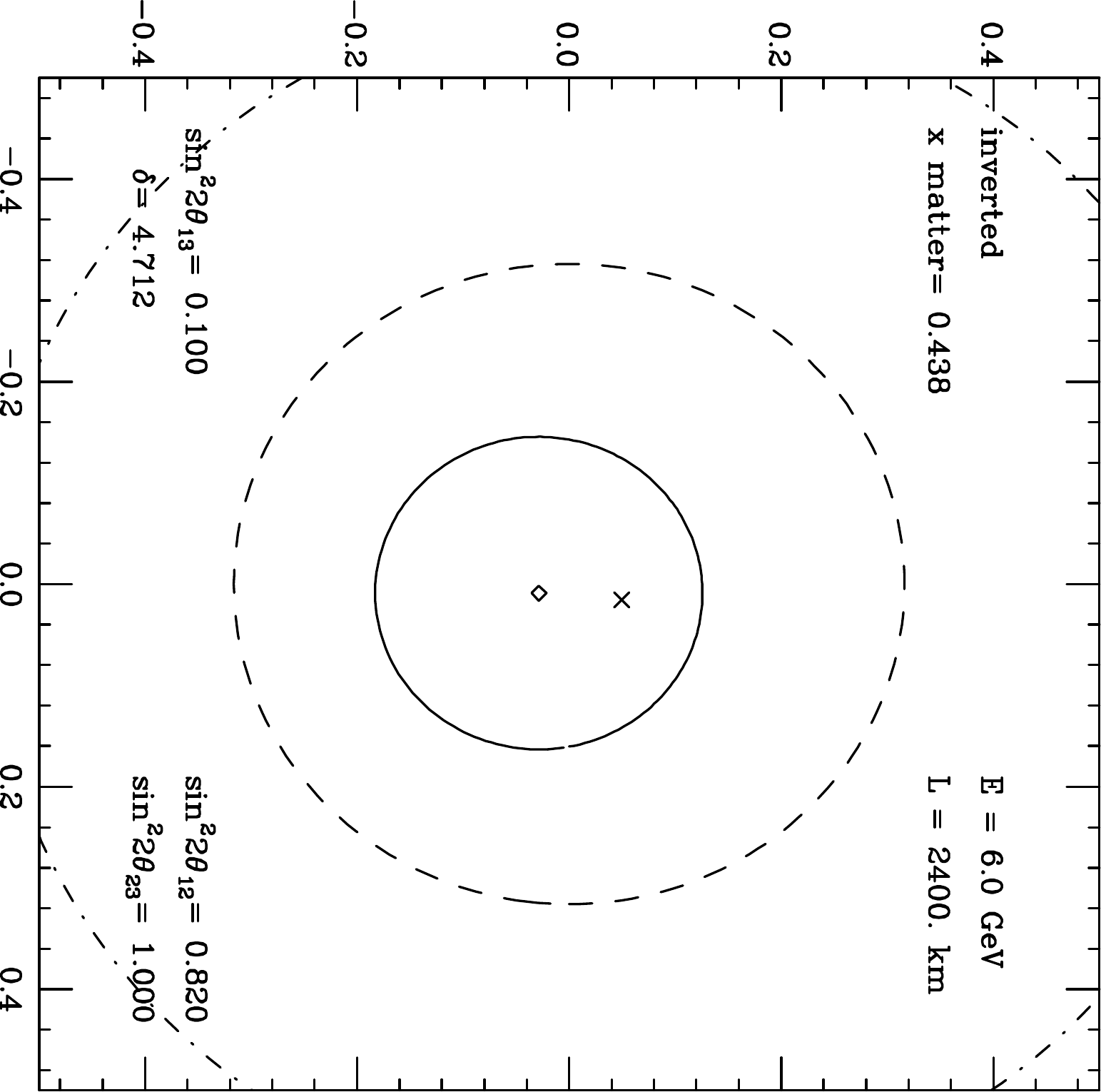}
\caption{ Plots for a hypothetical site 2400 km from the source.  Upper, with true hierarchy being normal, with $\delta_{CP}=\pi/2$.  Lower, again with true hierarchy being normal, but now with $\delta_{CP}=3\pi/2$. The dot dash circles show the constraints from $\nu_\mu$ scattering, while the solid circles are for $\nubar_\mu$.   The dashed curve shows the constraint from reactor neutrino experiments. \label{fig_three}}
\end{center}
\vspace{-1.4cm}
\end{figure}

The European version of the long-baseline neutrino oscillation
experiment is known as LAGUNA-LBNO~\cite{nonus:LAGUNA,
  nonus:EuroNuStrategy} (Large Apparatus studying Grand 
Unification and Neutrino Astrophysics and Long-Baseline Neutrino
 Oscillations).   The current cycle of design studies
(2011-2014) is focused on the conventional neutrino beam originating
at CERN, aiming at a deep underground site in Pyh\"asalmi, Finland, at
a baseline of 2300~km.  Fig.~\ref{fig_three}
illustrates the advantages of such a long baseline. 

Three detector concepts are being developed for
that site: GLACIER (Giant Liquid Argon Charge Imaging ExpeRiment), a
LAr dual-phase TPC scalable to a total mass of up to 100 kTon; MIND,
a 25 kTon magnetized iron-scintillator calorimeter similar to
MINOS~\cite{nonus:LBNO_LOI}; and
LENA (Low Energy Neutrino Astronomy), a $\sim$50 kTon liquid
scintillator detector. In addition, a MTon-scale water Cherenkov
detector MEMPHYS (MEgaton Mass PHYSics) is being considered for a site
in an extended 
Modane Laboratory at a very short baseline of 130 km from CERN. All
cases include a near detector (upgraded SHINE/NA61), a deep
underground location (2500 mwe for GLACIER, 4000 mwe for LENA, 4800
mwe for MEMPHYS), and a range of neutrino beam options, starting with
an 800 kW wide-band beam (for GLACIER and LENA) to an upgrade to 4 MW
(for MEMPHYS), to high-power $\beta$-beams at CERN (for MEMPHYS), and
ultimately to a neutrino factory.

Each detector offers a different set of
complementary physics measurements: GLACIER and LENA, with a very long
baseline and large underground detector, will be able to resolve the
neutrino mass hierarchy quickly, while also providing excellent CP
reach and opportunities to search for proton decay and supernova
neutrinos. The physics case for such detectors is well described in
Section~\ref{s:lb}. In addition, LENA, by virtue of its low energy
threshold, would be a large-scale solar neutrino experiment. MEMPHYS offers a different tradeoff: a clean
measurement of the CP phase $\delta_{CP}$ without significant matter
effects, increased detector mass (compared to SuperK) for proton decay
searches, and sensitivity to the mass hierarchy via measurements
of atmospheric neutrinos. The physics reach of MEMPHYS is similar to that of
HyperK. 

\begin{figure}[b]
\vspace{-0.1cm}
\begin{center}
\includegraphics[width=0.47\textwidth]{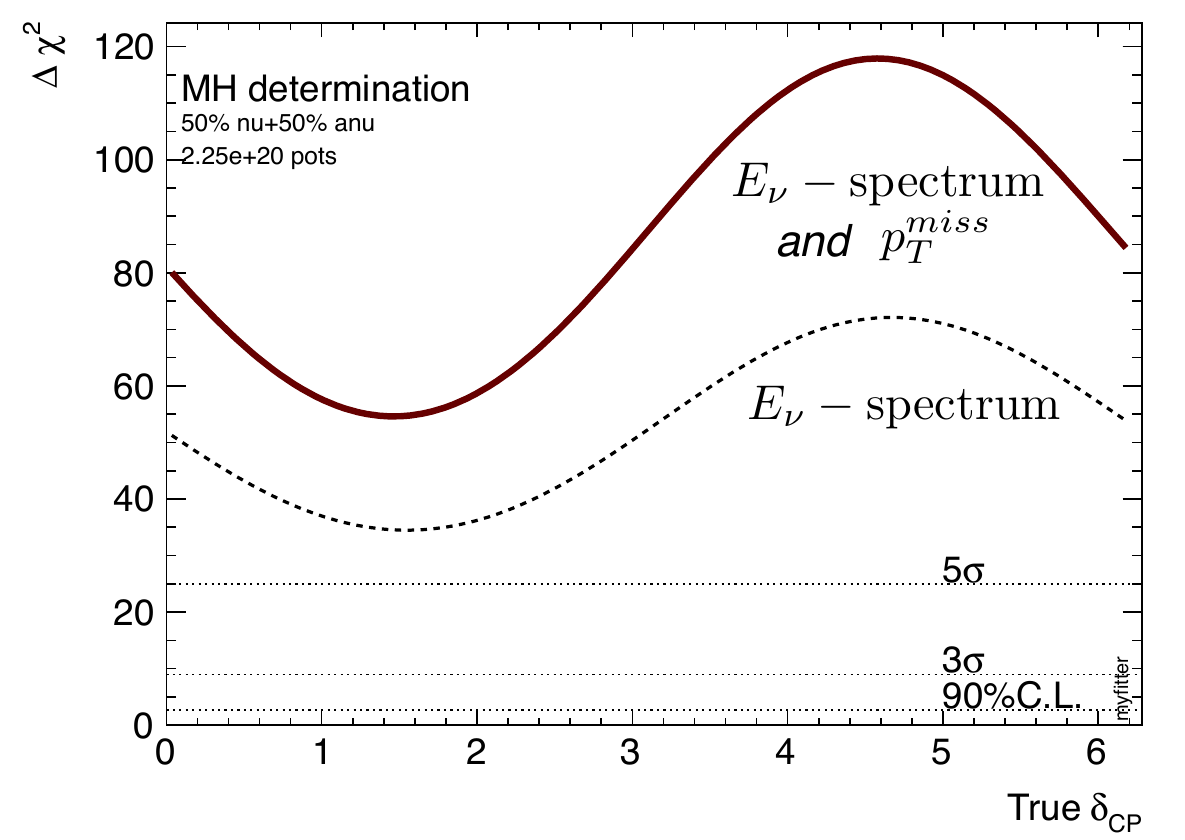}
\includegraphics[width=0.47\textwidth]{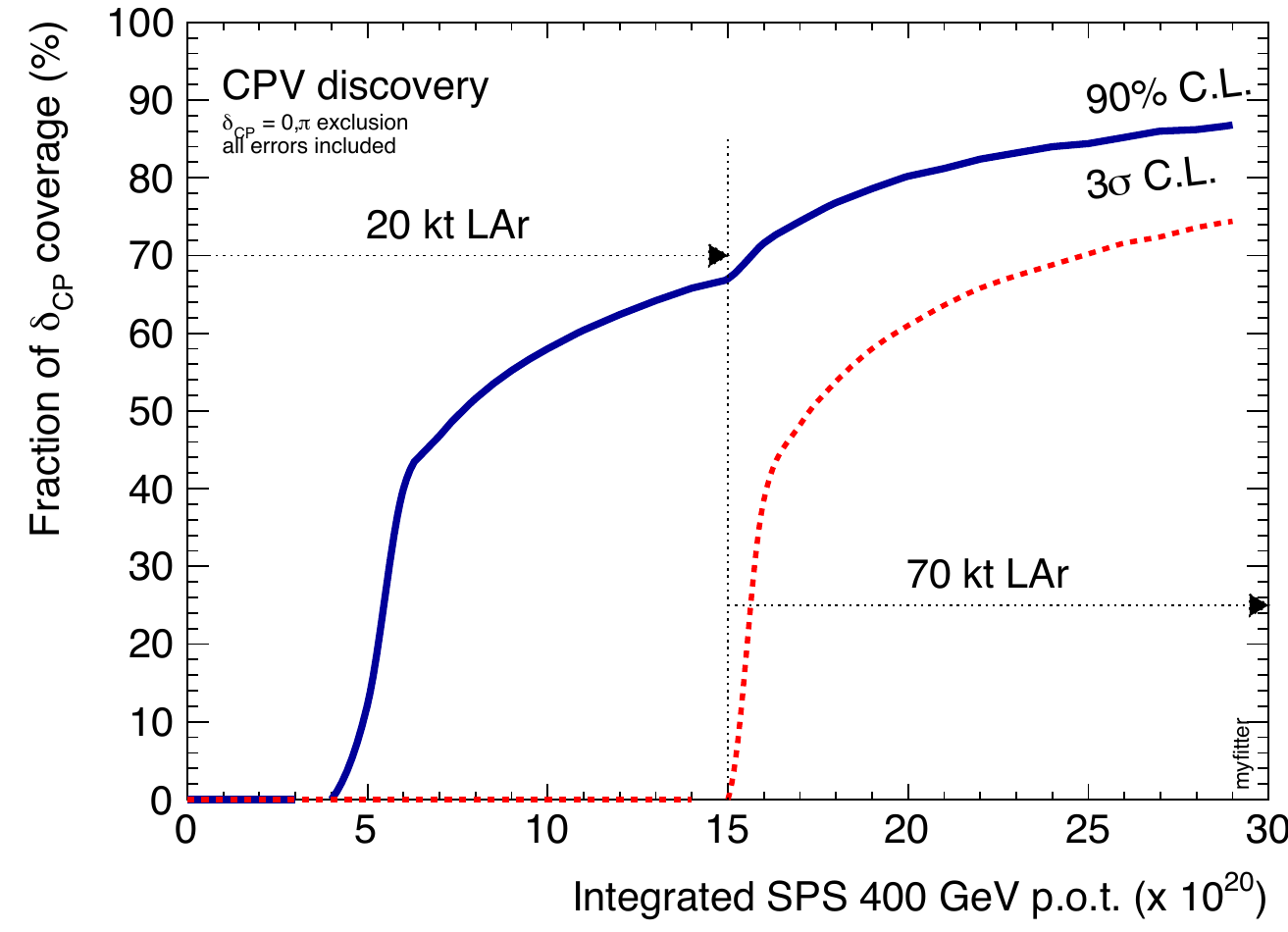}
\caption{LAGUNA sensitivity to the mass hierarchy as a function of
  $\delta_{CP}$
  (left)~\cite{nonus:LBNO_LOI}. Sensitivity to
  $\delta_{CP}$ (right) for staged detector construction. } 
\label{fig:LAGUNA_sens}
\vspace{-0.35cm}
\end{center}\end{figure}

GLACIER, by the virtue of its excellent tracking capabilities, large
mass, and a very long baseline, can offer an extremely sensitive measurement 
of the mass hierarchy (Fig.~\ref{fig:LAGUNA_sens}). It would be able
to resolve the mass 
hierarchy with a significance of more than $5\sigma$
 in a few years of operation with a
conventional SPS beam ($\approx 2\times10^{20}$ protons on target). If
a decision on construction is reached by 
2015, the proponents estimate a $5\sigma$ measurement by
2023~\cite{nonus:EuroNuStrategy,nonus:LBNO_LOI}. The high sensitivity
would possibly enable a staged 
approach to the project: a smaller (20 kTon) detector that starts operating
quickly, followed by a larger detector, near detector upgrades, beam
intensity upgrades, etc, aiming at a 30-year program for precision
measurement of $\delta_{CP}$, neutrino astrophysics, and proton
decay searches.  
LENA would require about 10 years of operations to reach $>5\sigma$
sensitivity to the mass hierarchy, with the completion timescale similar
to that of LBNE and HyperK. 

While LAGUNA-LBNO claims the best sensitivity to the mass hierarchy
among the options we have considered, it is not yet a fully
fleshed-out project. The current study of the detector configuration
options is funded through 2014, at which point the proponents are
planning to submit a full proposal to
CERN~\cite{nonus:LAGUNA,nonus:EuroNuStrategy}. As of 
this writing, no cost estimates are available, although one could
scale from LBNE and HyperK estimates and come up with numbers in the range of a
few billion Euro. At that scale, the neutrino projects will be
in direct competition with the LHC upgrades, any proposals for a
high-energy $e^+e^-$ collider, etc. The recently updated 2013 European
Strategy for Particle Physics document~\cite{nonus:EuroNuStrategy}
lists neutrino physics as priority \#4 (last) and states: 
\begin{quotation} ``CERN should 
develop a neutrino programme to pave the way for 
a substantial European role in future long-baseline 
experiments. Europe should explore the possibility 
of major participation in leading long-baseline 
neutrino projects in the US and Japan''.\end{quotation} Taken at face value, this
indicates a luke-warm interest for an expensive neutrino project in
Europe, although other opinions exist. 

\subsubsection{Open Concerns}

The biggest concern for the long-baseline projects in Japan and Europe
is the fact that these projects are not fully funded in
their host countries, and so they remain virtual for now. It is
difficult to imagine the world-wide scientific community supporting
more than one expensive long-baseline effort, especially in the face
of other priorities in high energy physics. The European
Strategy document seems to openly acknowledge this fact. However,
should the US LBNE effort falter, or should the political and economic
climates change, the off-shore projects may provide viable
alternatives (and in some scenarios stiff competition) to LBNE in the
race to unambiguously and decisively determine the neutrino mass
hierarchy.

\section{Reactor Neutrinos}\label{s:reac}
A signature of the neutrino mass hierarchy is present in the
oscillation of anti-neutrinos from nuclear power
reactors~\cite{jlearned_PRD_2008}.  A very precise, high-statistics
measurement of the reactor anti-neutrino energy spectrum is required
to determine the hierarchy.  JUNO (formerly known as Daya Bay II) is
the only experiment 
proposed to measure the hierarchy by this
method~\cite{yfwang_INPA_2013}.  


The
three-flavor electron anti-neutrino survival probability is given by:
\begin{eqnarray} \label{eq:elecSurv}
P(\overline{\nu}_e\rightarrow \overline{\nu}_e) = 1&-&\cos^4\theta_{13}\sin^2 2\theta_{12} \sin^2 \Delta_{21} \nonumber \\
&-&\sin^2 2\theta_{13} (\cos^2\theta_{12}\sin^2 \Delta_{31}+\sin^2\theta_{12}\sin^2{\Delta_{32}}),
\end{eqnarray}
where $\Delta_{ij} = 1.27{\Delta}m^2_{ij}[\rm{eV}^2]L[{\rm m}]/E[{\rm
    MeV}]$.  
    
 The hierarchy signature is contained in the term proportional to $\sin^2 2\theta_{13}$ and can be written as
 \begin{equation}
\frac 12\left[1-\cos(\Delta_{31}+\Delta_{32})\cos\Delta_{21}+\cos 2\theta_{12}\sin(\Delta_{31}+\Delta_{32})\sin\Delta_{21}\right]
 \end{equation}  
 The quantity $\Delta_{21}$ is unambiguous.  If the hierarchy is normal, then   $\Delta_{31}+\Delta_{32}$ is positive, while it is negative for the inverted hierarchy.  But this quantity is only the phase, whose sine must be evaluated.  By increasing or decreasing $\Delta_{31}+\Delta_{32}$ slightly, one hierarchy can be made to emulate the other over a modest range of energy, outside which the two solutions will no longer be in phase.  Thus to distinguish the inverted from the normal hierarchy we must measure the oscillations, suppressed by $\sin^2 2\theta_{13}$ over many cycles.  Additional discrimination can come from precision measurements of combinations of $\Delta m^2_{31}$ and $\Delta m^2_{32}$.
 
Disappearance measurements determine the effective mass-squared differences for different channels of neutrino oscillation:
\begin{eqnarray}
\Delta m^2_{ee} &\approx& \cos^2\theta_{12} \Delta m^2_{31} + \sin^2\theta_{12} \Delta m^2_{32} \\
\Delta m^2_{\mu\mu}&\approx& \sin^2\theta_{12} \Delta m^2_{31} + \cos^2\theta_{12} \Delta m^2_{32} + \sin 2 \theta_{12}\sin\theta_{13}\tan\theta_{23}\cos\delta_{CP}\Delta m^2_{21},
\end{eqnarray}
where terms of order $\sin^2\theta_{13} \Delta m^2_{21}$ have been neglected for simplicity.  Improvements here could constrain the fits to JUNO data.

\begin{figure}[b]
\centering
\includegraphics[width=0.7\textwidth]{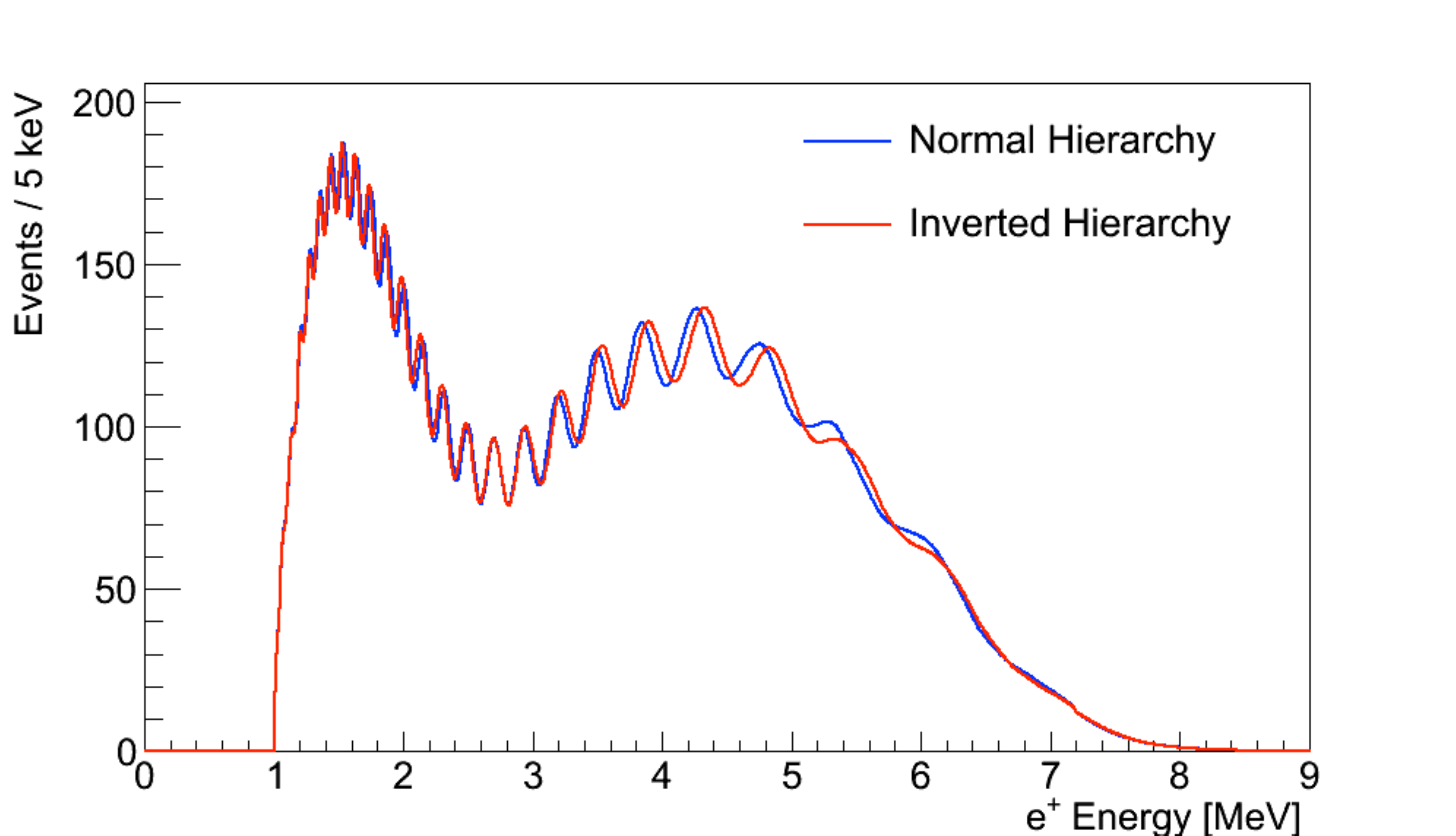}
\caption{Estimated positron energy spectrum of the inverse-beta decay reaction for
  an ideal reactor anti-neutrino experiment (20~kton detector,
  40~GW$_{th}$ reactor power at 58~km, 5~years operation).  The solar
  oscillation (${\Delta}m^{2}_{21}$) causes the broad deficit between
  2-4~MeV.  The normal (blue) versus inverted (red) hierarchy 
results in an
  effective shift in the phase of the high-frequency oscillation.  In this example, $\Delta m^2_{31}$ has been changed to  $-\Delta m^2_{31}$ in inverting the hierarchy.}
\label{fig:reactorTrue_NoDegeneracy}
\end{figure}

Reactor anti-neutrinos are commonly detected via the inverse beta-decay reaction,
$\bar{\nu}_{e} + p \rightarrow e^{+} + n$.  The outgoing positron
energy preserves information of the original anti-neutrino energy,
$E_{e+} \simeq E_{\bar{\nu}} - 0.8 MeV$.
Fig.~\ref{fig:reactorTrue_NoDegeneracy} shows the positron energy
spectrum (including annihilation) for an ideal reactor experiment
(20~kton detector, 40~GW$_{th}$ reactor power at 58~km, 5~years of
operation, ${\Delta}m^2_{32}$=$\pm2.32\times10^{-3}$~eV$^{2}$).  The
primary deficit between 2 and 4~MeV is due to solar oscillation at
58~km.  The high-frequency oscillation is due to the atmospheric
mass difference ${\Delta}m^{2}_{3X}$.  A choice of the mass hierarchy
appears as a difference
in the phase of the oscillation.

JUNO is a liquid-scintillator detector very similar to KamLAND in
design, but twenty times larger in mass (20 kton)~\cite{yfwang_INPA_2013}.  It
is proposed to be built in the Jiangmen city of Guangdong province in
southern China, roughly 70 km from Macau.  The Yangjiang
(17.4~GW$_{th}$) and Taishan (18.4~GW$_{th}$) reactor facilities serve
as the electron anti-neutrino source.  The detector is located at the first
solar oscillation minimum, approximately 58~km from both reactor
facilities.

An effort based in Korea also intends to build a large scintillating
reactor anti-neutrino experiment (RENO-50). 
The original design had a marginal sensitivity to
the mass hierarchy, due to a smaller 10~kton target 
mass and poorer detector resolution than JUNO.  Recent changes to the
design goals~\cite{KimRENO} aim at a detector size of $\mathcal{O}(20)$~kton and sensitivity
similar to JUNO. RENO-50 plans to submit a Letter of Intent to funding
agencies in 2013.  

To determine the neutrino mass hierarchy, JUNO and RENO-50 will
need to improve over current standards for energy resolution and
calibration.  The JUNO collaboration's goal 
is to achieve a detector resolution of $\sigma_{E}/E$=3\%/$\sqrt{E_{e+}{\rm [MeV]}}$
in order to measure the hierarchy.  For
KamLAND, the energy resolution was limited to 6.5\%/$\sqrt{E}$ by
photoelectron statistics.  An improved resolution requires more
photons per unit of positron energy be detected.  The JUNO
experiment aims to increase photon statistics by increasing the
scintillator light yield ($\times$1.5), increasing the total
photocathode coverage ($\times$2.3), and by increasing the
efficiency of each photomultiplier ($\times$2.0) relative to those of KamLAND.
Improving the light yield and PMT efficiency have not yet been
demonstrated; both require significant technological advances.
Increasing the photocathode coverage is straightforward, only
requiring sufficient funds to purchase $\sim$15,000 20"-diameter PMTs.

The current state of the art in liquid scintillator is $\sim$2\% for gamma-ray calibration, as
demonstrated by the KamLAND experiment, although sub-percent level precision has been achieved in water Cherenkov detectors such as SNO.  
In liquid scintillator, particle-dependent scintillator quenching introduces a comparable
systematic uncertainty when estimating the positron calibration from
gamma-ray data.  The JUNO group has addressed the problem of non-linearity by fitting the actual observed energy spectrum and parameterizing the non-linearity, a process of self-calibration.

\begin{figure}[tb]
\centering
\includegraphics[width=0.7\textwidth]{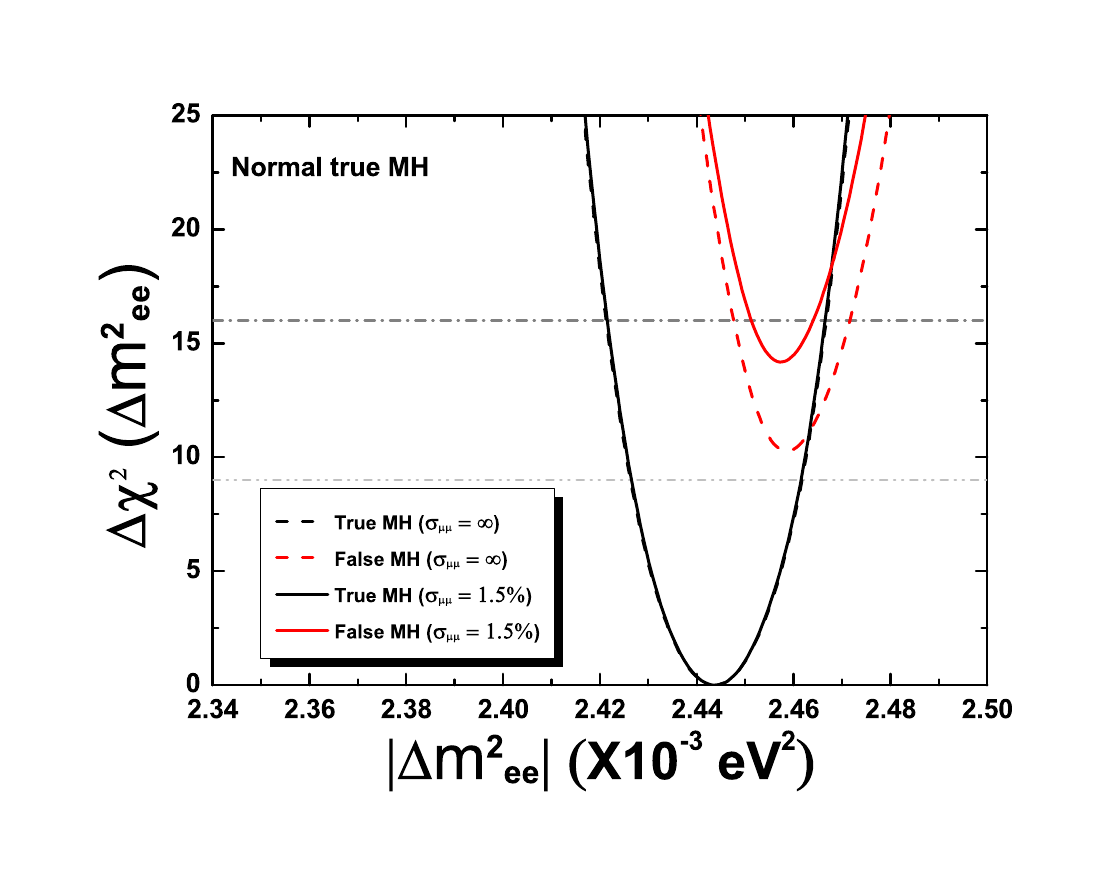}
\caption{${\Delta}\chi^2$ distributions obtained for the normal
  (black) and inverted (red) hierarchy models as a function of
  ${\Delta}m^2_{ee}$, taken from Ref.~\cite{liyf_dyb2_2013}.  The
  degeneracy due to uncertainty in the mass-squared difference is
  visible as the false minimum for the inverted hierarchy at a value
  of ${\Delta}m^2_{ee}$ 0.5\% from the true value.  Including a
  penalty based on a 1.5\% global uncertainty in
  ${\Delta}m^2_{32}$ disfavors the inverted hierarchy at an
  additional $\sim$0.5$\sigma$ (solid red versus dashed red). }
\label{fig:dyb2Sens}
\vspace{-0.3cm}
\end{figure}

The JUNO group has produced an estimate of their sensitivity in Ref.~\cite{liyf_dyb2_2013}.  Fig.~\ref{fig:dyb2Sens} shows the
${\Delta}\chi^2$ distributions obtained for both the normal and
inverted hierarchy models as a function of ${\Delta}m^2_{ee}$.  The
degeneracy due to uncertainty in the mass-squared difference is
visible as the false minimum for the inverted hierarchy at a value of
${\Delta}m^2_{ee}$ 0.5\% from the true value.  
The collaboration finds a sensitivity to the hierarchy of more than $3\sigma$ under the assumption of a 3\%$\sqrt E$ detector energy resolution, 2\% correlated and 0.8\% uncorrelated uncertainties in reactor flux, 1\% uncertainty in the reactor flux spectrum, and 1\% uncertainty in the detector response.  This analysis takes the best-fit oscillation parameters from the most recent global analysis, and takes into account the true spatial distribution of reactor cores.  When allowing for potential future improvements in the uncertainty in $\Delta m^2_{32}$ this significance can be improved to more than $4\sigma$.  
This sensitivity has been confirmed by an independent study~\cite{reac:indep} under the same assumptions for detector performance and future precision on oscillation parameters.

\begin{figure}[htb]
\centering
\includegraphics[width=0.7\textwidth]{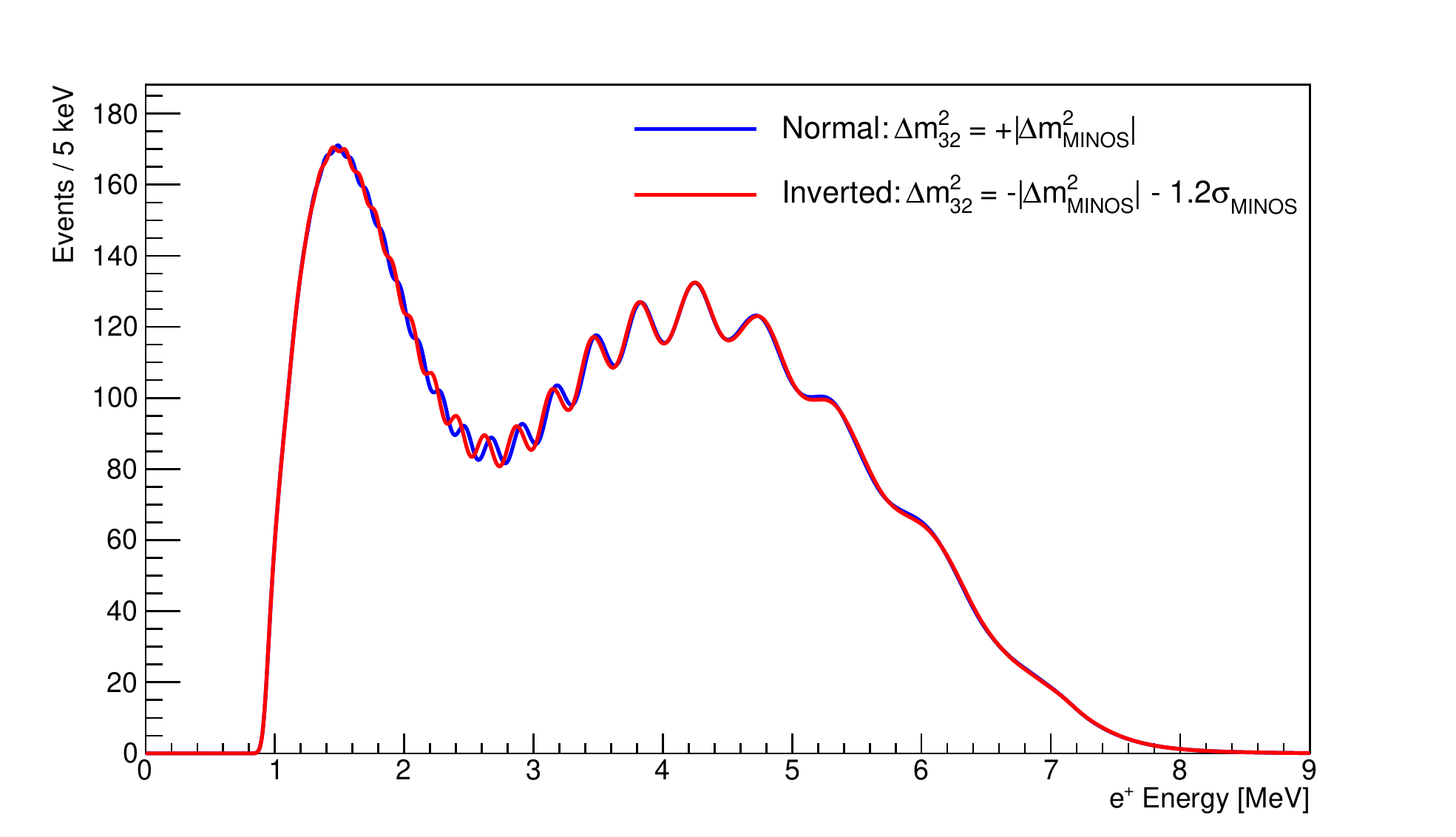}
\caption{Estimated positron energy spectrum of the inverse beta-decay reaction 
  for an ideal reactor anti-neutrino experiment (20~kton
  detector, 40~GW$_{th}$ reactor power at 58~km, 5~years operation).
  A detector energy resolution of
  $\sigma_{E}/E$=3\%/$\sqrt{E_{e+}{\rm [MeV]}}$ has been applied,
  reducing the oscillation signal below 2~MeV.  The mass difference
  for the inverted case has been shifted by 1.2$\sigma$ from the current MINOS
  best estimate (a shift of 5\%), removing the hierarchy discrimination above 3~MeV.
 }
\label{fig:reactorDet_WithDegeneracy}
\vspace{-0.3cm}
\end{figure}

The effect of finite detector resolution is to 
smear out the oscillation signal at lower positron energies.
Fig.~\ref{fig:reactorDet_WithDegeneracy} shows the expected positron
spectra, including
a detector resolution of $\sigma_{E}/E$=3\%/$\sqrt{E_{e+}{\rm [MeV]}}$ and a maximally-ambiguous 5\% shift in ${\Delta}m^2_{32}$ ($\sim 1.2\sigma$ based on the current global best fit).  The residual differences in the spectra of the
two hierarchies are limited to the 2-3~MeV region.  
Ref.~\cite{liyf_dyb2_2013} considers the effect of an improved uncertainty in ${\Delta}m^{2}_{32}$ and finds that reduction of the uncertainty to 1.5\% would result in a hierarchy significance of $3.7\sigma$, and $4.4\sigma$ if a 1\% uncertainty could be achieved.  The additional penalty to $\Delta \chi^2$ is illustrated in Fig.~\ref{fig:dyb2Sens}.

Current accelerator neutrino
experiments (T2K, NO$\nu$A) will, it is hoped, reduce the uncertainty on ${\Delta}m^{2}_{32}$ by
a factor of two.   Estimates give a global uncertainty of
1.5\% in ${\Delta}m^{2}_{32}$ by 2025.  Further reduction in this uncertainty will come from future long-baseline experiments (Section~\ref{s:lb}).

The JUNO project has an aggressive schedule~\cite{yfwang_INPA_2013}.
It has already received the Chinese equivalent of CD-1 approval.
Civil construction is targeted for 2014-2017.  Detector assembly is
planned for 2018-2019.  Data taking would commence in 2020, with a
target of 5-6~years of operation for the hierarchy measurement.

 The main challenges for JUNO are technological.  To obtain
sufficient detector resolution requires multiple factors of two
improvements in detector technology: improved scintillator light yield,
attenuation length, and PMT efficiency.  Constraints on detector
uniformity and linearity are demanding, and will likely require the
development of new methods to calibrate the detector response to
positrons.   These challenges must be met successfully if the subtle effect of the neutrino mass hierarchy on the observed positron signal is to tell us whether the neutrino mass hierarchy is normal or inverted.

\section{Atmospheric Neutrinos}\label{s:atm}

Atmospheric neutrinos have the potential to resolve the neutrino mass
hierarchy via matter-enhanced oscillation within the Earth.  Resonant
oscillation occurs either for neutrinos in the case of the normal
hierarchy, or antineutrinos for the inverted hierarchy.  Determination
of the hierarchy requires measurement of the energy and direction of
Earth-crossing atmospheric neutrinos with energies in the range of 2
to 10~GeV. Massive detectors ($\gtrsim$Mton) are required to obtain
sufficient signal statistics within a few years of operation.
Existing proposals use either water Cherenkov (PINGU, ORCA, HyperK),
liquid Argon TPC (LBNE, LBNO), or magnetized iron calorimeter (INO)
detectors.  Discrimination of neutrinos from antineutrinos enhances
hierarchy sensitivity.  Hierarchy determination has some dependence on
the oscillation parameters, in particular ${\Delta}m^2_{31}$, but is largely
insensitive to the 
Earth density
profile.  Primary concerns are detector properties such as total mass,
energy resolution, and angular resolution.

\subsection{Signature of the neutrino mass hierarchy}\label{atm:sign}

The two possible neutrino mass hierarchies predict distinctly
different oscillation probabilities for Earth-crossing neutrinos.  Due
to interactions with electrons within the Earth, resonant flavor
conversion occurs at a specific pattern of neutrino energies and
Earth-crossing paths.  This resonant conversion only occurs for
neutrinos in the case of the normal hierarchy, while only for
antineutrinos for the inverted hierarchy.  A detector capable of
discriminating $\nu$ interactions relative to $\overline{\nu}$ needs
only to demonstrate for which state the resonance occurs.  Detectors which
only distinguish neutrino flavor rely on the intrinsic difference in
the atmospheric flux between $\nu$ and $\overline{\nu}$ as well as
differences in interaction cross-sections to discriminate the
hierarchy.

\begin{figure}[tp]
\begin{center}
\includegraphics[width=7.5cm]{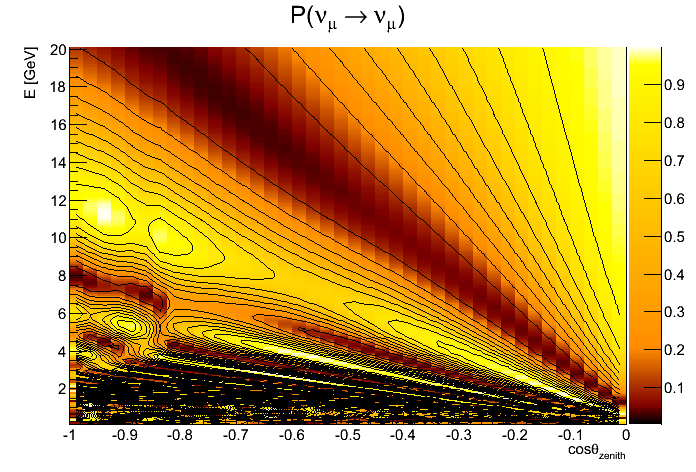}
\includegraphics[width=7.5cm]{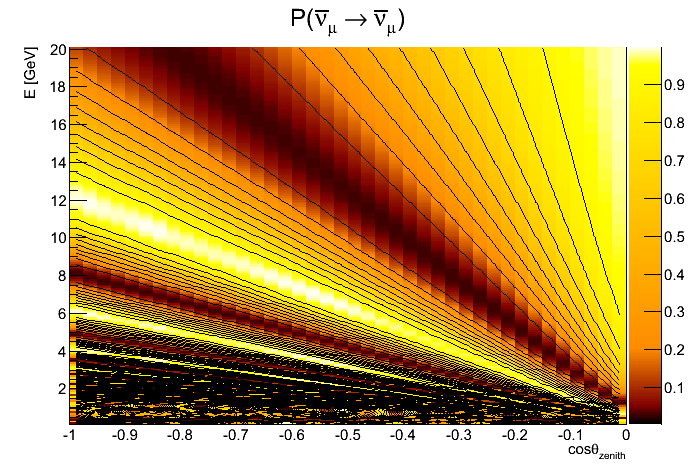}
\caption{The estimated oscillation survival probability for
  Earth-crossing $\nu_{\mu}$ (left) and $\overline{\nu}_{\mu}$
  (right), assuming the normal neutrino mass hierarchy.  Resonant
  oscillation in the mantle is most significant in the region
  $\sim$6~GeV and cos$\theta_Z\simeq -0.7$, as indicated by the difference between the two panels.  For the inverted hierarchy,
  the resonance occurs for $\overline{\nu}_{\mu}$ instead of
  $\nu_{\mu}$.}
\label{fig:atmoNuProb}
\end{center}
\end{figure}

Resonant oscillation of Earth-crossing atmospheric neutrinos has been
extensively examined~\cite{atm:Petcov1998, atm:Akhmedov2008,
atm:Akhmedov}.  Oscillation probabilities are commonly presented as
Earth {\em oscillograms}, which show probabilities as a function of
neutrino energy and zenith angle $\theta_{Z}$.\footnote{Downward-going
neutrinos have cos$\theta_{Z}$=1, while upward-going neutrinos which
have crossed the Earth's core have cos$\theta_{Z}$=-1.}
Fig.~\ref{fig:atmoNuProb} shows the calculated oscillation
probabilities assuming the normal hierarchy.  

\begin{figure}[tbp]
\begin{center}
\includegraphics[width=7.5cm]{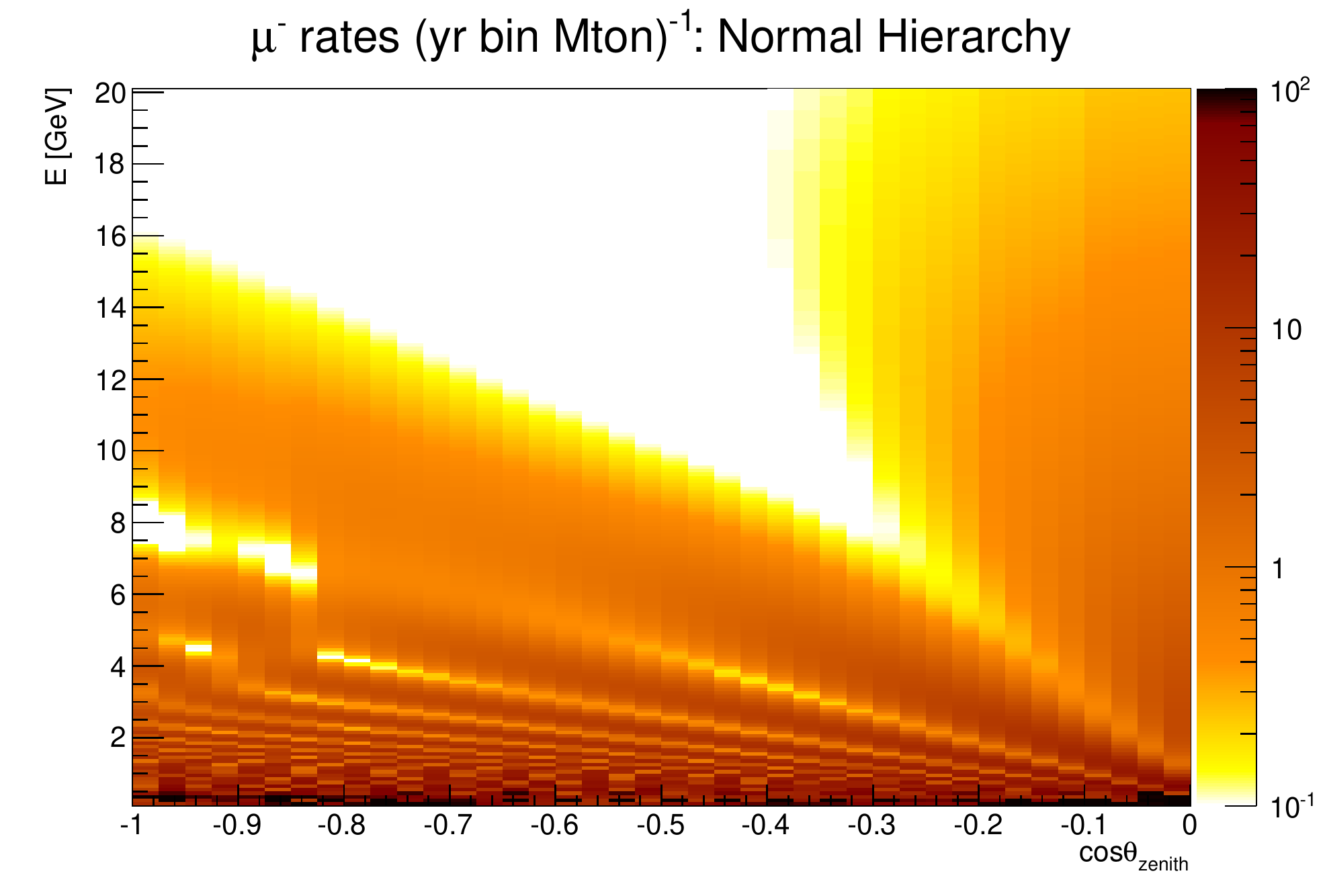}
\includegraphics[width=7.5cm]{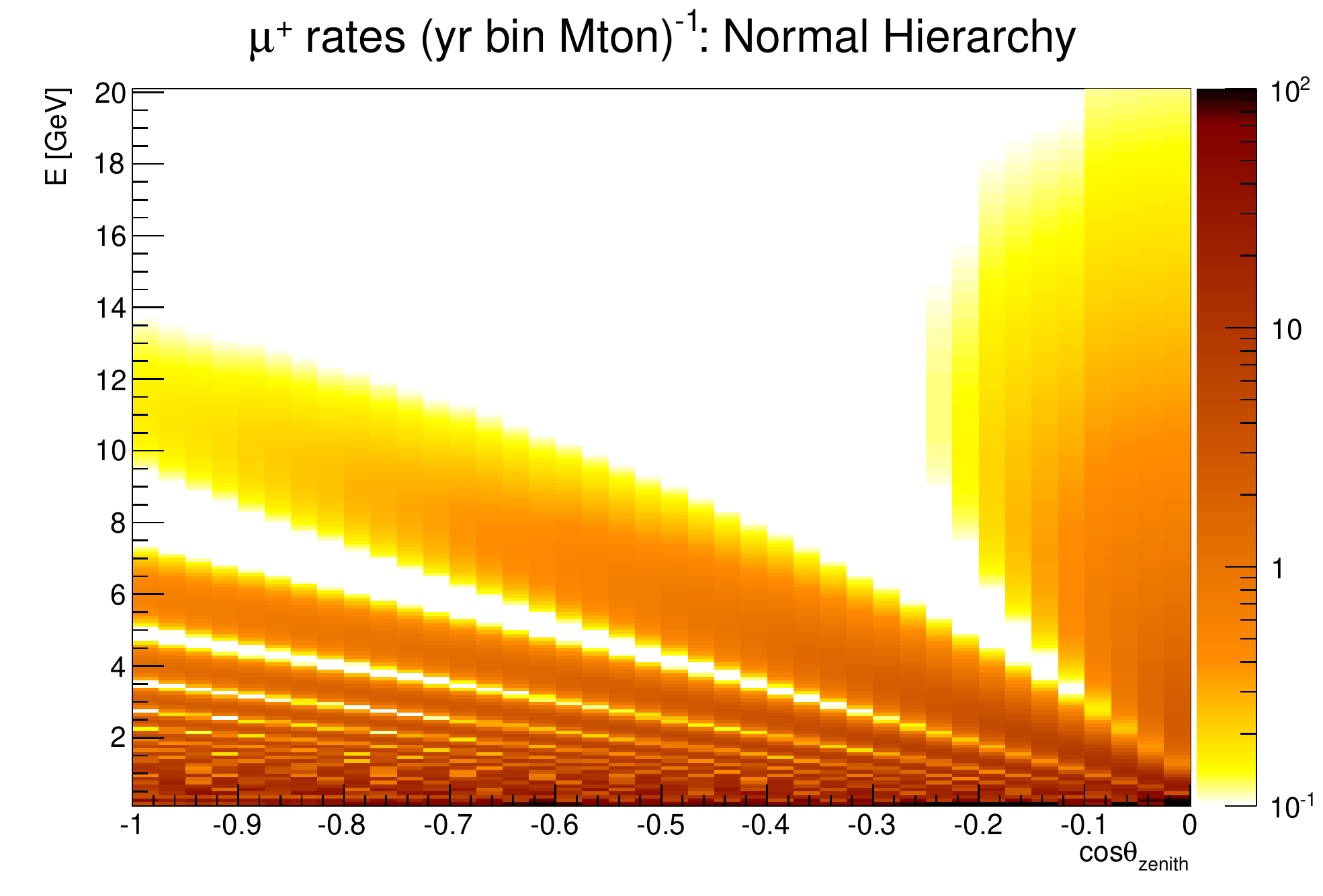}
\caption{The estimated rate of atmospheric neutrino generated $\mu^-$
  (left) and $\mu^+$ (right) per Mton-yr, assuming the normal neutrino
  mass hierarchy.  The rate is shown as a function of the incoming
  neutrino energy and direction.  The hierarchy is most significantly
  visible as a resonant enhancement of $\mu^-$ at $\sim$6~GeV and
  cos$\theta_Z\simeq -0.7$.  For the inverted hierarchy, the resonance
  occurs for $\mu^+$ instead of $\mu^-$.  For a detector insensitive
  to muon charge, the hierarchy produces a change in the rate
  according to the intrinsic $\nu$/$\overline{\nu}$ ratio of $\sim$1.3
  in this resonance region.}
\label{fig:atmoMuRates}
\end{center}
\end{figure}

 Assuming a 1~Mton detector,
$\sim$4000 muons per year are generated in the energy range of 2 to
10~GeV.  Fig.~\ref{fig:atmoMuRates} shows the rate of muon production
per Mton-yr assuming the normal hierarchy.  The major feature of the
hierarchy is the resonant excess of muons at $\sim$6~GeV and
cos$\theta_Z\simeq -0.7$.  The intrinsic $\nu$/$\overline{\nu}$ ratio of
$\sim$1.3 in the resonance region provides an observable signal of the
hierarchy, even for a detector insensitive to muon charge.

Sensitivity to the hierarchy is primarily driven by how well the
initial neutrino energy and direction can be determined.  It is clear
that as the energy resolution increases beyond a few GeV, the
resonance feature is obscured.  Intrinsic resolution is introduced by
the kinematics of the outgoing muon and nuclear effects for GeV $\nu$
interactions.   Of concern is the uncertainty in
${\Delta}m^{2}_{31}$, which has the strongest correlation with the
predicted oscillation pattern for muon-charge insensitive detectors.

\subsection{Proposed Experiments}\label{atm:expts}

\subsubsection{INO and MIND}

The India-based Neutrino Observatory (INO) will be located in the Bodi-West Hills in Pottipuram village
near Bodinayakanur in the Tamil Nadu State, with an overburden of about
1.3 km~\cite{atm:Dighe}.  Three
detector modules will form a 50-kT magnetic iron calorimeter.   The prominent features of the INO detector are its
capability of determining the charge and momentum of the the muon in
the charged-current interaction of an atmospheric muon
(anti-)neutrino, and the energy of the hadronic shower.  These
features would allow for discrimination between neutrinos and
anti-neutrinos as well as their energy.

MIND~\cite{nonus:LBNO_LOI} is a proposed 25~kT magnetized
iron-scintillator calorimeter, to be constructed in conjunction with
the European long-baseline effort in an underground lab in
Pyh\"asalmi, Finland. Its sensitivity to atmospheric neutrino
oscillations can be inferred from the INO numbers below. 

From a GEANT4-based simulation, the energy resolution of the hadronic
component, $\sigma/E$, is about 40\% for hadronic energies greater
than 4 GeV and is quite independent of the zenith angle of the
incident (anti-)neutrino.  The efficiency of correctly identifying the
charge of a track is better than 95\% for cos$\theta>0.35$ and
momentum greater than 1 GeV.  For momentum greater than 1 GeV, the
momentum resolution, $\sigma_p/p$, is better than 22\% and the
resolution in $\cos\theta$ is better than 0.045 for for
$\cos\theta>0.35$.

For sin$^2(2\theta_{13})$=0.1, the hierarchy can be resolved at 2
(2.7) standard deviations with 5 (10) years of running. The
sensitivity is limited by statistics. 

Civil construction for INO began in February 2013.  The first three detector
modules will be installed and commissioned in the underground
experimental hall by 2017. According to the schedule presented in Ref.~\cite{atm:Dighe},
excavation will take 1.5 years for a 2-km-long 7.5-m-wide tunnel,
which is quite an aggressive schedule when taking uncertainties in the
geotechnical conditions into account.

\subsubsection{PINGU}

Precision IceCube Next Generation Upgrade (PINGU) is a proposed
multi-megaton detector to be located below the dust layer in the inner
core of the existing IceCube and DeepCore detectors.  PINGU consists
of closely separated vertical strings of Digital Optical Modules
(DOMs) for detecting atmospheric neutrinos with energies down to a few
GeV.  In this configuration, IceCube and DeepCore serve as vetoes for
rejecting cosmic-ray muons both online and in the offline analysis.  The
DOM-to-DOM spacing is shorter than that of DeepCore.  The design
of the DOM is taken to be the one used in DeepCore.  The trigger
efficiency is expected to be much higher at lower energies for PINGU
than for DeepCore. 
The cost of constructing PINGU is about \$10 M for start up and \$1.25 M per string based on the 
IceCube experience.

\paragraph{Sensitivity}\label{atm:PINGUsen}

\begin{figure}[bp]
\begin{center}
\includegraphics[width=0.6\textwidth]{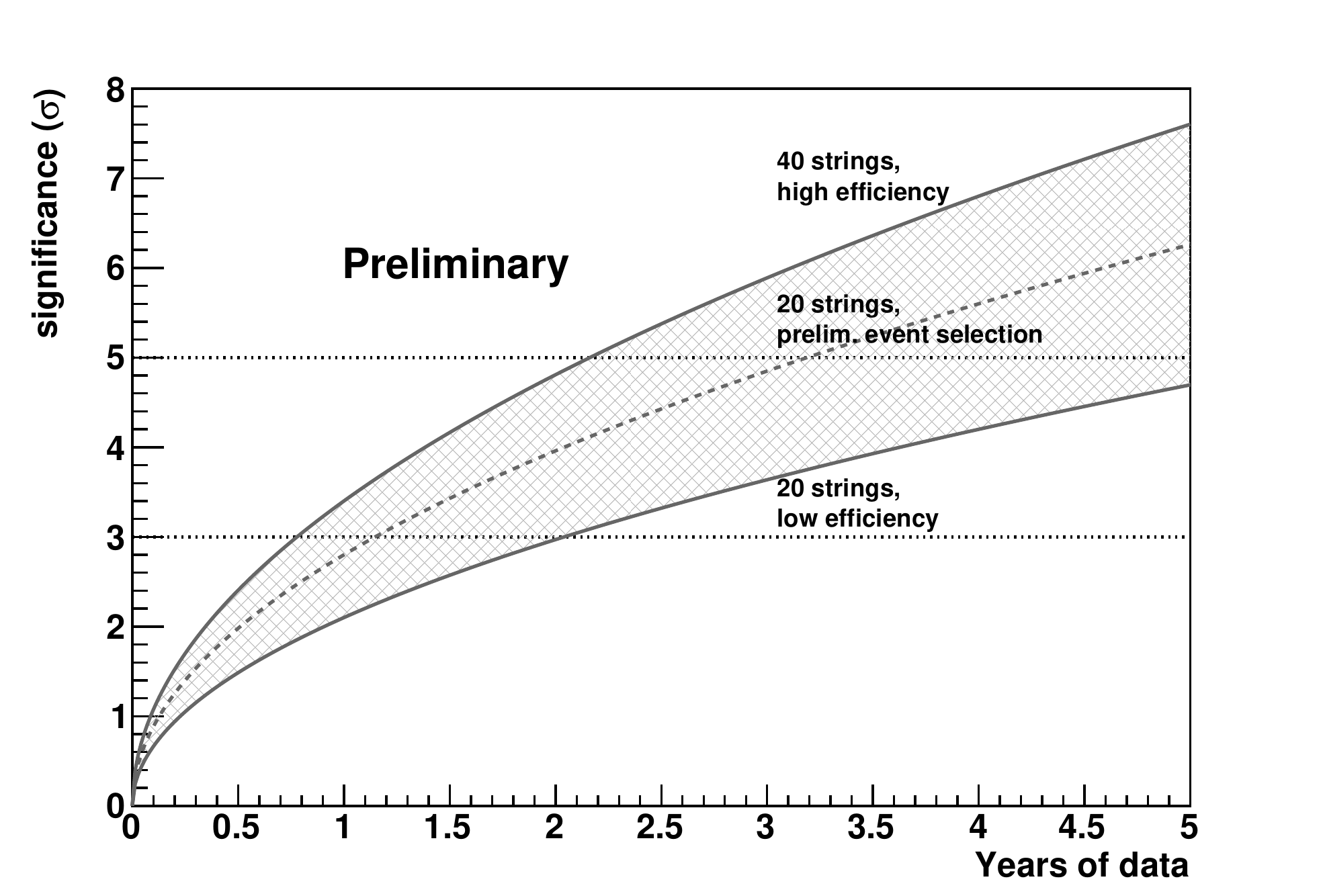}
\caption{Significance of determination of the mass hierarchy using
PINGU as a function of run time, for different detector configurations, by the proponents of the experiment~\cite{atm:pingu}.  }
\label{PINGU:senown}
\end{center}
\end{figure}

\begin{figure}[tbp]
\begin{center}
\includegraphics[width=0.52\textwidth]{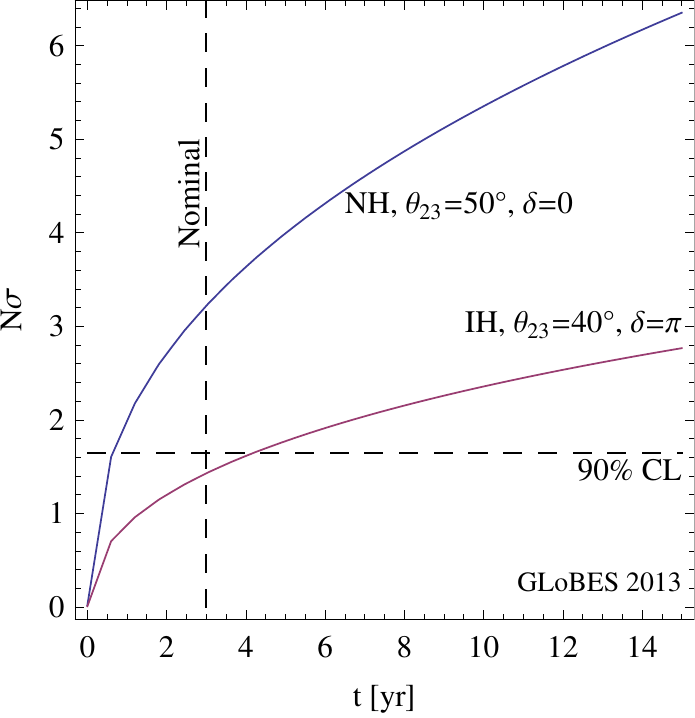}
\caption{Significance of determination of the mass hierarchy using
PINGU as a function of run time, for the best and worst case of the
true oscillation parameters, from an independent study~\cite{atm:Winter}.}
\label{PINGU:sen}
\end{center}
\end{figure}

The sensitivity
for determination of the mass hierarchy by PINGU has been
estimated both by proponents of the experiment~\cite{atm:pingu} and
independently~\cite{atm:Winter}, as shown in Figs.~\ref{PINGU:senown} and~\ref{PINGU:sen}, respectively.  
According to
the proponents, PINGU can achieve a sensitivity of $\sim 2.1$-$ 3.4 \sigma$ per year of data taking, resulting in a $3 \sigma$ 
discrimination between the normal and inverted hierarchies 
 by 2020, assuming initial deployment in 2016/17.  
 Details on the underlying assumptions and statistical techniques used to evaluate the sensitivity are somewhat limited at this stage, although a more detailed  Letter of Intent is in preparation~\cite{atm:pc}.  
The conclusions of Ref.~\cite{atm:Winter} are more modest, presumably because of differing assumptions in detector performance or statistical methodology.  Ref.~\cite{atm:Winter} points
out the complementarity between PINGU and the current accelerator and
reactor oscillation experiments (\NOvA, T2K, Daya Bay, RENO, Double
Chooz). 
As identified in Ref.~\cite{atm:Winter}, the key factors
influencing the experimental sensitivity are the current uncertainty
on $\Delta m^2_{31}$, the total active detector mass, energy
threshold, energy and 
angular resolution, and mis-identified cascades.  The existing studies
have not incorporated possible muon charge discrimination via
inelasticity of the neutrino interaction. %


It will take 1.5 years to procure and ship the detector components to
the South Pole.  Installation of all strings is expected to complete in
2-3 years.  If the proposal is submitted to the funding agencies in a
year, the full PINGU detector could be deployed by 2020.  Note that PINGU
can begin taking data as PMT strings are installed, slowly reaching
the full target mass by 2020.

Besides having relatively poor energy and directional resolutions, 
lack of charge determination and minimal particle identification 
could be an issue in achieving the scientific goals. 
Of primary concern is whether PINGU can achieve the required detector
performance characteristics.  These concerns with detector performance may be
mitigated by improvements in detector design, or even with incremental
extension of the detector after initial data has been obtained.

\subsubsection{ORCA}

Oscillation Research with Cosmics in the Abyss (ORCA) is a proposed
dense instrumentation of the central region of the KM3NeT detector in
the Mediterranean Sea~\cite{atm:Coyle}.  In principle, this is similar
in concept to PINGU, except the target is water instead of ice.  Funding of 40M Euros for KM3Net phase 1 has been secured, and 50-70
lines of sensors are expected to be ready for deployment by the end of
2016.  This comprises a total of 1200 pressure-vessel sensor modules,
each containing 31 3'' photomultiplier tubes.

\subsubsection{Hyper-Kamiokande}

Hyper-Kamiokande (HyperK) is a proposed multi-purpose water Cherenkov
detector, described in detail in Section~\ref{s:nonus}. 
The detector is based on well-known technology, and
is primarily a scale up of the existing Super-Kamiokande detector.  In
this respect, the expected performance in event reconstruction and
particle identification in the sub-GeV region is well-characterized.
The detector is  sensitive to and can discriminate both muons
and electrons.  However, it will have limited capability to determine
the sign of the lepton generated by the neutrino charged-current
interaction.

\begin{figure}[tp]
\begin{center}
\includegraphics[width=0.95\textwidth]{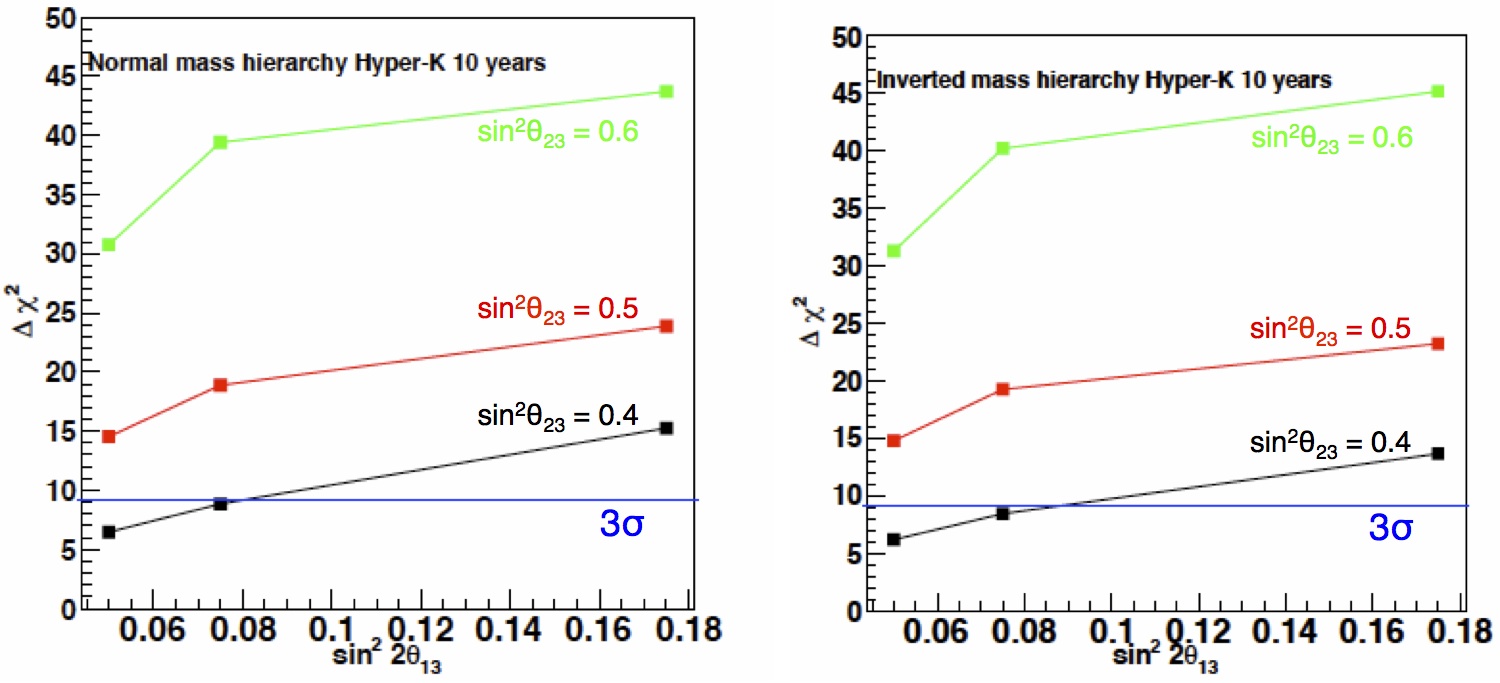}
\caption{Significance of discriminating the mass hierarchy when the true hierarchy is (Left) normal and (Right) inverted, as a function of sin$^22\theta_{13}$ and sin$^2\theta_{23}$, after 10 years of operation of HyperK~\cite{nonus:HyperK}.}
\label{HyperK:sen}
\end{center}
\end{figure}

With excellent particle identification, HyperK can tackle the mass hierarchy
problem using both disappearance of muon neutrinos and appearance of
electron neutrinos. 
The expected statistical significance in settling the mass hierarchy of HyperK
is presented in Fig.~\ref{HyperK:sen}.
For sin$^2\theta_{23}=0.5$ and sin$^22\theta_{13}=0.1$, the wrong hierarchy
could be ruled out at almost three standard deviations with about five years of running.  
The discriminating power is weakly dependent on the value of the CP-phase 
$\delta_{CP}$.  Qualitatively, a small $\delta_{CP}$ will yield better discrimination 
for a given value of sin$^2\theta_{23}$.

In January 2013, HyperK was included in the future plan of
KEK~\cite{atm:Nakaya}. If funded, the project would start in JPY2016,
with the two-year construction of access tunnels beginning in
2016. Excavation of the underground cavern is expected to start in
2018 and beneficial occupancy would take place in 2021. Operation of
the HyperK detector modules would start in January 2023.
The main concern with HyperK is cost, which is estimated at  
\$500-700M (US).  It is unclear if Japan will have sufficient funding for
this experiment.  Given the drive for liquid argon-based long-baseline
accelerator programs in both the US and Europe, it is unlikely that
foreign partners will contribute significantly to the cost.

\subsubsection{Atmospheric neutrino experiments using liquid argon detectors}

Besides providing excellent particle identification, 
liquid-argon detectors are superior to iron-scintillator calorimeters, water Cherenkov
detectors, or ice-based detectors in energy and angular resolutions over the energy
range of MeV to multi-GeV.  
Furthermore, since the energy of the hadronic component of the charge-current
interaction is measured, the energy of the incident (anti-)neutrino can be determined. 
One could even imagine magnetizing a large liquid-argon detector for charge
discrimination. Thus, liquid-argon detectors could be an ideal tool for addressing
the mass hierarchy problem with atmospheric neutrinos. 

\begin{figure}[t]
\begin{center}
\includegraphics[width=0.95\textwidth]{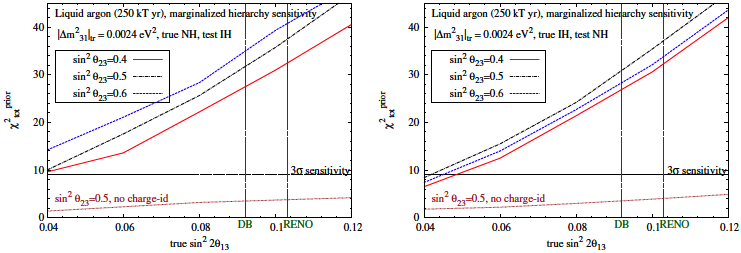}
\caption{Significance of discriminating the mass hierarchy as a function of sin$^22\theta_{13}$ and sin$^2\theta_{23}$ after an exposure of 250 kT-yr for liquid-argon detectors~\cite{atm:LAr}.}
\label{LAr:sen}
\end{center}
\end{figure}

Assuming the energy resolution to be $\sqrt{(0.003)^2+(0.15)^2/E_{had}}$ 
for hadronic showers, 0.01 for charged leptons in the GeV range, and the angular 
resolution to be 0.03 (0.04) rad for electrons (muons or hadronic
showers), an estimated 
sensitivity in delineating the mass order as a function of
sin$^22\theta_{13}$ and  
sin$^2\theta_{23}$ for an exposure of 250 kT-yr is given in
Fig.~\ref{LAr:sen}~\cite{atm:LAr}. This exposure corresponds to: 25
years of operation of a Phase-I sized LBNE detector; 7.5 years of the full-scale 34-kT detector (either would need to be located underground); 12.5 years of Phase I of
LBNO-GLACIER; and only 2.5 years of the full GLACIER detector. Discrimination power of over
$3\sigma$ for the neutrino mass hierarchy seems achievable using liquid argon detectors.

\section{Cosmology }\label{s:cosmo}

\newcommand{\summnu}{\Sigma m_\nu}
\newcommand{\FoM}{{\rm FoM}}
\newcommand{\cm}{\ \text{cm}}
\newcommand{\pc}{\ \text{pc}}
\newcommand{\Mpc}{\ \text{Mpc}}
\newcommand{\hMpc}{\ h^{-1}\text{Mpc}}
\newcommand{\hMpcc}{\ h^{-3}\text{Mpc}^3}
\newcommand{\ihMpc}{\ h\text{Mpc}^{-1}}
\newcommand{\iMpc}{\ {\rm Mpc}^{-1}}
\newcommand{\Ms}{\ M_\odot}
\newcommand{\hMs}{\ h^{-1} M_\odot}
\newcommand{\eh}[1]{\exp{\left[#1\right]}}
\newcommand{\tim}[1]{\times 10^{#1}}
\newcommand{\unit}[1]{\ \text{#1}}
\newcommand{\tb}[1]{\textcolor{blue}{#1}}
\newcommand{\tr}[1]{\textcolor{red}{#1}}
\newcommand{\tsc}[1]{#1}
\newcommand{\derd}{\,\mathrm{d}} 
\newcommand{\ddir}{\delta^\text{(D)}}
\newcommand{\dkron}{\delta^\text{(K)}}
\newcommand{\be}{\begin{equation}}
\newcommand{\ee}{\end{equation}}
\newcommand{\la}{\left\langle}
\newcommand{\ra}{\right\rangle}
\newcommand{\derivd}{\text{d}}
\renewcommand{\vec}{\bm}
\newcommand{\dqc}{\frac{\derivd^3q}{(2\pi)^3}}
\newcommand{\dkpc}{\frac{\derivd^3k'}{(2\pi)^3}}
\newcommand{\dqcp}{\frac{\derivd^3q'}{(2\pi)^3}}
\newcommand{\dqcpp}{\frac{\derivd^3q''}{(2\pi)^3}}
\newcommand{\cyc}{\, \text{cyc.}}
\newcommand{\Dz}{\Delta z}
\newcommand{\Dv}{\Delta v}
\newcommand{\Dx}{\Delta x}
\newcommand{\Dtheta}{\Delta \theta}
\newcommand{\bz}{\bar{z}}
\newcommand{\kms}{{\rm km~s^{-1}}}

\newcommand{\bdelta}{b_\delta}
\newcommand{\bdtwo}{b_{\delta^2}}
\newcommand{\bstwo}{b_{s^2}}
\newcommand{\msdtwo}{\left[\delta^2\right]}
\newcommand{\msstwo}{\left[s^2\right]}
\newcommand{\vtheta}{\mathbf{\theta}}
\newcommand{\vv}{\mathbf{v}}
\newcommand{\vnabla}{\mathbf{\nabla}}
\newcommand{\vk}{\mathbf{k}}
\newcommand{\vr}{\mathbf{r}}
\newcommand{\vx}{\mathbf{x}}
\newcommand{\br}{\mathbf{r}}
\newcommand{\Tr}{\mathrm{Tr}}
\newcommand{\vxperp}{\mathbf{x_\perp}}
\newcommand{\vkperp}{\mathbf{k_\perp}}
\newcommand{\vrperp}{\mathbf{r_\perp}}
\newcommand{\vR}{\mathbf{r}}

\newcommand{\lya}{Ly$\alpha$}
\newcommand{\lyb}{Ly$\beta$}
\newcommand{\lyaf}{Ly$\alpha$ forest}
\newcommand{\vdf}{{\mathbf \delta_f}}
\newcommand{\lr}{\lambda_{{\rm rest}}}
\newcommand{\PF}{$P_F^{\rm 1D}(k_\parallel,z)$}
\newcommand{\bF}{\bar{F}}

\newcommand{\mnras}{{\em Mon. Not. Roy. Astron. Soc. }}
\newcommand{\apjl}{{\em Astrophys. J. Let. }}
\newcommand{\apj}{{\em Astrophys. J. }}
\newcommand{\apjs}{{\em Astrophys. J. Sup. }}
\newcommand{\jcap}{{\em JCAP }}
\newcommand{\physrep}{{\em Phys. Rept. }}
\newcommand{\aap}{{\em Astron. Astrophys. }}
\newcommand{\aj}{AJ }
\newcommand{\pasp}{PASP}
\newcommand{\prd}{PRD}
\def\pr{{Phys.\ Rev.\ }}
\def\astropart{{Astro-particle Phys.~}}
\def\rvmp{{Rev.\ Mod.\ Phys.\ }}

\def\pvm#1{[PM: {\it #1}] }
\def\af#1{[AF: {\it #1}] }

Forseeable  cosmological measurements can determine the sum of the 
neutrino masses, but not directly the separate contributions of the three species. However, if the hierarchy is normal with the sum of 
masses near the minimum of about 0.06 eV,
strong evidence for this should accumulate through the next decade, 
as expected large-scale structure/gravitational lensing experiments 
(Euclid, LSST, MS-DESI/BigBOSS, etc) come online. However, should the sum turn out to be about 0.10 eV or more, no conclusion could be drawn about the hierarchy.
Fortunately, the constraints on the neutrino masses are a byproduct of
experiments primarily motivated by dark energy studies, so they
will happen independent of neutrino-science considerations.

Neutrinos decouple from other particles very early and become
non-relativistic quite late, $z_{\rm nr}\sim 83~(m_\nu / 0.05~ {\rm
eV})$, relative to the time CMB was imprinted, $z=1100$.  The masses
of the neutrinos affect the fundamental cosmological observable, the
power spectrum.  The power spectrum is the Fourier transform of the
two-point correlation function between mass density at one point and
mass density at another point.  To be sure, we are talking about
cosmological separations.  The density of galaxies can stand as a
proxy for the density of matter, both ordinary and dark.

\begin{figure}[bp]
\vspace{-0.6cm}
  \centering
\includegraphics[width=0.6\textwidth]{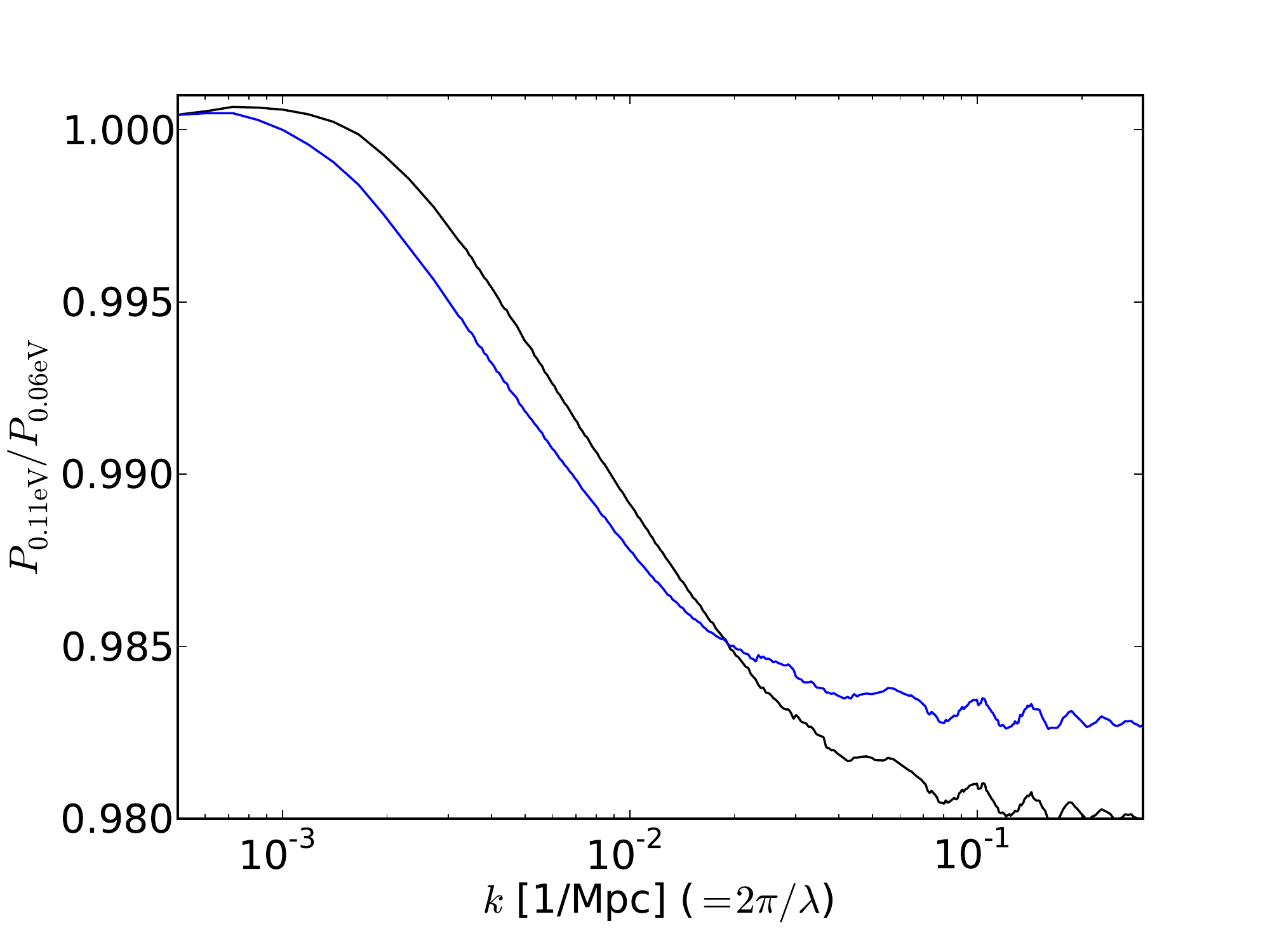}
\vspace{-0.2cm}
\caption{
Ratio of linear power for $\summnu=0.11 {\rm eV}$ in the inverted (black) or
normal (blue) hierarchy, to $\summnu=0.06{\rm eV}$ in the normal hierarchy.
}
\label{fig:powerratio}
\end{figure}

Fig.~\ref{fig:powerratio} shows the
ratio of power for $\summnu=0.11$ eV to $\summnu=0.06$ eV as a function of 
$k$, the Fourier transform variable measured in ${\rm Mpc}^{-1}$.
Fig.~\ref{fig:powerratio} illustrates the small differences in power spectra for different 
distributions of
the total mass between particles, but these are not expected to be observable
on the time scale under consideration~\cite{2006PhRvD..73l3501S}.
The dominant effect that can be measured is the suppression in power shown
in Fig.~\ref{fig:powerratio}. Note, however, that it is not necessary to rely 
entirely on measuring the effect on the {\it shape} of the power spectrum shown
in Fig.~\ref{fig:powerratio}. Redshift-space 
distortions, lensing, and other methods can be used to measure the overall 
suppression
in amplitude relative to the CMB measurement of the primordial perturbations 
(this relies, however, on GR calculations of the growth of structure).

There are numerous ways to detect the large-scale power suppression by 
neutrinos, with varying degrees of statistical power and expected 
systematic uncertainties. 
The focus here is on galaxy redshift surveys, which appear to offer the best 
combination of statistical power and projected control of systematic effects.

The first major redshift survey that may seriously test the hierarchy is 
MS-DESI (approximately equivalent to BigBOSS \cite{2011arXiv1106.1706S}). 
Led by LBNL, the survey will cover $\sim 14000$ square 
degrees, 
taking spectra of $\sim 20$ million galaxies and quasars at $z\sim 0-3$. 
MS-DESI is likely to run on the Mayall 4~m telescope at Kitt Peak National 
Observatory near Tucson, AZ, over a $\sim5$ year period from 
2018 to 2022. It is possible that the spectrograph could then be moved to the
twin Blanco telescope in Chile to cover another $\sim 10000$ square degrees.

Another big redshift survey that appears certain to happen is the 
European-led Euclid satellite \cite{2012SPIE.8442E..0ZA}. 
Euclid will measure redshifts for 
$\sim 50$ million galaxies over $\sim 15000$ square degrees.  
The target survey period is $\sim 2020-2026$. In addition to redshifts, 
Euclid will do imaging for gravitational lensing measurements.

While not a redshift survey (meaning it has limited radial resolution), LSST
will be a major US cosmological experiment running in the 2022-2032 time 
period \cite{2009arXiv0912.0201L}.
It will image $\sim 20000$ square degrees for gravitational lensing,
which can add  power 
to the redshift survey measurements of neutrino masses.
Before LSST, and even before MS-DESI, there will be a smaller, 5000 sq. deg.
lensing survey, DES
(\url{http://www.darkenergysurvey.org}). 
We add DES to all projections that do not include LSST. 

Another experiment that {\em might} happen is the NASA/WFIRST satellite, 
intended for a 2022 launch \cite{2012arXiv1208.4012G}. 
It would be qualitatively similar to MS-DESI and
Euclid, but complementary ($i.e.$ adding statistical power) because it would
take a different strategy of going deeper over a smaller area.

Projections are given in Table~\ref{tablemnulensing}.
\begin{table}[b]
\caption{ 
Potential constraints on 
$\Sigma m_{\nu}$ for the minimal parameter set.
``P'' means Planck CMB data has been included.
Numbers in parentheses are maximum $k$ used for galaxy clustering, in units of
$\iMpc$. 
BigBOSS14 / 24 means 14000 or 24000 square degrees (BigBOSS is later shortened to BB). 
From the Euclid satellite (sometimes shortened to Euc) 
only the redshift space clustering information is used, not lensing.
The $\sigma_{0.04~{\rm eV}}$ column shows the detection significance for a
mass difference of $0.04$ eV, corresponding to the hierarchy detection 
significance {\em if} the total mass is absolutely minimal.
DES and LSST stand for the lensing and galaxy clustering components of these 
surveys. 
\label{tablemnulensing}}
\vspace{-0.3cm}
\begin{center}
\begin{tabular}{lcccc}
\hline
$ $ & $k_{\rm max}$ & $\sigma_{\Sigma m_{\nu}}$ & $\sigma_{0.04~{\rm eV}}$ & 
Year \\
$ $ & $\left[\iMpc\right]$ & $\left[{\rm eV}\right]$ & &  \\
\hline
P+BigBOSS14+DES & 0.07 & $0.021$ & 1.9 & 2022 \\
P+Euclid+DES & 0.07 & 0.019 & 2.1 & 2026 \\
P+BigBOSS24+DES & 0.07 & $0.019$ & 2.1 & 2026 \\
P+BB24+Euc+DES & 0.07 & $0.016$ & 2.5 & 2026 \\
P+BB24+Euc+LSST & 0.07 & $0.014$ & 2.9 & $\lesssim 2030$  \\
P+BB14+DES & 0.14 & $0.017$ & 2.4 & 2022 \\
P+Euclid+DES & 0.14 & $0.015$ & 2.9 & 2026 \\
P+BB24+DES & 0.14 & $0.015$ & 2.7 & 2026 \\
P+BB24+Euc+DES & 0.14 & $0.013$ & 3.1 & 2026 \\
P+BB24+Euc+LSST & 0.14 & $0.011$ & 3.6 & $\lesssim 2030$ \\
\hline
\end{tabular}
\end{center}
\vspace{-0.3cm}
\end{table}
As shown, cosmology can generally achieve or at least approach the 
$0.01$ eV RMS error level needed to probe the hierarchy. 
These calculations are consistent with similar projections made for
Euclid~\cite{2013JCAP...01..026A}.

As seen in Table~\ref{tablemnulensing}, cosmology is 
likely to reach the $2-2.5~\sigma$ level for distinguishing the minimal normal 
from minimal inverted hierarchies by the end of MS-DESI (in the North at 
least) in 2022. At that point
Euclid and LSST will be running, and significance will accumulate, probably 
reaching the $\sim 3.5~\sigma$ level around 2026 (LSST will only be half-done 
at that
point, but most of the gain from it will likely be extractable already).
Whether these measurements will determine the neutrino mass hierarchy is contingent on the hierarchy being normal and the masses minimal.

\section{Conclusions}
This report has considered several approaches to a measurement of the neutrino mass hierarchy, including long-baseline experiments (Section~\ref{s:lb}), reactor neutrinos (Section~\ref{s:reac}), atmospheric neutrinos (Section~\ref{s:atm}) and cosmology (Section~\ref{s:cosmo}).
It was outside the scope of this study to evaluate the importance of determining the neutrino mass hierarchy relative to other opportunities on a similar timescale. 


The question of the confidence level needed to ``decisively'' determine the mass hierarchy is a subjective one.  One approach is to consider the impact of such a determination, for example on the field of neutrinoless double beta decay.  Were the hierarchy determined to be inverted, the next generation of experiments, with the ability to cover the inverted-hierarchy region of parameter space, will become decisive.  To motivate these experiments, a 2--3$\sigma$ indication of an inverted hierarchy would be sufficient.  However, before claiming a Dirac nature of neutrinos on the basis of no signal in such experiments, a much higher significance determination of the hierarchy would be required.  Likewise, if cosmological and terrestrial determinations disagree, a high significance determination will be needed before interpreting such a discrepancy as a violation of cosmological theories.

We note that many current studies rely on simplified calculations of confidence levels, assuming $\chi^2$ distributions of test statistics. These assumptions are not always valid, and care must be taken to correctly interpret quoted significance levels, and future experimental results.  In the following, we quote sensitivities as stated by the authors in each case, with the here-noted caveat that these are not always directly comparable.

\begin{figure}[!b]
\begin{center}
\includegraphics[width=0.95\textwidth]{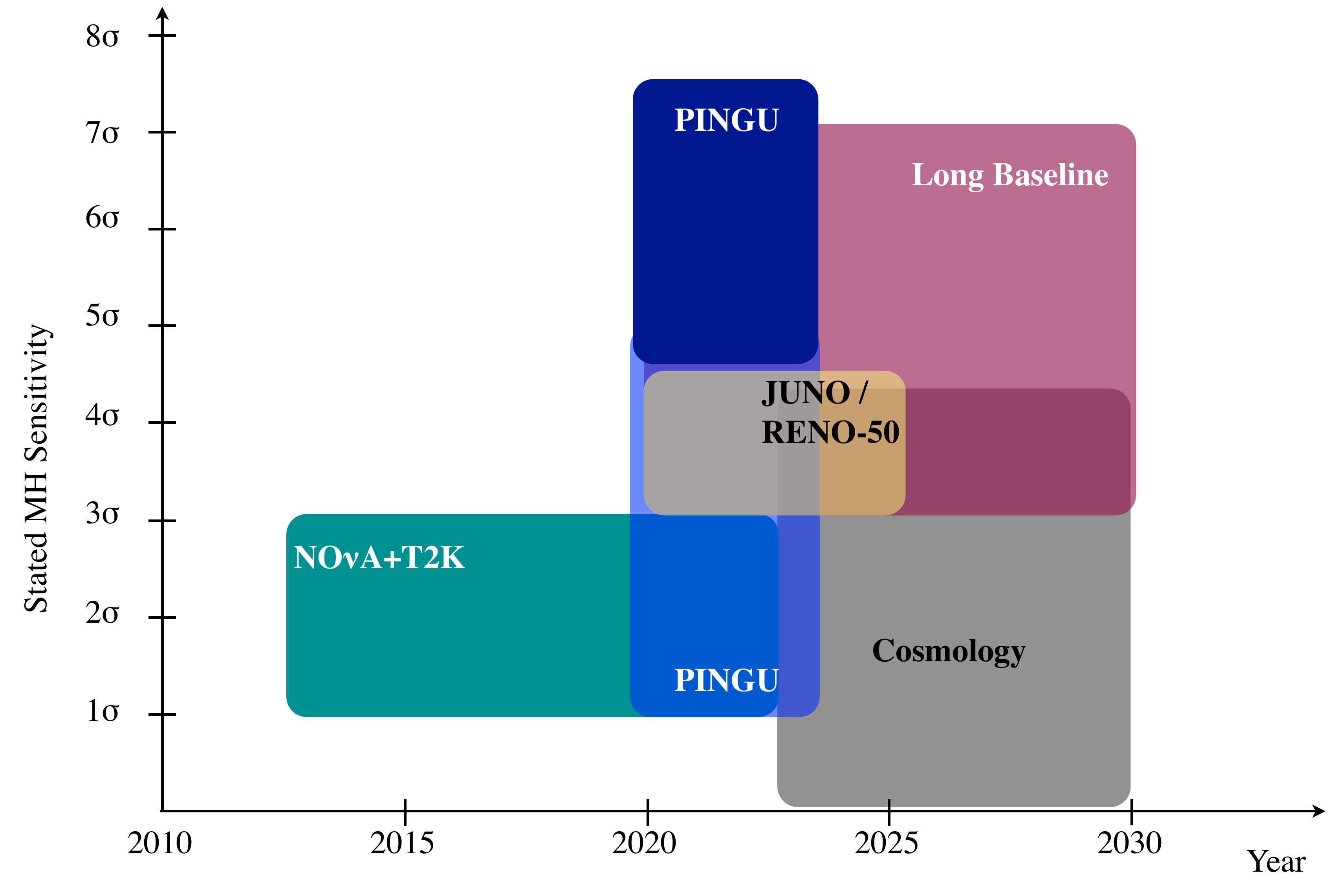}
\caption{ Summary of sensitivities to the neutrino mass hierarchy for various experimental approaches, with timescales, as claimed by the proponents in each case.  In the case of PINGU, for which multiple studies exist, the proponents' stated sensitivity~\cite{atm:pingu} is shown in the dark blue region, with the larger blue region representing the independent analysis of Ref.~\cite{atm:Winter}.  One difference between the two is the consideration of a wider range of oscillation parameters in~\cite{atm:Winter}  (see Section~\ref{s:atm} for details).
%
The vertical scale of each region represents the spread in the expected sensitivity after the full exposure.  We do not attempt to project the natural increase in sensitivity over time. Note: the ``long baseline'' region represents the inclusive range of sensitivities for individual long-baseline experiments (LBNE, HyperK, and LBNO) rather than a combined sensitivity.
\label{f:bubble}}
\end{center}
\end{figure}

Table~\ref{t:MH2} summarizes the potential sensitivity, timescale, and open questions for each approach.
The claimed sensitivity and timescale are also summarized in Fig.~\ref{f:bubble}.  The spread in the displayed sensitivity includes both the projected experimental uncertainties (where evaluated by the proponents), and also the underlying limitations due to currently unknown physics parameters, such as the CP-violating phase and the overall neutrino mass scale.

Of the experiments we have considered, only the long-baseline 
experiments (LBNE in combination with T2K/\NOvA, the European-based LBNO, and HyperK's combined long-baseline and atmospheric data) have demonstrated the
ability to measure the mass hierarchy with a statistical significance
of at least $4\sigma$ regardless of the
value of $\delta_{CP}$ and other oscillation parameters. This statement is not without some
caveats. So far, LBNE sensitivities are based on parameterized
(``toy'') simulations, rather than a complete understanding of
the complicated physics involved in neutrino-nucleus interactions at low
energies, or even detailed reconstruction algorithms in LAr
TPC. Addressing these deficiencies is very important. Also, there is a
finite risk that with the long
timescale of LBNE construction (the 10-year run is planned to start in
about 2020), some other experiment(s) will determine the hierarchy
before LBNE produces a result. This would put a costly project in an
unfortunate position to confirm someone else's discovery. 

In addition to LBNE, both Japan and Europe are considering
long-baseline neutrino oscillation experiments. HyperK is a
Megaton-scale water Cherenkov detector, to be constructed in the
Kamioka mine, with the same baseline (about 300 km) from the neutrino
source at J-PARC as T2K. With the short baseline, HyperK needs
atmospheric neutrino data to break the degeneracy between the matter
effects and $\delta_{CP}$, and to determine the mass hierarchy with
a significance of at least $3\sigma$. The timescale and costs for
HyperK are comparable to that of LBNE. 

European long-baseline projects (LAGUNA-LBNO) involve an intense neutrino
source at CERN, a near detector, and a (phased) 
100 kT underground LAr detector at 
Pyh\"asalmi in Finland, at a baseline of 2300~km.  The long
baseline, large detector mass, underground location, near detector,
and a broad-band neutrino beam from a 2~MW proton source make LAGUNA-LBNO
an ultimate neutrino oscillation experiment, with outstanding
sensitivity to both the neutrino mass hierarchy and
$\delta_{CP}$. However, the timescale, costs, and priority to host such
an experiment in Europe are not well defined at present. 

JUNO is a 20~kT liquid scintillator detector to be located at
the solar oscillation maximum, approximately 60~km away from two
nuclear power plants in China. This experiment plans to exploit subtle
distortions in the neutrino energy spectrum sensitive to the sign of
$\Delta m^2_{32}$.  RENO-50 is a similar experiment proposed in South Korea.  
This measurement is extremely challenging, both technologically and in terms of required experimental precision.  
Successful determination of the neutrino mass hierarchy depends
critically on achieving unprecedented energy resolution
and controlling the energy scale systematics to about 1\%. 
If these challenges can be met, then the hierarchy could be measured to $>3\sigma$ ($>4\sigma$) assuming current (future $1.5\%$-level) uncertainties on $\Delta m^2_{32}$.  

Another proposed experiment that could in principle rapidly determine
the neutrino 
mass hierarchy is PINGU, a dense array of phototube strings in the
middle of the IceCube detector at the South Pole. The measurement relies
on polar-angle dependent distortions in the neutrino energy
spectrum due to matter-induced neutrino
oscillations. With the copious samples of upward-going electron and
muon neutrinos and large target mass, PINGU 
has excellent statistical sensitivity to the mass hierarchy.
However,
disentangling the hierarchy-dependent effects from the data requires
an excellent energy resolution and understanding the energy scale
systematics in the detector.  The sensitivity also depends on Nature's choice of oscillation parameters, including the hierarchy itself.
As evaluated by the proponents,  the sensitivity of PINGU could be $\sim 4.8$-$7.6\sigma$ with 5 years of data.  
While the potential for a decisive,
inexpensive, and fast measurement is there, we do not feel the
systematic issues have been fully addressed by the proponents to
date. These studies are ongoing as of this writing. 
An independent study finds a larger range of potential sensitivities ($1-5\sigma$), including a lower bound for the case of an inverted hierarchy, which can be partially mitigated by combination with T2K/\NOvA\ data. 

Future dark energy experiments such as MS-DESI (formerly BigBOSS),
Euclid, and LSST, in combination with the Cosmic
Microwave Background measurements, have the capability to measure the
sum of the neutrino masses with a precision relevant to the neutrino
mass hierarchy.  Should the hierarchy 
be normal and the neutrino masses minimal, global cosmological fits
could discern the neutrino mass hierarchy on a timescale comparable
to that of LBNE. Early indication can be obtained from  currently
running or near-future measurements. These measurements
rely on the present best cosmological model ($\Lambda$CDM), which is
supported by the wealth of astrophysical data. 

While any individual measurement in the next decade may be susceptible to large uncertainties, either statistical or systematic, a combination of results from multiple experiments and techniques could yield a greater confidence in a determination of the mass hierarchy.  We have not explicitly considered the potential improvements from such combinations, although independent studies exist~\cite{atm:Winter, combo}.  It is  our hope that several of the experiments here described will be pursued.
  Ultimately, a cross check between multiple techniques will be required for any decisive, unambiguous determination of the mass hierarchy.

\begin{sidewaystable}[!htdp]
\begin{center}
\caption{Comparison of mass hierarchy experiments.  ``NH'' refers to the normal hierarchy. 
In the case of PINGU, for which multiple studies exist, both the sensitivities claimed by the proponents~\cite{atm:pingu} and the independent analysis of Ref.~\cite{atm:Winter} are presented.
\label{t:MH2}}
\begin{tabular}{ m{3cm}p{3cm}p{3.cm}p{6cm}p{0cm}}
\hline \hline
{\bf Technique} $\qquad$ Experiment  &  \centering{MH sensitivity} & \centering{Timescale for  results}    &  \centering{Major concerns} &   \\
\hline 
\bf Accelerator \\
T2K+\NOvA		&  \centering{1--3$\sigma$}
& \centering{$\sim$2020}		& \centering{Non-optimal baselines} & \\
HyperK 	&   \centering{$>3\sigma$ with atmospheric} 								& \centering{$\sim$2030} 		 &  \centering{Likelihood of going ahead} &   \\
LAGUNA-LBNO 	&   \centering{$>5\sigma$} 								& \centering{$\sim$2025} 		 &  \centering{Likelihood of going ahead} &    \\
 LBNE phase I 	&  \centering{3$\sigma$ (2$\sigma$) over 80\% (100\%) of $\delta_{CP}$}																& \centering{$\sim$2030} 	 	&    \\
 LBNE 34~kT 	&  \centering{$>6\sigma$}																& \centering{$\sim$2030} 	 	& \centering{Assuming Phase-I timescale}  & \\
\hline
\bf Reactor \\
 JUNO 		&  \centering{3--4$\sigma$} 				& \centering{$\sim$2025}  		&  \centering{Energy scale \& resolution} & \\
\hline 
\bf Atmospheric \\
 PINGU 			&  \centering{$4.8-7.6\sigma$~\cite{atm:pingu}  $1-5\sigma$~\cite{atm:Winter} (dependent on oscillation parameters)}
 & \centering{$\sim$2023} 		&  \centering{Energy scale \& resolution, correlated
 parameters} & \\
\hline
\bf Cosmology \\
 All 
&  \centering{0--4$\sigma$ $\qquad\qquad$(3--4$\sigma$ for NH and minimal masses)}
& \centering{$\sim$2025}   & \centering{Measures sum of masses -- can only determine hierarchy for minimal masses} & \\
\hline \hline
\end{tabular}
\end{center}
\end{sidewaystable}%

\section{Acknowledgments}
We would like to thank Mark Strovink for contributions to this document, and many members of the Nuclear Science and Physics divisions at LBNL for enlightening conversations.

\end{document}